\documentclass[reprint,aps,pra,showkeys,showpacs]{revtex4-1}  

\usepackage{graphicx}
\usepackage{epstopdf}
\usepackage{epsfig}
\usepackage{xcolor}   
\usepackage{amsmath}

\begin{document}

\title{Josephson Dynamics of 2D Bose-Einstein Condensates in Dual-Core Trap: Homogeneous, Droplet-Droplet, and Vortex-Vortex Regimes }
\author{Sherzod R. Otajonov$^{1, 2}$}
\author{Fatkhulla Kh. Abdullaev$^{1, 3}$}
\address{$^1$ Uzbekistan Academy of Sciences S. A. Azimov Physical-Technical Institute, Chingiz Aytmatov Str. 2-B, 100084, Tashkent, Uzbekistan}
\address{$^2$ National University of Uzbekistan, Department of Theoretical Physics, 100174, Tashkent, Uzbekistan}
\address{$^3$ Institute of Theoretical Physics, National University of Uzbekistan, 100174, Tashkent, Uzbekistan}

\begin{abstract}
The dynamics of a two-dimensional Bose-Einstein condensate mixture, loaded into a dual-core trap, when the beyond-mean-field effects are taken into account, are considered.
The effects of quantum fluctuations are described by the Lee-Huang-Yang correction terms in the extended coupled Gross-Pitaevskii equations. 
The spatially uniform and inhomogeneous condensate cases are studied. In the first case, the parameter regimes associated with macroscopic quantum tunnelling, macroscopic self-trapping, and revival-like localisation dynamics are found. The Josephson oscillation frequencies for both the zero-phase and the $\pi$-phase modes are derived. As the total atom number is varied, the dynamics exhibit a nontrivial bifurcation structure: along the zero-phase branch, two successive pitchfork bifurcations generate bistability and hysteresis, while the $\pi$-phase branch shows a single pitchfork bifurcation.

In the second case, the Josephson dynamics for quantum droplets and vortices are investigated. The analytical predictions for the oscillation frequencies of the atomic population between quantum droplets are found, and results are validated by direct numerical simulations of the system of coupled extended GP equations. The existence of the Andreev-Bashkin nondissipative drag through simulations of droplet-droplet interactions is shown. 
The Josephson dynamics of vortex states are studied. It was found that vortices with topological charge $S$ and sufficiently small particle number are typically unstable, breaking up into $S+1$ (and occasionally $S+2$) fundamental, non-vortical fragments, with the time to breakup increasing as the particle number grows. It is also found that unstable asymmetric vortices exhibit the splitting and/or crescent-like instability. For vortices with sufficiently large norms, long-time simulations confirm robust stability against small perturbations. In this stable regime, the properties of Josephson oscillations and Andreev-Bashkin-type entrainment for vortex states with charges $S=1, 2$, and $3$ are investigated.
\end{abstract}
\maketitle

\section{Introduction}
\label{intro}

Investigation of Josephson phenomena in atomic Bose-Einstein condensates (BECs) is now an active area of research. The fundamental interest represents the demonstration of macroscopic quantum effects in ultracold quantum gases. The theoretical description is mainly based on the mean-field approach for BECs loaded into double-well or double-core trapping potentials~\cite{Raghavan1999}. Experimentally, macroscopic quantum tunnelling and self-trapping have been observed for a BEC in a double-well potential in Ref.~\cite{Albiez}.

The effects of quantum fluctuations are addressed by going beyond the mean-field approximation~\cite{LHY}. The corresponding theory is based on the extended Gross-Pitaevskii equation (GPE) with Lee-Huang-Yang (LHY) correction terms~\cite{Petrov-2015, PA-2016, Rev1}. The LHY correction introduces additional nonlinear interaction terms into the GPE, whose form depends on the dimensionality of the system and on the underlying mean-field interactions. For example, for a Bose-Bose mixture in 3D the correction to the energy density is repulsive and scales as $\sim n^{5/2}$, where $n$ is the condensate density; in 2D it has a logarithmic form $\sim n^{2}\ln n$ (attractive at low and repulsive at high densities); and in 1D it yields an effective attraction $\sim -n^{3/2}$. If the residual mean-field interactions are sufficiently small so that the contribution of the quantum-fluctuation terms becomes comparable, quantum stabilization against collapse becomes possible~\cite{Petrov-2015, PA-2016}. A key prediction is the existence of quantum droplets (QDs), which have been observed experimentally in Bose-Bose mixtures~\cite{BB_mixture} and in dipolar BECs~\cite{DP-BEC}.

The influence of quantum fluctuations on Josephson phenomena can be essential, because the dynamics of BECs in such systems are highly sensitive to the effective nonlinearity produced by the combined action of the residual mean-field and LHY interaction terms. The role of quantum fluctuations in macroscopic quantum tunneling and self-trapping in double-well potentials has therefore become a subject of active investigation. In Refs.~\cite{QF1, Abdullaev-2023, Abdullaev-2024, Wysocki2024, Liu-2024}, quasi-one-dimensional BECs in double-well traps were analyzed, and quantum revivals, LHY-fluid tunneling, and self-localization driven solely by quantum fluctuations were reported.

A separate problem concerns Josephson phenomena between quantum droplets and vortices. In 3D, the combined action of Josephson oscillations and droplet motion can lead to nontrivial effects, such as the Andreev-Bashkin drag~\cite{Pylak}. Analogues of this effect for quantum droplets in one- and two-dimensional two-core traps were analyzed in Refs.~\cite{Abdullaev-2025, Otajonov2026}. Most of those studies considered very narrow 1D and 2D traps, which lie outside the range of parameters of current experiments.

It should also be noted that comparatively little is known about Josephson oscillations between vortices. Mass exchange between two rotating, massive vortices in a two-component Bose-Einstein condensate was investigated in Ref.~\cite{Bellettini-2024}, where macroscopic quantum tunnelling was shown to enable realisation of a bosonic Josephson junction. Josephson oscillations between 2D chiral solitons of semivortex and mixed-mode types in spin-orbit coupled BECs loaded into a two-core trap were studied in Ref.~\cite{Chen-2020}, and oscillations between solitons in spin-orbit coupled BECs were considered in Ref.~\cite{Abdullaev-2018}.

The subject of this work is to consider Josephson phenomena between quantum droplets and vortices in a two-core, relatively thick two-dimensional trap, which allows the use of a quasi-2D description in an experimentally relevant parameter domain~\cite{Shamriz-2020, Lin-2021, Santos-2025}.

The structure of the work is as follows: In Sec.~\ref{sec:model}, we introduce the extended 2D GPE model and its dimensionless reduction. In Sec.~\ref{sec:dimer}, we derive an effective Hamiltonian for the spatially uniform state, obtain expressions for the Josephson frequencies for the zero- and $\pi$-phase modes, classify oscillatory versus running-phase (self-trapped) dynamics, and derive separatrix conditions in terms of the critical linear coupling (or the initial imbalance). In Sec.~\ref{sec:bifurcation}, we analyse how quantum fluctuations, two-body interactions, and linear coupling generate symmetry-breaking pitchfork bifurcations. In Sec.~\ref{sec:QDinterac}, we study Josephson dynamics of coupled quantum droplets using a variational approach (VA) validated against numerical simulations, determine the frequencies for the zero- and $\pi$-phase regimes, and analyse droplet-droplet interactions numerically, revealing $\pi$-mode separation and Andreev-Bashkin drag effects. In Sec.~\ref{sec:VQDinteraction}, we address vortex states by delineating instability-driven fragmentation at low norms and a robust stability domain at higher norms; within the unstable and stable regimes, we demonstrate Josephson population transfer for vortex charges $S=1,2,3$ and Andreev-Bashkin type entrainment of moving vortices. In Sec.~\ref{sec:estimations}, we estimate experimentally relevant parameters, and we conclude in Sec.~\ref{sec:Conc}.

\section{The model}
\label{sec:model}

	When studying the Josephson effect in BECs, two canonical experimental realizations of coupling are typically employed: internal and external (spatial) coupling. In the internal-coupling scheme, two hyperfine states of the same atomic cloud are coherently coupled by a radiofrequency or microwave drive, giving rise to Rabi oscillations. In this case there is no spatial barrier, and population transfer between the components is often described, by analogy, as “tunneling” between internal states. In the external (spatial) coupling scheme, the condensate is split by a narrow barrier (e.g., in a double-well potential), and atoms tunnel through the barrier between the two spatially separated BECs.
	Here, we consider the dynamics of a two-component BEC confined in a spatially coupled dual-core, disk-shaped trap, taking into account beyond-mean-field effects via the LHY correction term, which represents quantum fluctuations.  We focus on the symmetric configuration, in which the atom numbers and the coupling constants of the two components are equal in each core. The model we consider can be implemented experimentally using a quasi-two-dimensional double-well potential. In this configuration, the two wells play the role of parallel coupled cores, and the linear tunnelling coefficient $\tilde{\kappa}$ is controlled by the barrier between them. In particular, by adjusting the height or width of the separating potential barrier, one can effectively tune the strength of interwell tunnelling: a higher barrier leads to weaker coupling, while a lower barrier results in stronger coupling. This provides a practical mechanism for exploring different dynamical regimes of the system in a controllable manner. Figure~\ref{fig:model} provides a schematic illustration of the system under consideration, including the two parallel pancake-shaped traps, their linear coupling. In this case, the system is governed by a pair of weakly linearly coupled modified Gross-Pitaevskii equations of the form~\cite{Petrov-2015,PA-2016,Malomed-2019,Chen2017}:

\begin{figure*}[htbp]
  \centerline{ \includegraphics[width=5.5cm]{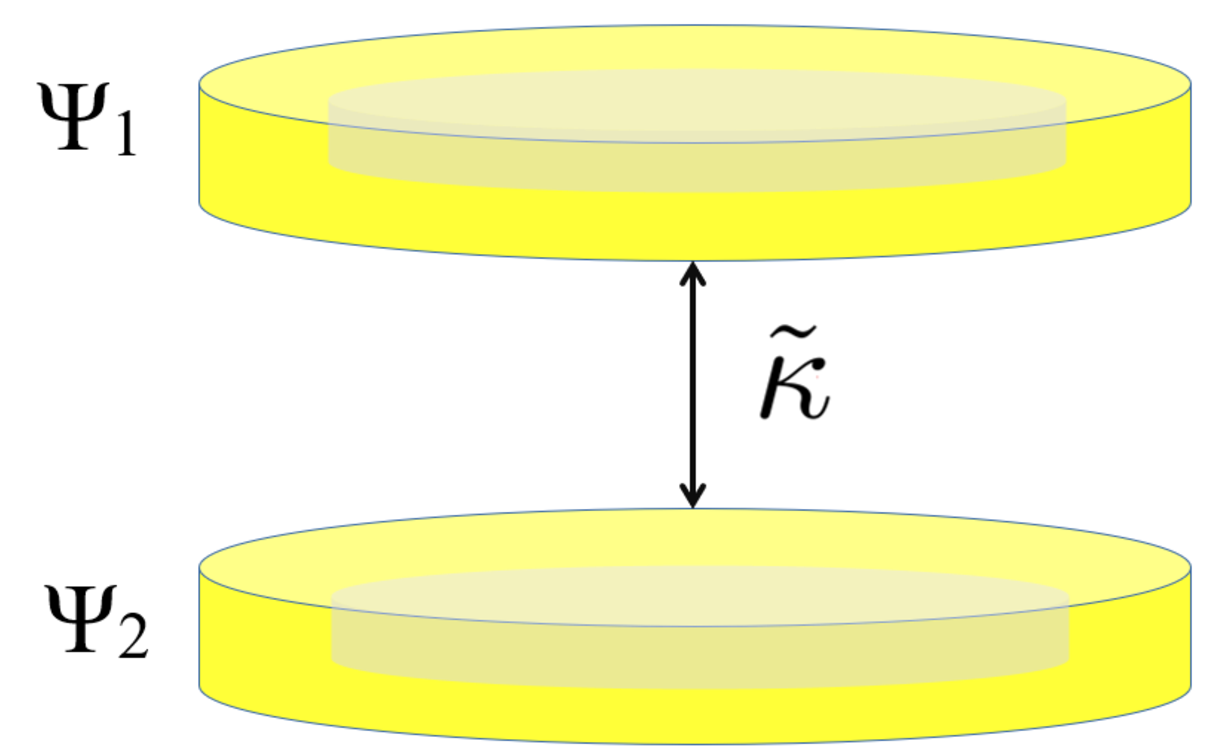}}
\caption{Schematic illustration of the model. The yellow regions represent two parallel pancake-shaped core traps confining atomic clouds that form BECs. In each core, the same atomic species occupies two distinct hyperfine states. In the symmetric configuration, the wavefunctions in the first core satisfy $\Psi_1$, whereas those in the second core satisfy $\Psi_2$. The two cores are coupled through a linear tunnelling term with strength $\tilde{\kappa}$.}
\label{fig:model}
\end{figure*}

\begin{eqnarray}
 \nonumber
&i \hbar \Psi_{j, \mathrm{T}} + \cfrac{\hbar^2}{2 m_0}\,\,\nabla^2 \Psi_{j} +
\cfrac{2\sqrt{2\pi}\,\delta a\,\hbar^{2}}{l_{z}\,m_{0}}
\,\, |\Psi_{j}|^2 \Psi_{j} - \\
&\cfrac{512\,a^{5/2}\,\hbar^{2}}{3\sqrt{5}\,\pi^{1/4}\,l_{z}^{3/2}\,m_{0}}
 |\Psi_{j}|^3 \Psi_{j} + \tilde{\kappa} \,\Psi_{3-j}=0,
\label{eq:DimGPE}
\end{eqnarray}
here $\Psi_j(X, Y, T)$ ($j=1,2$) are the macroscopic wave functions of the two cores, $T$ is time, and the subscript $T$ denotes the time derivative. The operator
$\nabla^2=\partial_X^2+\partial_Y^2$ is the two-dimensional Laplacian. The atomic mass is $m_0$, and
$l_{z}=\sqrt{\hbar/(m_0\omega_z)}$ is the transverse harmonic-oscillator length associated with the tight confinement frequency $\omega_z$. The coefficient $\tilde{\kappa}$ accounts for the linear inter-core (Josephson) coupling due to tunnelling. The parameter $\delta a=-|a_{12}|+\sqrt{a_{11}a_{22}}$ represents the residual two-body (mean-field) interaction, expressed via the $s$-wave scattering lengths: $a_{11}=a_{22}\equiv a$ are the intra-species scattering lengths, while $a_{12}=a_{21}$ is the inter-species one. We focus on the physically relevant regime of weak intra-species repulsion ($g_{11},g_{22}>0$) combined with inter-species attraction ($g_{12}<0$), which, together with the LHY correction, supports self-bound quantum droplets and vortices. Derivation of the model equation is provided in Appendix~\ref{appen}.

Introducing the dimensionless variables $\psi=\Psi/\psi_s$, $t=T/t_s$ and $(x,y)=(X,Y)/r_s$, where the scaling parameters $\psi_s$, $t_s$, and $r_s$ are chosen as: 
$$
\psi_{s}=\cfrac{3 \sqrt{5} \pi^{3/4} l_{z}^{1/2} a g }{128 \sqrt{2} q a^{5/2}}, \qquad r_{s}=l_z,
$$

$$
t_s=\omega_{z}^{-1}, \qquad \kappa=\frac{\tilde{\kappa}}{\hbar \omega_z}\, .
$$
Using the scaling introduced above, Eq.~(\ref{eq:DimGPE}) can be written in a dimensionless form:
\begin{equation}
i \psi_{j,t} + \frac{1}{2}\nabla^2 \psi_{j} + q |\psi_{j}|^2 \psi_{j} - g |\psi_{j}|^3 \psi_{j} + \kappa \psi_{3-j}=0, 
\label{eq:gpe}
\end{equation}
where $\nabla^2=\partial_x^2+\partial_y^2 \,$, $\kappa$ is the dimensionless linear-coupling coefficient, and the parameters $q$ and $g$ quantify, respectively, the residual mean-field and the LHY contribution strength. The case $q=0$ ($\delta a=0$) corresponds to the LHY fluid~\cite{LHYfluid} loaded into a two-core trap.

\section{Homogeneous Condensate}
\label{sec:Homo}

\subsection{Dimer limit}
\label{sec:dimer}

We first consider a spatially uniform BEC. In this homogeneous setting, the derivative terms in Eqs.~(\ref{eq:gpe}) can be neglected, which yields the LHY dimer equations:
\begin{equation}
i \psi_{j,t} + q |\psi_{j}|^2 \psi_{j} - g |\psi_{j}|^3 \psi_{j} + \kappa \psi_{3-j}=0, 
\label{eq:LHYdimer}
\end{equation}
This reduction is equivalent to projecting Eqs.~(\ref{eq:gpe}) onto the zero-momentum mode (or, in a two-core trap, onto the lowest spatial mode in each core), following the same logic that underpins the standard two-mode model of a bosonic Josephson junction~\cite{Raghavan1999}. 

Expressing the fields in the form, $\psi_1=A_1 e^{i\theta_1}, \, \psi_2=A_2 e^{i\theta_2}$, and define the local norms of the components $N_i=A_i^2$ with total local norm $N=N_1+N_2$. Introducing the population imbalance $Z$ and the relative phase $\theta$, $Z=(N_2-N_1)/N, \theta=\theta_2-\theta_1$, the governing dynamics can be written as the following system of equations:
\begin{eqnarray}
\nonumber
&Z_t=2 \kappa \sqrt{1-Z^2}\sin(\theta), \\
\nonumber
&\theta_t =-\cfrac{2 \kappa Z}{\sqrt{1-Z^2}}\cos(\theta)+ q N Z\\
&  -g(N/2)^{3/2}\left[(1+Z)^{3/2}-(1-Z)^{3/2} \right] \equiv F(Z).
\label{eq:dimerZtThetat}
\end{eqnarray}
In addition to the relative phase, one may introduce the total phase
$\Phi=\theta_1+\theta_2,$ which is canonically conjugate to the total particle number $N$. Because the underlying conservative GPE system is invariant under a global $U(1)$ phase transformation, the reduced Hamiltonian depends only on the relative phase and not on $\Phi$. Hence, $\Phi$ is a cyclic variable and $N_t=-\frac{\partial H}{\partial \Phi}=0$. Its evolution only produces an overall phase rotation of the condensate, which does not affect the Josephson dynamics of the population imbalance and relative phase. Therefore, for a fixed total particle number, the dynamics can be consistently reduced to the pair $(Z,\theta)$.

For convenience, we denote the right-hand side of the equation for the relative phase by $F(Z)$. This $(Z,\theta)$ system has a Hamiltonian structure,
$$
Z_t=-\frac{\partial H}{\partial \theta}, \qquad \theta_t=\frac{\partial H}{\partial Z},
$$
with the corresponding Hamiltonian given by
\begin{eqnarray}
&H=2 \kappa \sqrt{1-Z^2}\cos\theta + \cfrac{q N}{2} Z^2 - \nonumber\\
&\cfrac{2g}{5} (N/2)^{3/2} \left[(1-Z)^{5/2} + (1+Z)^{5/2} \right].
\label{HamiltonianPW}
\end{eqnarray}
This Hamiltonian coincide in the form with that obtained early in the works~\cite{Abdullaev-2023,QF1}.
We consider small-amplitude Josephson oscillations for the zero- and $\pi$-phase modes. 
The explicit Josephson frequencies are, for the zero-phase mode,
\begin{equation}
\omega_J^{(0)} = 2\kappa \sqrt{1-\frac{1}{2\kappa} \left[ qN-3g (N/2)^{3/2} \right]}\,\, ,
\label{eq:Jfzero}
\end{equation}
and for the $\pi$-phase mode,
\begin{equation}
\omega_J^{(\pi)} = 2\kappa \sqrt{1+\frac{1}{2\kappa} \left[ qN-3g (N/2)^{3/2} \right]}\,\, .
\label{eq:Jfpi}
\end{equation}
In the self-trapping regime, the population imbalance $Z(t)$ does not change sign throughout the time evolution. In phase-space terms, this means that the constant energy contour $H(Z,\theta)=H_0$ does not intersect the line $Z=0$. Using this condition, one can determine the critical coupling constant $\kappa_{\mathrm{cr}}$,
\begin{equation}
\kappa_{\mathrm{cr}}
=\frac{\sqrt{2}\,g\,N^{3/2}\Big((1-Z)^{5/2}+(1+Z)^{5/2}-2\Big)
-5Nq\,Z^{2}}{20\left(\pm 1+\sqrt{1-Z^{2}}\right)}. 
\label{eq:KcrPW}
\end{equation}
that separates the Josephson oscillation regime from the macroscopic self-trapping regime for the initial values of the population imbalance $Z_0$ and the relative phase $\theta_0$. This critical value defines the threshold above which the system is no longer able to exhibit complete population exchange between the two cores; instead, a finite population imbalance persists over time, indicating the onset of macroscopic self-trapping.
For a fixed value of the coupling constant $\kappa$, Eq.~(\ref{eq:KcrPW}) also allows one to determine the critical initial imbalance $Z_{0\mathrm{cr}}$. For the parameter values ($q,g,\kappa,N$) used in Fig.~1, $Z_{0\mathrm{cr}}$ is computed as the value corresponding to the separatrix that separates the closed (bounded) trajectories associated with Josephson oscillations from the open (unbounded) trajectories corresponding to the self-trapping regime in the ($Z,\theta$) phase plane.
\begin{figure}[htbp]
  \centerline{ \includegraphics[width=4.4cm]{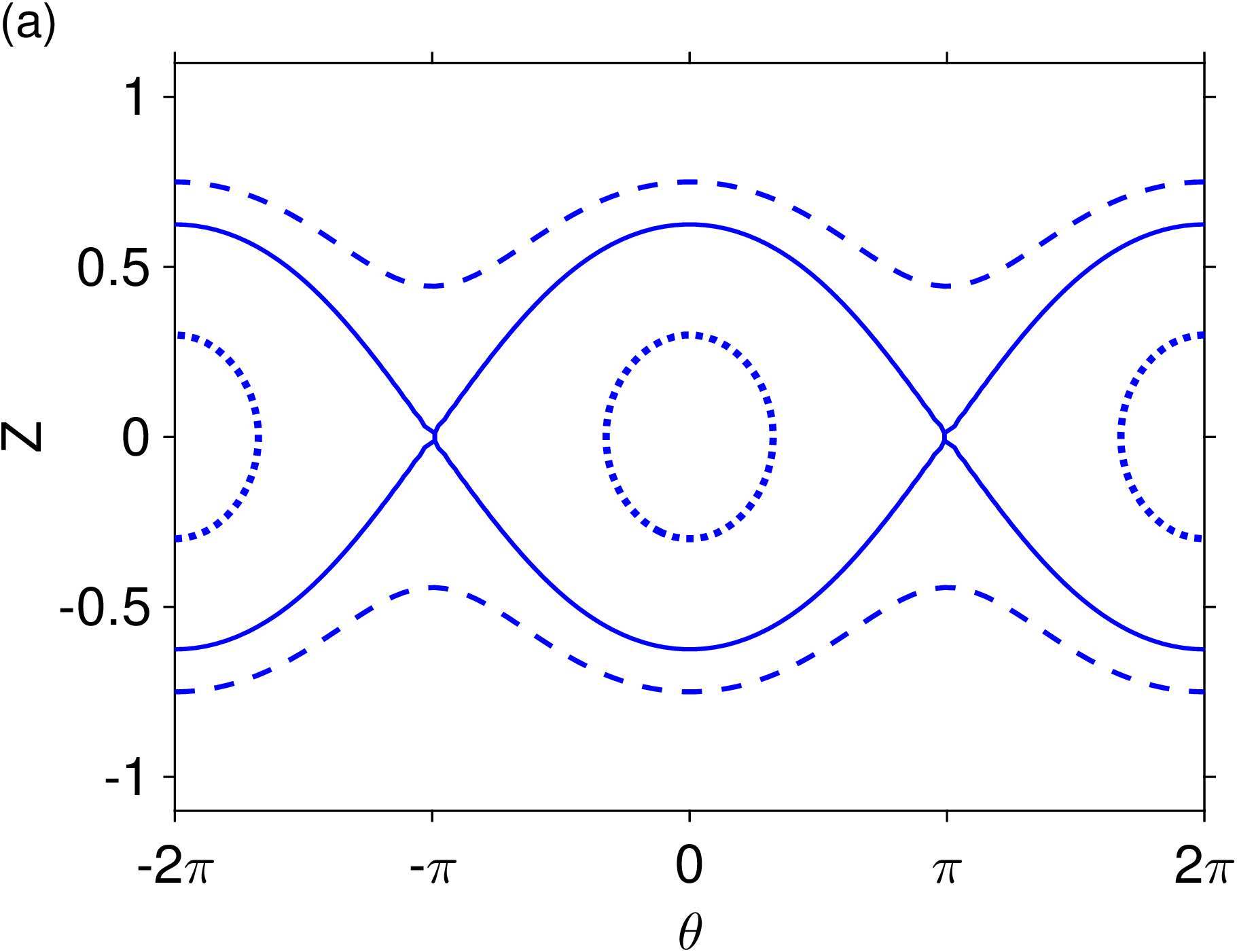}
  \includegraphics[width=4.4cm]{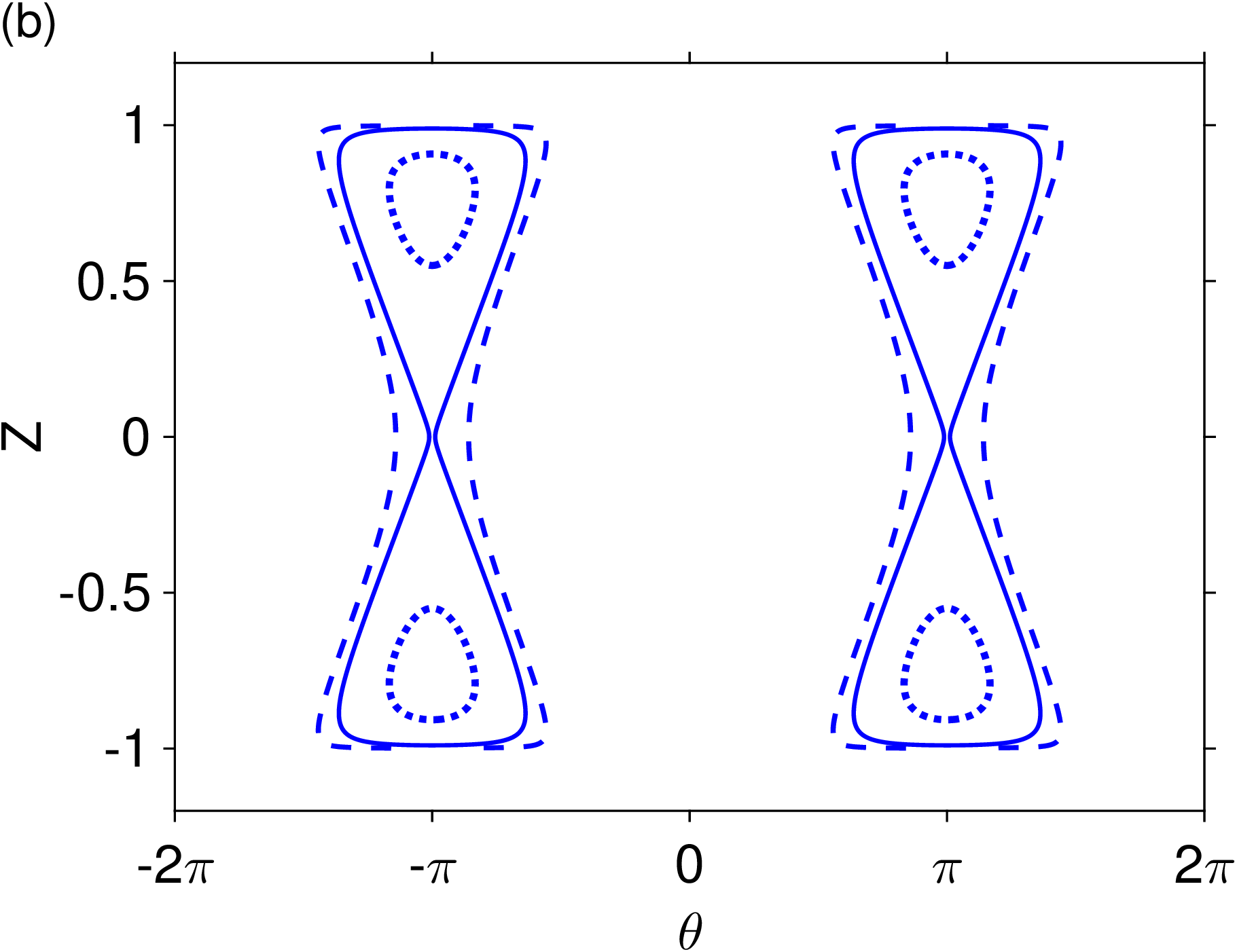} }
  \includegraphics[width=4.4cm]{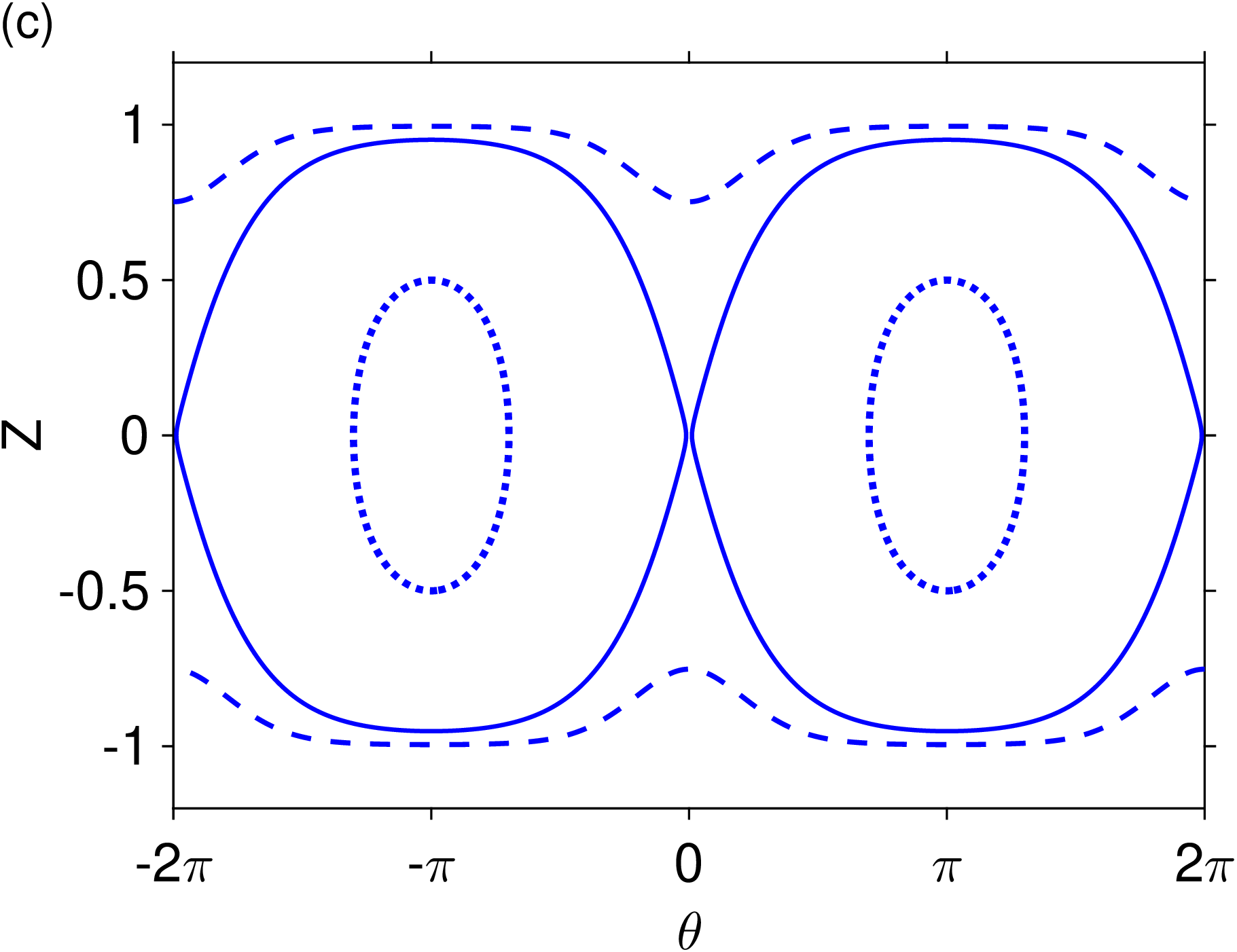}
\caption{Phase portraits in the $(Z,\theta)$ phase plane are shown for different atom numbers $N$ and phase modes. In each panel, the dotted curve corresponds to bounded trajectories (finite motion) representing Josephson oscillations, the separatrix is traced by the critical initial imbalance $Z_0=Z_{0\mathrm{cr}}$ (solid line), and the dashed curve represents unbounded trajectories (infinite motion) associated with the self-trapping regime. 
(a) Zero-phase mode, $N=1.2$: $Z_0=0.3$ (dotted, Josephson oscillations), $Z_0=Z_{0\mathrm{cr}}\simeq 0.6251$ (solid, separatrix), and $Z_0=0.75$ (dashed, self-trapping).
(b) $\pi$-phase mode, $N=1$: $Z_0=0.55$ (dotted), $Z_0=Z_{0\mathrm{cr}}\simeq 0.9898$ (solid, separatrix), and $Z_0=0.999$ (dashed).
(c) $\pi$-phase mode, $N=0.8$: $Z_0=0.5$ (dotted, Josephson oscillations), $Z_0=Z_{0\mathrm{cr}}\simeq 0.9517$ (dashed, separatrix), and $Z_0=0.995$ (solid, self-trapping). Other parameters $q=1$, $g=1$ and $\kappa=0.01$.}
\label{fig:phaseport}
\end{figure}

	Figure~\ref{fig:phaseport} shows the phase portraits obtained from the Hamiltonian in Eq.~(\ref{HamiltonianPW}). Panel~(a) focuses on the zero-phase mode and represents trajectories for different initial population imbalances $Z_0$. For an initial imbalance below the threshold, e.g., $Z_0=0.3<Z_{0\mathrm{cr}}$, the system exhibits Josephson oscillations between the two cores (dotted curve). The trajectory with $Z_0\simeq 0.6251$ lies on the separatrix (solid curve), which separates bounded Josephson motion from the self-trapped dynamics. Increasing the initial imbalance further, e.g., to $Z_0=0.75$, drives the system into the self-trapping regime, where the atoms remain predominantly localized in one core (dashed curve).
Panel~(b) illustrates the case $N=1$. At $Z_0=0.9898$, the dynamics enters a localization-like regime that is distinct from conventional self-trapping. This behavior corresponds to the localization-revival mode (observed for an LHY quantum fluid in a double-well potential~\cite{Wysocki2024}), in which the imbalance $Z(t)$ oscillates over an extended range, nearly spanning the interval $[0,1]$.
 In contrast to the self-trapped case, the relative phase does not grow without bound, instead, it executes small oscillations around the $\pi$ state, see Fig.~\ref{fig:LocRev} (a). 
Panel~(c) presents the analogous phase-space structure for the $\pi$-phase mode at $N=0.8$. The same qualitative transition from Josephson oscillations to self-trapping occurs as $Z_0$ increases. The corresponding critical value is $Z_{0\mathrm{cr}}\simeq 0.9517$, in agreement with the theoretical prediction given by Eq.~(\ref{eq:KcrPW}).

\begin{figure}[htbp]
  \centerline{ \includegraphics[width=4.4cm]{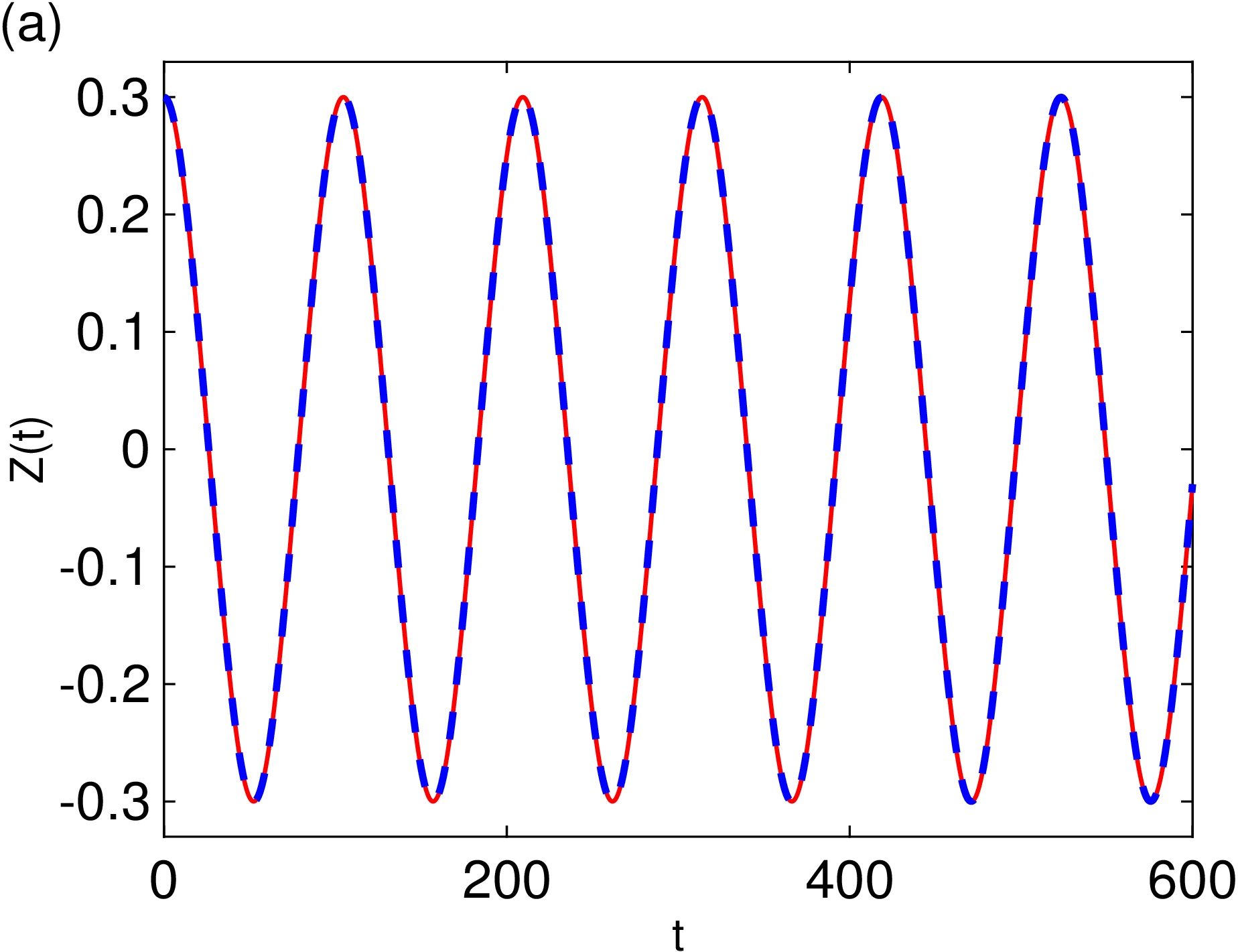}
  \includegraphics[width=4.4cm]{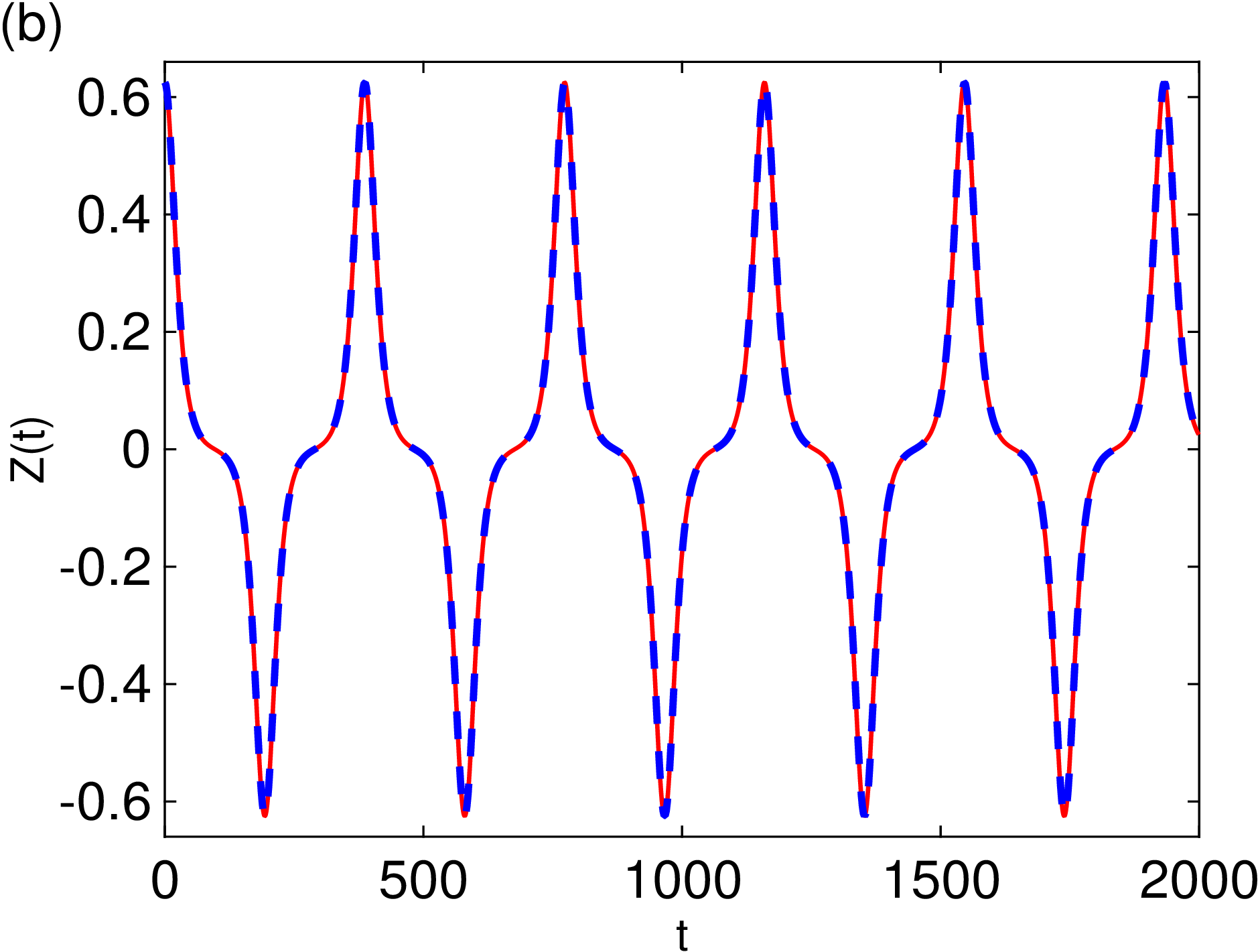}}
\centerline{\includegraphics[width=4.4cm]{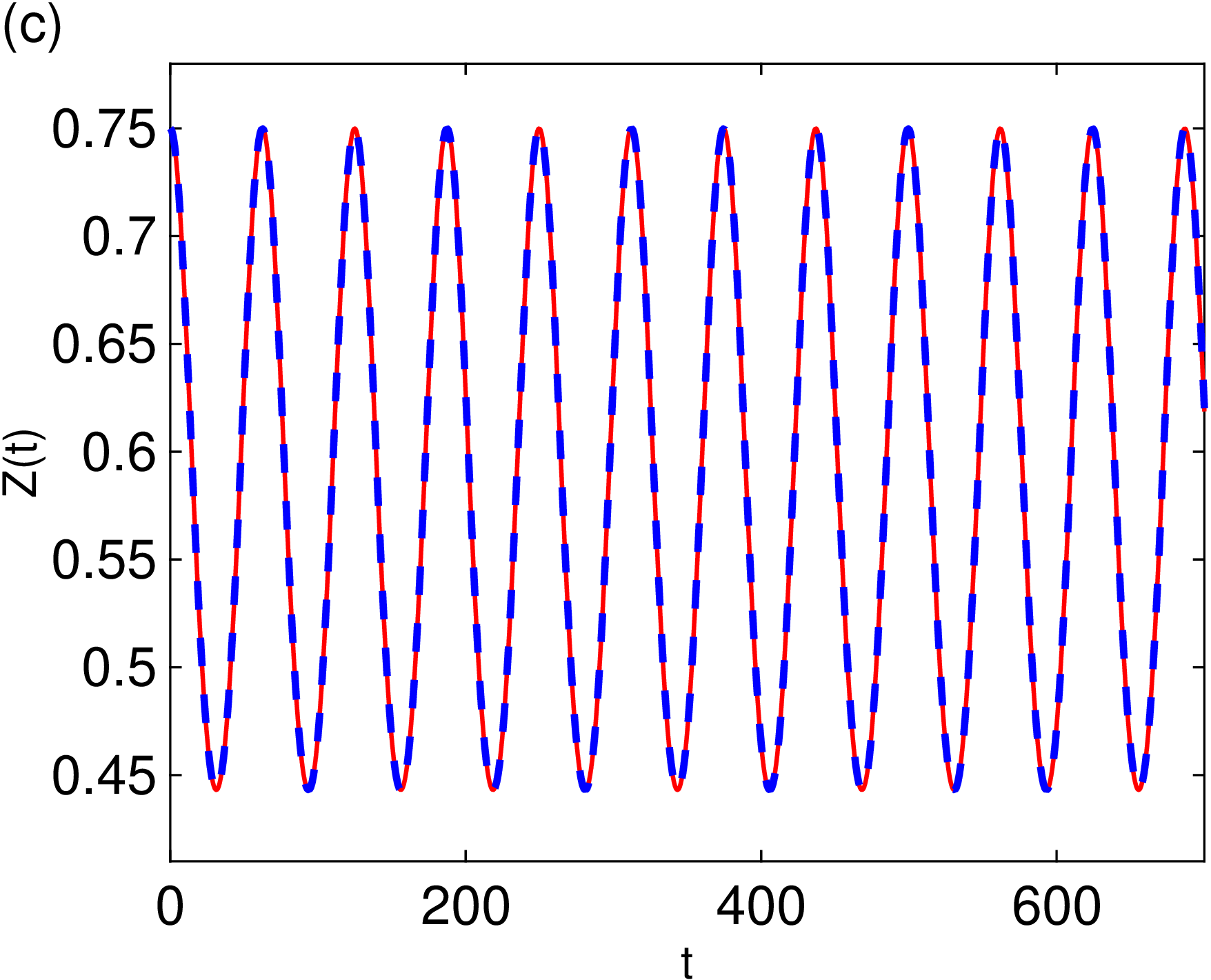} \includegraphics[width=4.4cm]{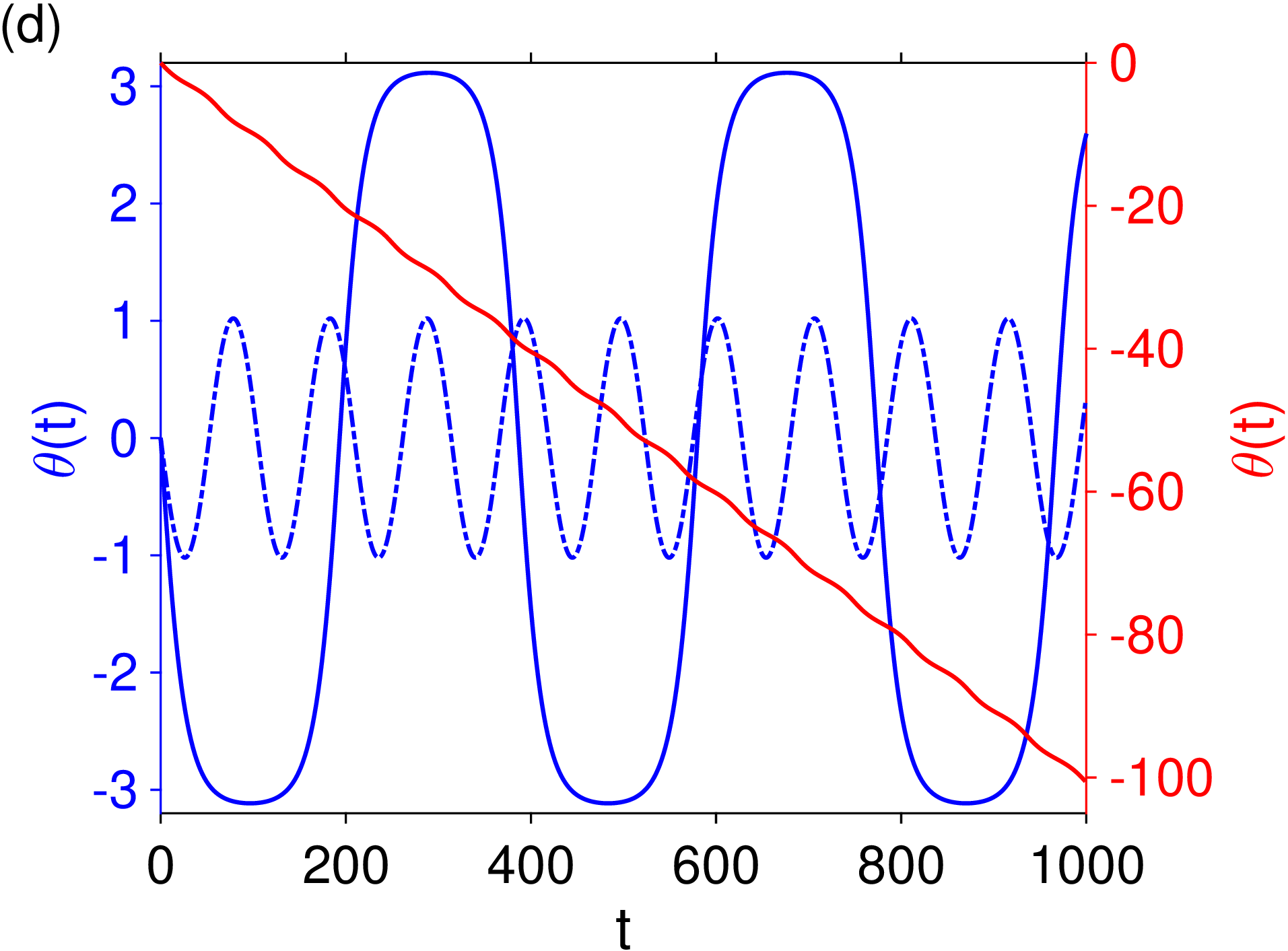}}
\caption{The time evolution of the population imbalance $Z(t)$ is shown for three representative initial conditions: (a) $Z_0=0.3$, which yields bounded motion corresponding to Josephson oscillations; (b) $Z_{0\mathrm{cr}}=0.6251$, which produces the critical trajectory on the separatrix; and (c) $Z_0=0.75$, which results in unbounded (running-phase) dynamics associated with the self-trapping regime. In panels (a)--(c), the red solid curves represent the theoretical predictions obtained from Eq.~(\ref{eq:dimerZtThetat}), whereas the blue dashed curves show the numerical results from simulations of Eq.~(\ref{eq:gpe}).
Panel (d) presents the corresponding phase dynamics, $\theta(t)$, calculated from Eq.~(\ref{eq:dimerZtThetat}) for the same initial imbalances considered in the previous panels. The Josephson-oscillation regime is identified by small-amplitude oscillations (blue dashed curve), while larger-amplitude oscillations occur on the separatrix (blue solid curve). In the self-trapping regime, $\theta(t)$ exhibits running-phase behaviour (red solid curve; note the right vertical axis). These time evolutions correspond to the trajectories shown in the $(Z,\theta)$ phase portrait in Fig.~\ref{fig:phaseport}(a).}
\label{fig:dynamPW}
\end{figure} 

Figure~\ref{fig:dynamPW} (a)--(c) compares the time evolution of the population imbalance $Z(t)$ for three representative initial imbalances $Z_0$, corresponding to those used in Fig.~\ref{fig:phaseport}(a). The solid curves show the theoretical predictions obtained by integrating the reduced system of ordinary differential equations~(\ref{eq:dimerZtThetat}), while the dashed curves are results of direct numerical simulations of the governing extended GP equation~(\ref{eq:gpe}). For the $\pi$-phase mode, the agreement between the two approaches is excellent for all three dynamical regimes: bounded Josephson oscillations, self-trapped (running-phase) dynamics, and the critical trajectory on the separatrix. We have also verified that similarly good agreement is obtained for other values of the initial relative phase.

Figure~\ref{fig:dynamPW}(d) shows the corresponding evolution of the relative phase $\theta(t)$ for the same initial imbalances. The Josephson-oscillation regime is characterized by oscillatory $\theta(t)$ with increasing amplitude as $Z_0$ approaches the threshold, whereas the separatrix (blue solid line) marks the critical trajectory separating bounded from unbounded motion. Beyond this threshold, $\theta(t)$ exhibits a running-phase behavior (red solid line, right axis), indicating the onset of the self-trapping regime.

Figure~\ref{fig:LocRev}(a) displays the coupled dynamics of the population imbalance and the relative phase. In this example, the imbalance $Z(t)$ undergoes large-amplitude oscillations, taking values between $0$ and $1$, while the phase difference $\theta(t)$ performs bounded oscillations about its initial value $\theta_0$ (here, $\theta_0=\pi$). The imbalance parameter has a transparent physical meaning: $Z=0$ corresponds to equal populations $N_1=N_2$; $Z=1$ implies complete transfer to the second core, $N_1=0$ and $N=N_2$; and $Z=-1$ corresponds to complete transfer to the first core, $N_2=0$ and $N=N_1$.
\begin{figure}[htbp]
  \centerline{ \includegraphics[width=4.4cm]{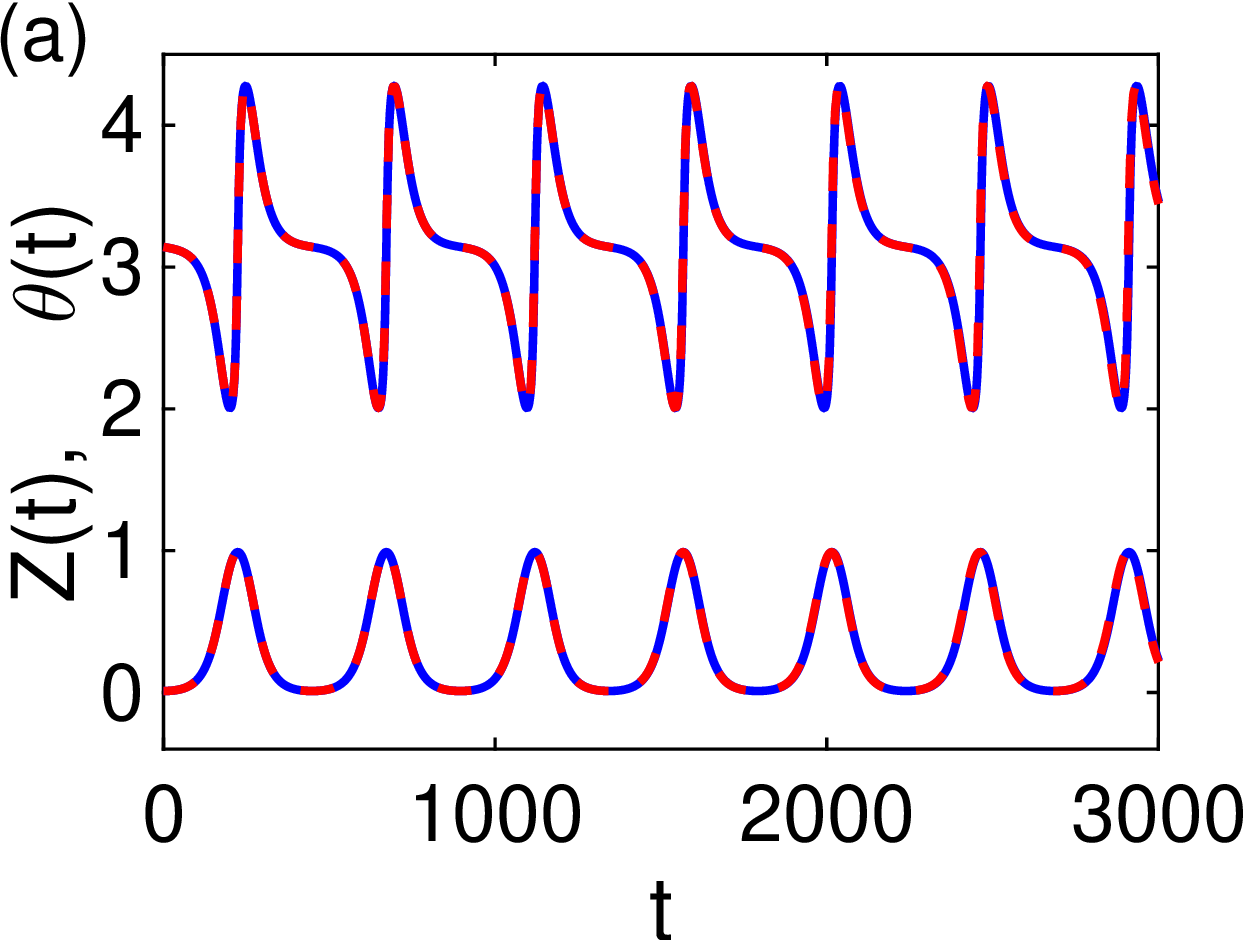}
  \includegraphics[width=4.4cm]{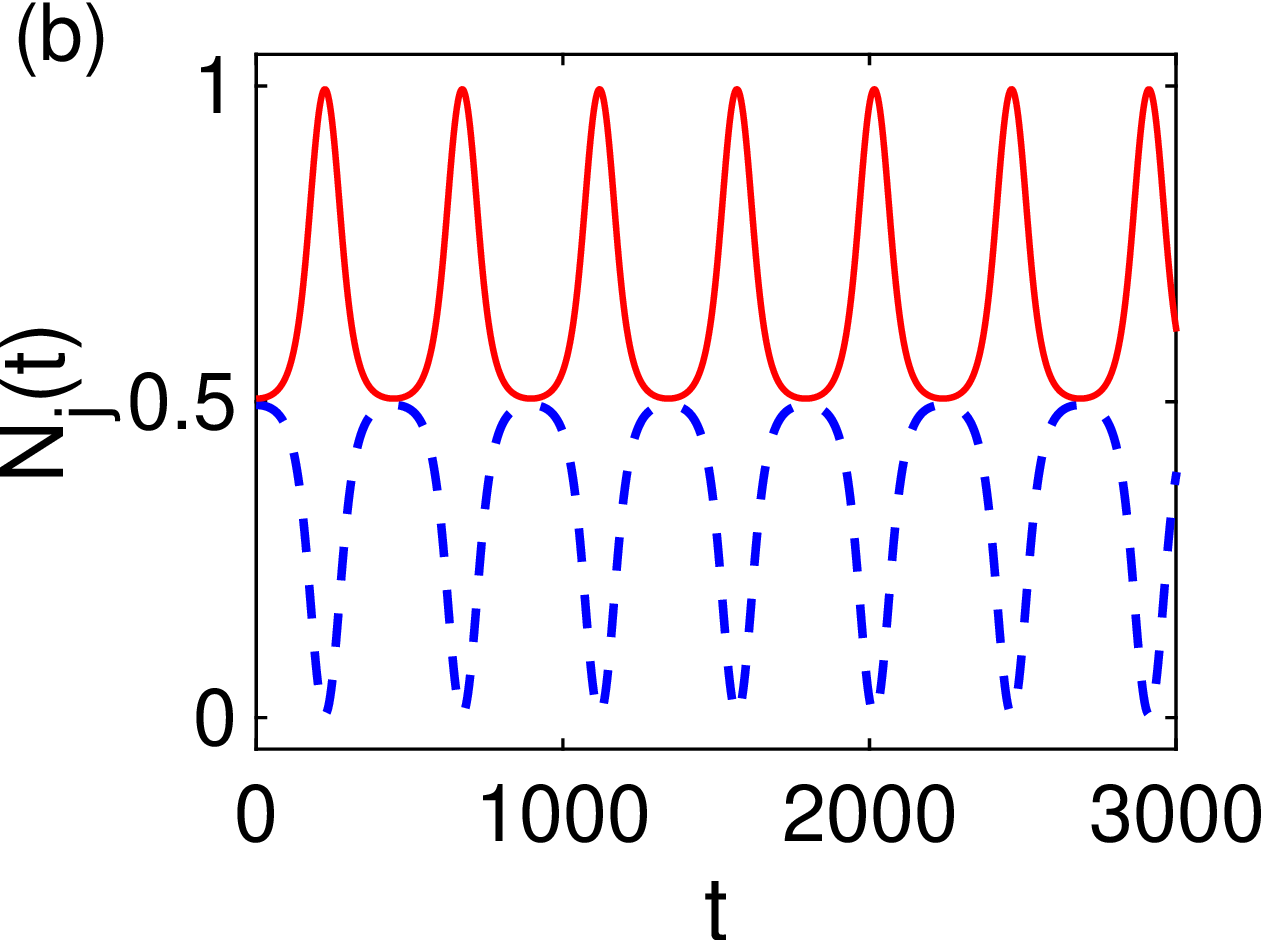} }
\caption{(a) The time evolution of the relative phase difference $\theta(t)$ (top panels) and the population imbalance $Z(t)$ (bottom panels) in the localization-revival regime is shown. Solid curves represent the theoretical predictions, while dashed curves correspond to direct numerical simulations. (b) The corresponding dynamics of the atom numbers in each component, $N_1(t)$ (lower curve) and $N_2(t)$ (upper curve), is displayed. The parameter set is the same as for the dashed trajectory in the phase portrait of Fig.~\ref{fig:phaseport}(b).}
\label{fig:LocRev}
\end{figure}
This behavior identifies the localization-revival regime. In this regime, atoms are transferred almost entirely from one core to the other and then return, with the populations periodically rebalancing, as illustrated in Fig.~\ref{fig:LocRev} (b). In contrast to the self-trapping regime, the relative phase remains bounded and does not exhibit running (diverging) dynamics. Similar nonlinear oscillations have been reported for a one-dimensional ultradilute quantum liquid in a double-well potential~\cite{Wysocki2024} and for a two-dimensional dual-core BEC with logarithmic LHY corrections~\cite{Otajonov2026}.

\begin{figure}[htbp]
  \centerline{ \includegraphics[width=4.4cm]{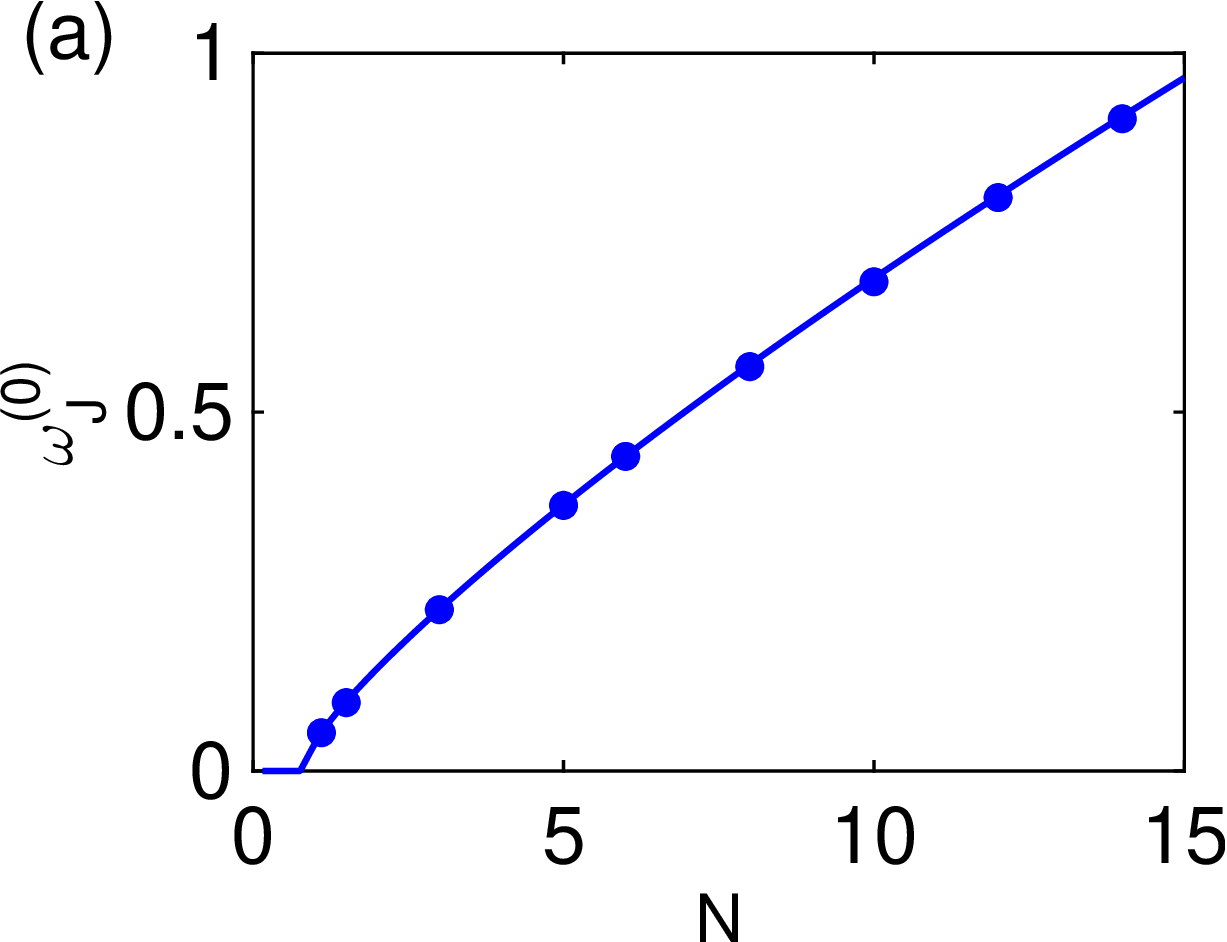}
  \includegraphics[width=4.4cm]{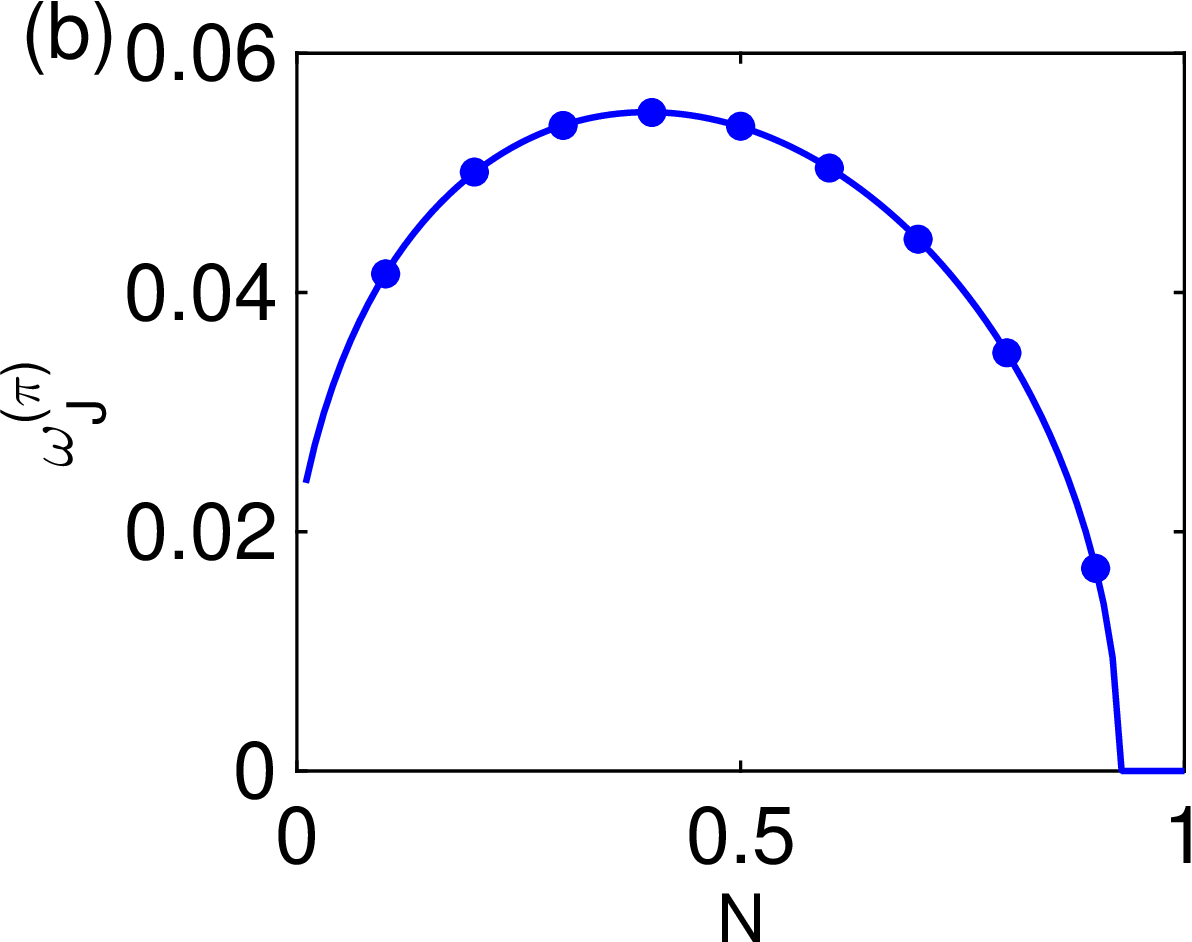} }
\caption{The dependence of the small-amplitude Josephson oscillation frequency on the total number of atoms $N$ is shown for (a) the zero-phase mode and (b) the $\pi$-phase mode. The solid curves correspond to the analytical predictions given by Eqs.~(\ref{eq:Jfzero}) and (\ref{eq:Jfpi}), while the points represent results obtained from direct numerical simulations of Eq.~(\ref{eq:gpe}). The remaining parameters are fixed, with $q=1$, $g=1$, and $\kappa=0.01$.}
\label{fig:freqPW}
\end{figure}

Figure~\ref{fig:freqPW} compares the Josephson frequency as a function of the total atom number $N$. The solid curves show the analytical predictions given by Eqs.~(\ref{eq:Jfzero}) and (\ref{eq:Jfpi}), whereas the symbols correspond to frequencies extracted from direct numerical simulations of Eq.~(\ref{eq:gpe}). Overall, the theory is in good agreement with the numerical results in both panels.

In panel~(a), which corresponds to the zero-phase mode, Eq.~(\ref{eq:Jfzero}) predicts a real oscillation frequency only for $N \geq 0.7576$. For $N \leq 0.7576$, the expression under the square root becomes negative, and the resulting complex value is plotted as zero; physically, this parameter range corresponds to the onset of the self-trapping regime, where small-amplitude Josephson oscillations cease to exist. For $N \geq 0.7576$, Josephson oscillations are supported and the frequency increases monotonically with $N$.

Panel~(b) shows the $\pi$-phase mode. In this case, Josephson oscillations persist only within a narrow interval, $N \leq 0.93$. When $N>0.93$, Eq.~(\ref{eq:Jfpi}) yields a complex frequency, indicating the transition to the self-trapped (running-phase) regime.

Notably, there exists an intermediate range of particle numbers in which Josephson oscillations can be observed in both the zero- and $\pi$-phase modes. The results shown in Fig.~\ref{fig:freqPW} are obtained for the fixed parameter set $(q,g,\kappa)=(1,1,0.01)$, and the same qualitative trends are retained for other parameter values. Variations in the total number of atoms induce structural changes in the system. A detailed analysis of these changes is presented in Sec.~\ref{sec:bifurcation}.

\subsection{Bifurcation}
\label{sec:bifurcation}

In this section, we analyse the bifurcation conditions of the system of ordinary differential Eq.~(\ref{eq:dimerZtThetat}) governing the population imbalance $Z$ and the relative phase $\theta$. To this end, we first determine the fixed points of the ordinary differential equation system given in Eq.~(\ref{eq:dimerZtThetat}). These fixed points, denoted by $(Z^\ast,\theta^\ast)$, are obtained from the stationary conditions $Z_t=0$ and $\theta_t=0$.
From the equation for $Z_t$, it follows that the fixed-point condition is satisfied when either $Z^\ast=\pm 1$ or $\theta^\ast=n\pi$, where $n=0,1,2,\ldots$. However, the cases $Z^\ast=\pm 1$ make the equation for $\theta_t$ singular; therefore, these solutions are excluded from the physically relevant fixed point set. Consequently, the admissible fixed points must satisfy $\theta^\ast=n\pi$. 
Substituting this condition into the equation for $\theta_t$, the corresponding values of $Z^\ast$ are determined by the roots of
\begin{eqnarray}
&F(Z^\ast) =-\cfrac{2 \kappa \sigma Z^\ast}{\sqrt{1-Z^{\ast 2}}}+ q N Z^\ast - \\
\nonumber
&  g(N/2)^{3/2}\left[(1+Z^\ast)^{3/2}-(1-Z^\ast)^{3/2} \right]=0.
\nonumber
\label{eq:F}
\end{eqnarray}
where $F(Z^\ast)$ denotes the right-hand side of the equation for the relative phase, and where $\sigma = \cos(\theta^*) = \cos(n\pi) = (-1)^n = \pm 1$. One immediately finds that $Z^\ast=0$ is a trivial fixed point of this equation. Additional nontrivial fixed points may also exist depending on the system parameters, and their appearance signals the onset of bifurcation. The nature and stability of these fixed points can then be examined by evaluating the Jacobian matrix at the fixed point $(Z^*, \theta^*)$
\begin{equation}
J^* =
\begin{pmatrix}
0 & 2 \kappa \sigma \sqrt{1 - Z^{*2}} \\
F_Z(Z^*)& 0
\end{pmatrix}.
\label{eq:Jacob}
\end{equation}

The trace and determinant of the Jacobian matrix are given by
$
\mathrm{Tr}(J^*) = 0,
$
\begin{equation}
\mathrm{det}(J^*) = -2 \kappa \sigma \sqrt{1 - Z^{*2}} F_Z(Z^*).
\label{eq:detJ}
\end{equation}

The sign of $\mathrm{det}(J^*)$ determines the type and stability of the fixed point. The eigenvalues $\lambda=\pm \sqrt{-\mathrm{det} J^*}$ are purely imaginary when $\mathrm{det}(J^*) > 0$; in this case, the fixed point is a center, which is neutrally stable. When $\mathrm{det}(J^*) < 0$, the eigenvalues are real and have opposite signs, indicating a saddle point, which is unstable. The case $\mathrm{det}(J^*) = 0$ corresponds to a bifurcation.

\begin{figure}[htbp]
  \centerline{ \includegraphics[width=4.55cm]{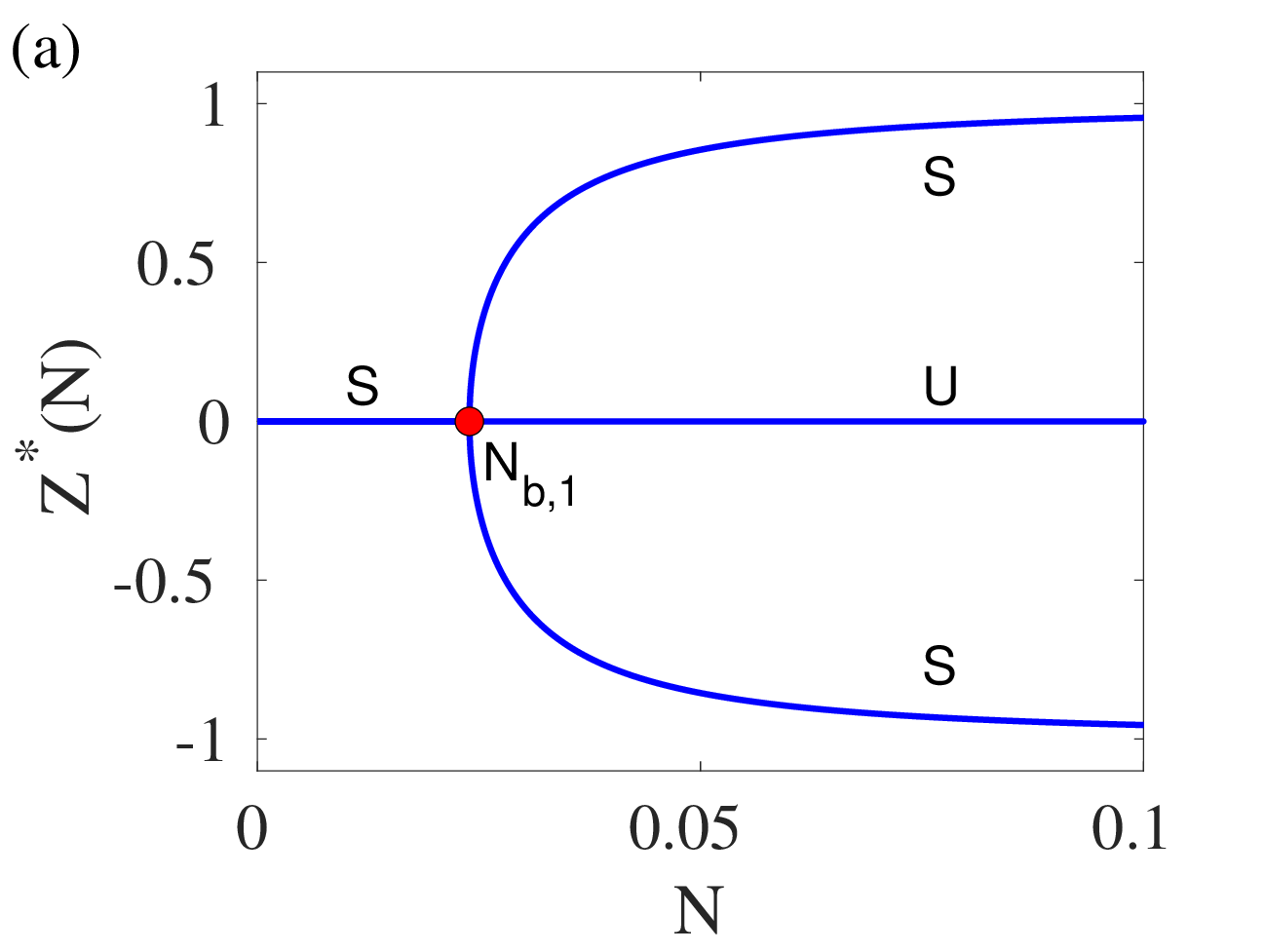}
  \includegraphics[width=4.55cm]{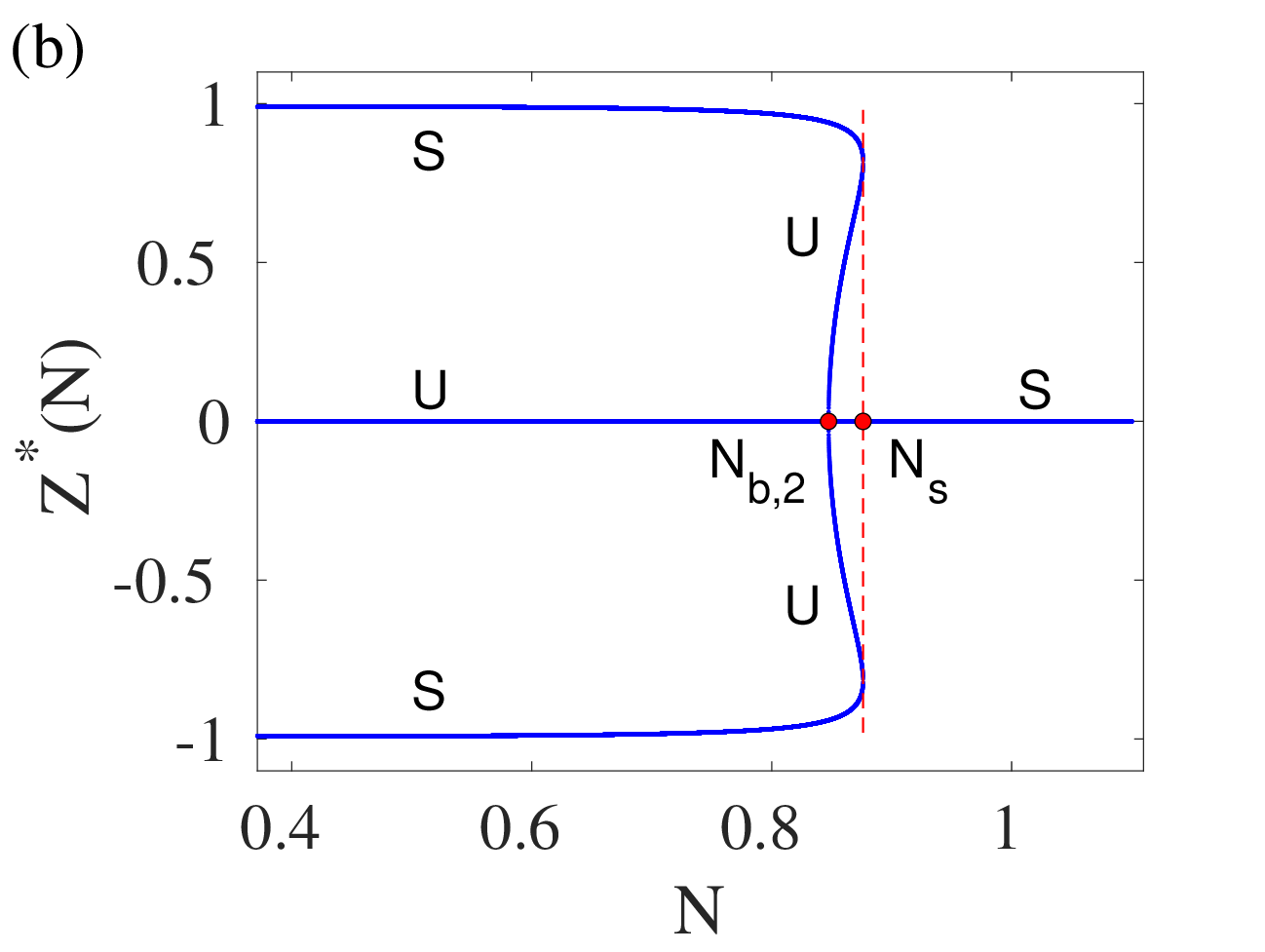}}
  \includegraphics[width=4.55cm]{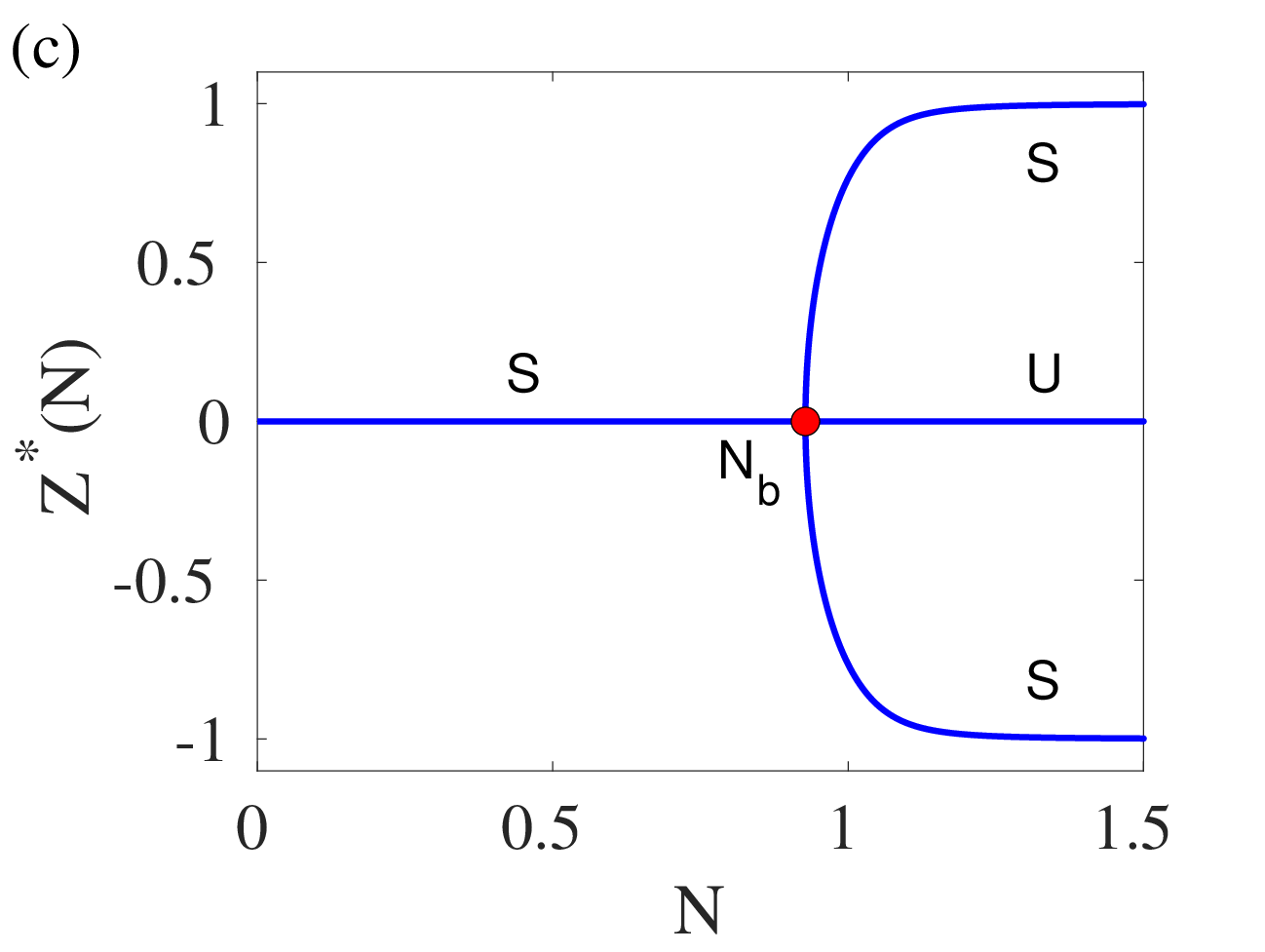}
\caption{Bifurcation diagram in the $(N, Z)$-plane. (a) In the zero-phase configuration $(\sigma = 1)$, the first bifurcation is supercritical and occurs at $N_{b,1} \simeq 0.0239252$, as determined by Eq.~(\ref{eq:Nb}). (b) As $N$ increases further in the zero-phase, a second bifurcation of subcritical type emerges at $N_{b,2} \simeq 0.847427$. Additionally, a saddle-node bifurcation is identified at $N = N_s \simeq 0.876157$, giving rise to a hysteresis loop within the interval $N_{b,2} < N < N_s$. (c) In the $\pi$-phase $(\sigma = -1)$, a single supercritical bifurcation takes place at $N_b \simeq 0.927631$, again as predicted by Eq.~(\ref{eq:Nb}). The remaining parameters are fixed as $\kappa = 0.01$, $q = 1$, and $g = 1$.}
\label{fig:bifdiag}
\end{figure}

The bifurcation condition \(\det(J^*) = 0\) yields the following equation for the bifurcation points \(N_{\mathrm{b}}\):
\begin{equation}
-\frac{3 g N_{\mathrm{b}}^{3/2}}{2 \sqrt{2}} - 2 \kappa \sigma + N_{\mathrm{b}} q = 0.
\label{eq:Nb}
\end{equation}
From Eq.~\eqref{eq:Nb}, it follows that for given parameters $q = 1$ and $g = 1$, the zero-phase ($\sigma = 1$) admits two real roots, whereas the $\pi$-phase ($\sigma = -1$) admits only a single real root. This behavior persists for other boundary values of $q$ and $g$ as well.
Consequently, there are two bifurcation points in the zero-phase and a single bifurcation point in the $\pi$-phase. The corresponding bifurcation diagrams, obtained by numerically solving Eq.~\eqref{eq:dimerZtThetat}, are shown in Fig.~\ref{fig:bifdiag}(a) and (b) for the zero-phase, and in Fig.~\ref{fig:bifdiag}(c) for the $\pi$-phase.

It can be observed that all bifurcations in the system exhibit the characteristic features of pitchfork bifurcations. To derive the corresponding normal form, we expand Eq.~(\ref{eq:F}) as a Taylor series around the symmetric equilibrium point $Z = 0$. Noting the odd symmetry $F(-Z) = -F(Z)$, the equilibrium $Z = 0$ exists for all $N > 0$ and is the unique solution that respects the $Z \to -Z$ symmetry.

Expanding $F(Z)$ up to fifth order yields:
\begin{equation}
F(Z) \simeq \alpha(N) Z + \beta(N) Z^3 + \gamma(N) Z^5 + \mathcal{O}(Z^7),
\end{equation}
where the coefficients depend on the system parameters as follows:
\begin{align}
\nonumber
\alpha(N) &= -\frac{3 g N^{3/2}}{2 \sqrt{2}} - 2 \kappa \sigma + N q, \\
\nonumber
\beta(N)  &= \frac{g N^{3/2}}{16 \sqrt{2}} - \kappa \sigma, \\
\gamma(N) &= \frac{3}{512} \left( \sqrt{2} g N^{3/2} - 128 \kappa \sigma \right).
\end{align}

To identify the bifurcation structure, it is often sufficient to truncate the expansion to cubic order:
\begin{equation}
F(Z) \approx \alpha(N) Z + \beta(N) Z^3,
\end{equation}
which corresponds to the canonical normal form of a pitchfork bifurcation~\cite{Strogatz2018}.

The emergence of nontrivial symmetry-breaking branches $Z^* = \sqrt{ -\alpha(N)/\beta(N) }$ requires the condition $\alpha(N)\beta(N) < 0$ to be satisfied. The nature of the bifurcation is governed by the sign of the cubic coefficient $\beta(N)$: the bifurcation is supercritical when $\beta(N) < 0$, leading to the emergence of stable off-axis equilibria; it is subcritical when $\beta(N) > 0$.
The critical bifurcation point $N_{\mathrm{b}}$, at which the symmetric state loses stability, is obtained by solving the condition $\alpha(N) = 0$. This coincides precisely with the expression given in Eq.~(\ref{eq:Nb}).

In the following analysis, we examine the bifurcation diagrams corresponding to the zero-phase and $\pi$-phase regimes separately, emphasising the unique features and stability characteristics of each case.
We begin with the zero-phase regime, which corresponds to the choice $\sigma = 1$. For the selected parameters, ($g = 1$ and $\kappa = 0.01$) the system undergoes two distinct bifurcations. The first bifurcation occurs at $N_{b,1} \approx 0.0239252,$ as indicated by a point in Fig.~\ref{fig:bifdiag} (a). Within the interval $N_{b,1} < N_{\mathrm{cr}} < 8 \left( \frac{\kappa}{g} \right)^{2/3} \approx 0.371$, the system parameters satisfy $\alpha(N) > 0$ and $\beta(N) < 0$. This configuration leads to a supercritical pitchfork bifurcation, in which the initially stable symmetric equilibrium at $Z^* = 0$ becomes unstable, giving rise to two new stable, symmetry-broken equilibria at $\pm Z^*$.
In the bifurcation diagrams, stable and unstable branches are denoted by ``S'' and ``U", respectively. Stable branches correspond to the Josephson-oscillation regime, whereas unstable branches indicate the onset of the self-trapping regime.

Beyond the critical atom number $N_{\mathrm{cr}}$, a second sub-critical pitchfork bifurcation appears because the coefficient $\beta(N)$ turns positive for $N > N_{\mathrm{cr}}$. This bifurcation occurs at $N_{b,2} \simeq 0.847427$ (see Fig.~\ref{fig:bifdiag} (b). When $N < N_{b,2}$ the system has three fixed points $Z^{*} = 0$ (unstable) and $Z^{*} = \pm Z_{large}$ (stable); in the intermediate range $N_{b,2} < N < N_{s} \simeq 0.876157$ the total fixed points increases to five (the origin stable, a pair of small-imbalance unstable, and a pair of large-imbalance stable states); and for $N > N_{s}$ only the symmetric equilibrium $Z^{*}=0$ persists and it is stable. Throughout the interval $N_{b,2} < N < N_{s}$, two qualitatively different stable equilibria coexist, the symmetric state $Z^{*}=0$ and a pair of large-imbalance fixed points. As a result, the ultimate state reached as $t \to \infty$ depends on the initial imbalance $Z_{0}$; the system settles into the basin of attraction that contains that initial condition. This multi-stability permits discontinuous jumps in $Z$ and produces a hysteresis loop when the parameter $N$ is swept slowly.
Assume the atom number is initially set at some value inside the interval $N_{\mathrm{cr}}<N_{0}<N_{b,2}$. In this range, the coefficient $\beta(N)$ is positive, the symmetric equilibrium $Z^*=0$ is linearly unstable, and the system therefore relaxes to one of the two symmetry-broken imbalanced branches $\pm Z^{*}$.
When $N$ is increased very slowly from $N_{0}$ toward larger values, the trajectory simply follows this large-imbalance branch because it remains stable up to the saddle-node bifurcation at $N_{s}$.  Crossing $N_{b,2}$ does not trigger a jump: although the origin $Z^{*}=0$ becomes stable again, the already occupied branch $\pm Z^{*}$ retains its stability, so the system continues to climb that branch. Only when the parameter reaches the saddle-node at $N_{s} \simeq 0.8762$ does the large-imbalance branches vanish in a saddle-node bifurcation; at that instant, the state must abandon the branch and drop abruptly to the only remaining stable equilibrium, the symmetric point $Z^*=0$.

Now reverse the sweep and decrease $N$ quasi-statically from a value just above $N_{s}$. Throughout the interval $N_{s}>N>N_{b,2}$, the symmetric equilibrium stays stable, so the system sticks to $Z^*=0$ even though the symmetry-broken equilibria reappear and are again stable in this interval. As soon as $N$ slips below $N_{b,2}$, the origin loses stability, an infinitesimal perturbation ejects the trajectory, and it lands back on one of the large-imbalance equilibria $\pm Z^{*}$. Continuing to decrease $N$ simply carries the state smoothly down this branch until the initial value $N_{0}$ is reached, closing the cycle.

Because the system experiences an irreversible jump at $N_{s}$ during the upward sweep and a second, oppositely directed jump at $N_{b,2}$ during the downward sweep, the path traced in the $(N, Z)$ plane encloses a finite loop. The presence of two distinct, coexisting attractors in the strip $N_{b,2}<N<N_{s}$ means that the branch followed on the way up is not retraced on the way down; the long-term state for the same value of $N$ depends on
whether the control parameter is being increased or decreased.  This path-dependent response is precisely the hysteresis associated with the Josephson-like dimer in the interval $N_{b,2}<N<N_{s}$. Similarly, a bifurcation diagram for a two-core system with a cubic-quintic nonlinear system was reported in~\cite{Albuch2007}.

Figure~\ref{fig:bifdiag}(c) shows the bifurcation diagram for the $\pi$-phase regime, corresponding to the case illustrated in Fig.~\ref{fig:bifdiag}(a). In this scenario, the bifurcation occurs at a critical atom number $ N_b\simeq 0.927631$, as determined from Eq.~(\ref{eq:Nb}) with $\sigma = -1$. The bifurcation is of the subcritical pitchfork type, since the coefficient $\beta(N_b)$ in the normal form is positive.

\begin{figure}[htbp]
  \centerline{ \includegraphics[width=6.cm]{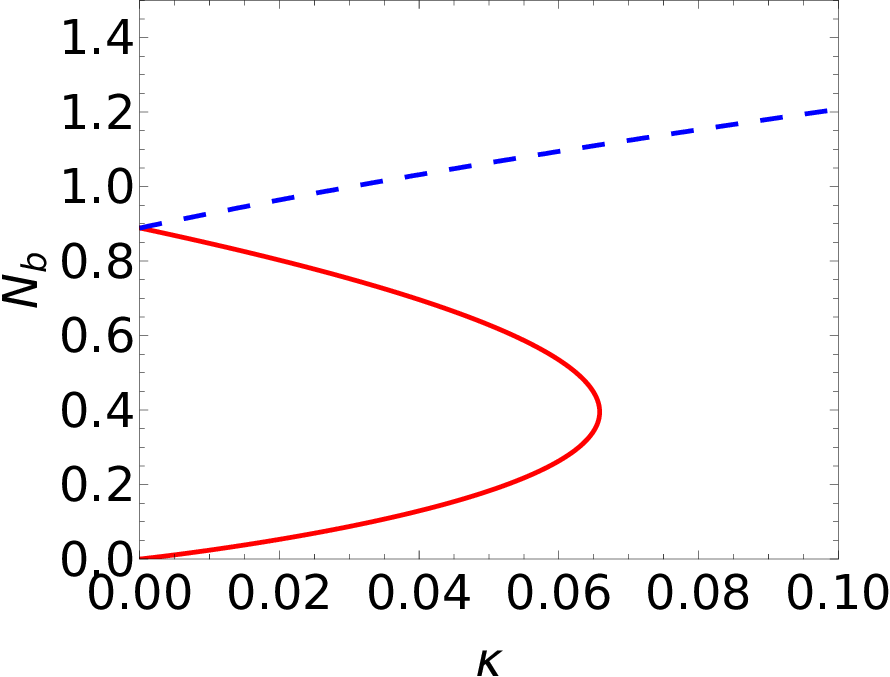}}
\caption{Dependence of the bifurcation points $N_b$ on the coupling constant $\kappa$ for fixed $q=1$ and $g = 1$.
The blue dashed line at the top corresponds to the $\pi$-phase mode ($\sigma = -1$) in Eq.~(\ref{eq:Nb}).
The red solid line represent the zero-phase mode ($\sigma = 1$), respectively; see Eq.~(\ref{eq:Nb}).}
\label{fig:NbvsK}
\end{figure}

Figure~\ref{fig:NbvsK} displays how the bifurcation point $N_b$ varies with the coupling constant $\kappa$, based on Eq.~(\ref{eq:Nb}). The solid red curve corresponds to the zero-phase, while the dashed blue curve represents the $\pi$-phase.
Notably, in the zero-phase, bifurcations occur only within a limited region of the $(N, \kappa)$ parameter space. In contrast, the $\pi$-phase supports bifurcations for any values of $\kappa$. Hence, pitchfork bifurcations in the zero-phase can be entirely suppressed by sufficiently strong tunnelling, whereas the $\pi$-phase continues to exhibit bifurcations for all values of $\kappa$. In the next section, we turn our attention to the dynamics of Josephson oscillations between quantum droplets.

\section{Interactions of the Quantum Droplets}
\label{sec:QDinterac}

Based on the approach introduced by Ref.~\cite{Pare1990, Otajonov2026}, we extend the variational framework to explore Josephson oscillations and macroscopic self-trapping phenomena in the context of quantum droplet dynamics. This method allows us to reduce the governing nonlinear equations to a set of effective dynamical equations, offering physical insights into coherent tunnelling and interaction-induced localisation regimes. 

The Lagrangian density of Eq.~(\ref{eq:gpe}) is given by
\begin{align}
&\mathcal{L} = \sum_{j=1}^2 \Biggl\{ \frac{i}{2}(\psi_{j,t}^{\ast} \psi_j - \psi^{\ast}\psi_{j,t}) \nonumber\\
&+\frac{1}{2}|\nabla \psi_j|^2 -\frac{q}{2}|\psi_j|^4 + \frac{2 g}{5}|\psi_j|^5 \Biggl\} \nonumber\\
&-\kappa(\psi_j^{\ast}\psi_{3-j} + \psi_{3-j}^{\ast} \psi_j)\, .
\label{eq:Lagr}
\end{align}

The super-Gaussian ansatz for the wavefunctions is~\cite{Otajonov2019Lavoine2021, Baizakov2011Otajonov20202024}:
\begin{equation}
\psi_j=A_j r^S \exp\left(-\frac{1}{2} (a r)^{2m} +i\theta_j + i S \phi \right),
\label{eq:ansatz}
\end{equation}
where $(r,\phi)$ are the polar coordinates, with $\phi$ being the angular coordinate. Its role is to generate the vortex phase $e^{i S \phi}$ where nonnegative integer $S> 0$ is the topological charge (or vorticity) of the quantum droplet. This factor ensures that the wavefunction carries angular momentum and has the correct phase winding around the origin. Here, $A_j$, $a$, $\theta_j$, and $m$ are variational parameters representing the amplitude, inverse width, initial phase, and super-Gaussian index, respectively. The fundamental, zero-vorticity state corresponds to $S=0$, whereas vortex states have $S>0$. By substituting the ansatz~(\ref{eq:ansatz}) into Eq.~(\ref{eq:Lagr}) and performing the integration $L = \int_{-\infty}^{\infty} \mathcal{L} dx dy$, we obtain the averaged Lagrangian as:
\begin{align}
L &= \pi\,M \sum_{j=1}^2
\Biggl\{
   g\,a^{-5S-2}
   \Bigl(\tfrac{2}{5}\Bigr)^{\frac{5MS}{2}+M+1}
   A_j^5\,\Gamma\!\Bigl[(\tfrac{5S}{2}+1)M\Bigr]
\nonumber\\[-1ex]
&\qquad
  -\,q\,a^{-4S-2}\,2^{-2MS-M-1}\,
    A_j^4\,\Gamma\bigl[(2S+1)M\bigr]
\nonumber\\[4pt]
&\qquad
  +\,\frac{\beta\,S}{2}\,a^{-2S-1}\,A_j^2
    \Bigl[
      \frac{a}{M}\,\Gamma(MS)
      + S\,\Gamma\!\Bigl[M\bigl(S+\tfrac12\bigr)\Bigr]
    \Bigr]
\nonumber\\[4pt]
&\qquad
  +\,a^{-2(S+1)}\,\Gamma[(S+1)M]\,A_j^2\,\dot\theta_j
\Biggr\}
\nonumber\\[6pt]
&\quad
-\,2\pi\,\kappa\,M\,
   a^{-2(S+1)}\,A_1\,A_2\,\cos\theta\,
   \Gamma\!\bigl[M(S+1)\bigr]\,,
\label{eq:avLagr}
\end{align}
with $M=1/m$.
The Euler-Lagrange equations give the system of ordinary differensial equations for population imbalance $Z$ and relative phase difference $\theta$:
\begin{eqnarray}
&Z_t=2\kappa \sqrt{1-Z^2}\sin{\theta},  \nonumber \\
&\theta_t=-\cfrac{2 \kappa  Z}{\sqrt{1-Z^2}} \cos (\theta ) + G Z + \nonumber \\
&F \left[ (1-Z)^{3/2}-(Z+1)^{3/2} \right]
\label{eq:QD_JosepEq}
\end{eqnarray}
\begin{eqnarray*}
&G\equiv \cfrac{2^{-M (2 S+1)} q N a^2 \Gamma (M (2 S+1))}{\pi  M \Gamma^2 (M (S+1))}\, , \nonumber \\
&F \equiv \left(\cfrac{N}{2 \pi  M}\right)^{\tfrac{3}{2}} \left(\cfrac{2}{5}\right)^{M \left(\tfrac{5 S}{2}+1\right)} \cfrac{g a^3 \Gamma \left(M \left(\frac{5 S}{2}+1\right)\right)}{\Gamma^{5/2}(M(S+1))}\, ,
\label{eq:GF}
\end{eqnarray*}
where $Z=(N_2-N_1)/N$ is the atomic imbalance, $\theta=\theta_2-\theta_1$ is relative phase, $N_j=2\pi\int_{0}^{\infty}|\psi_j|^2dr=\pi  M a^{-2 (S+1)} A_j^2 \Gamma (M (S+1))$ are the number of atoms in the cores and $N=N_1+N_2$ is the total number of atoms. We can write the Hamiltonian function for these new variables as:
\begin{eqnarray}
   & H= 2\kappa\sqrt{1-Z^2}\cos{\theta} + G \cfrac{Z^2}{2} - \nonumber \\
   & \cfrac{2F}{5}\left[ (1-Z)^{5/2}+(Z+1)^{5/2} \right]
\label{eq:QDHamiltonian}
\end{eqnarray}
Linearising the system (\ref{eq:QD_JosepEq}) about the fixed points yields the following expressions for the Josephson frequencies of small-amplitude oscillations. For the zero-phase mode ($\theta=0$),
\begin{equation}
    \omega_J^{(0)} = \sqrt{\,2\kappa \,\bigl(2\kappa - G + 3F \bigr)},
\label{JFzero}
\end{equation}
and for the $\pi$-phase mode ($\theta=\pi$),
\begin{equation}
    \omega_J^{(\pi)} = \sqrt{\,2\kappa \,\bigl(2\kappa + G - 3F \bigr)} .
    \label{JFpi}
\end{equation}
These formulas are valid in the small-amplitude limit and require the square-root arguments to be positive. The onset of the self-trapping regime is marked by the frequencies becoming complex, i.e., developing a nonzero imaginary part.
In the self-trapping regime, the population imbalance $Z(t)$ does not change sign; equivalently, the energy level set $H(Z,\theta)=H_0$ does not intersect the line $Z=0$. Since at $Z=0$ the Hamiltonian takes the values $H(0,\theta)=2\kappa\cos\theta-4F/5$, the set of energies accessible at $Z=0$ is the interval $[\, -2\kappa-4F/5,\, 2K-4F/5\,]$. A trajectory exhibits Josephson oscillations (with crossings of $Z=0$) if and only if $H_0$ belongs to this interval, otherwise it is self-trapped. Hence, the self-trapping condition is
\begin{equation}
\bigl|H_0+\cfrac{4F}{5}\bigr|>2\kappa,
\label{•}
\end{equation}
while the separatrix is given by $|H_0+4F/5|=2\kappa$ with $\kappa>0$ the linear coupling.

\begin{figure}[htbp]
  \centerline{ \includegraphics[width=4.5cm]{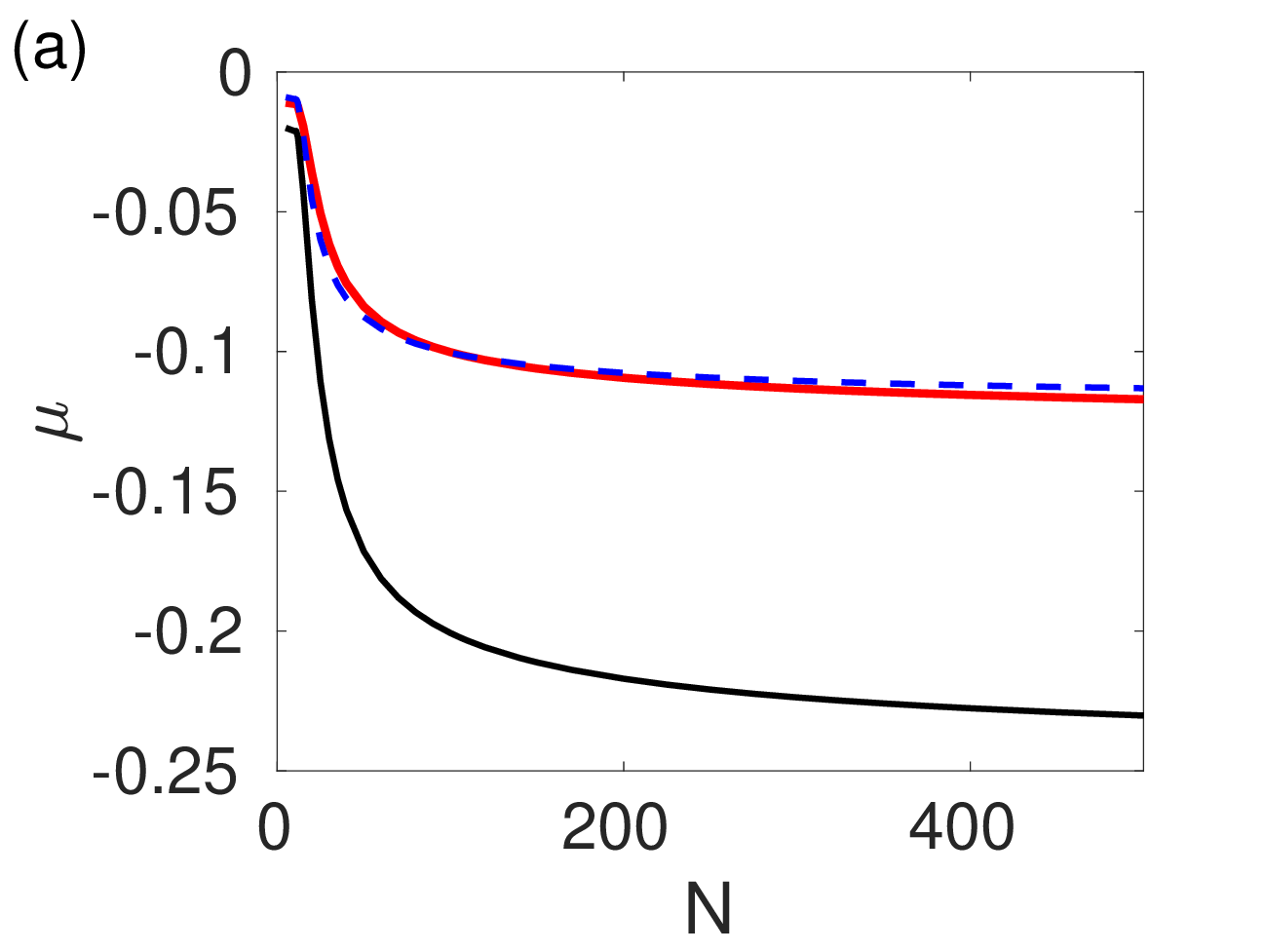}
  \includegraphics[width=4.5cm]{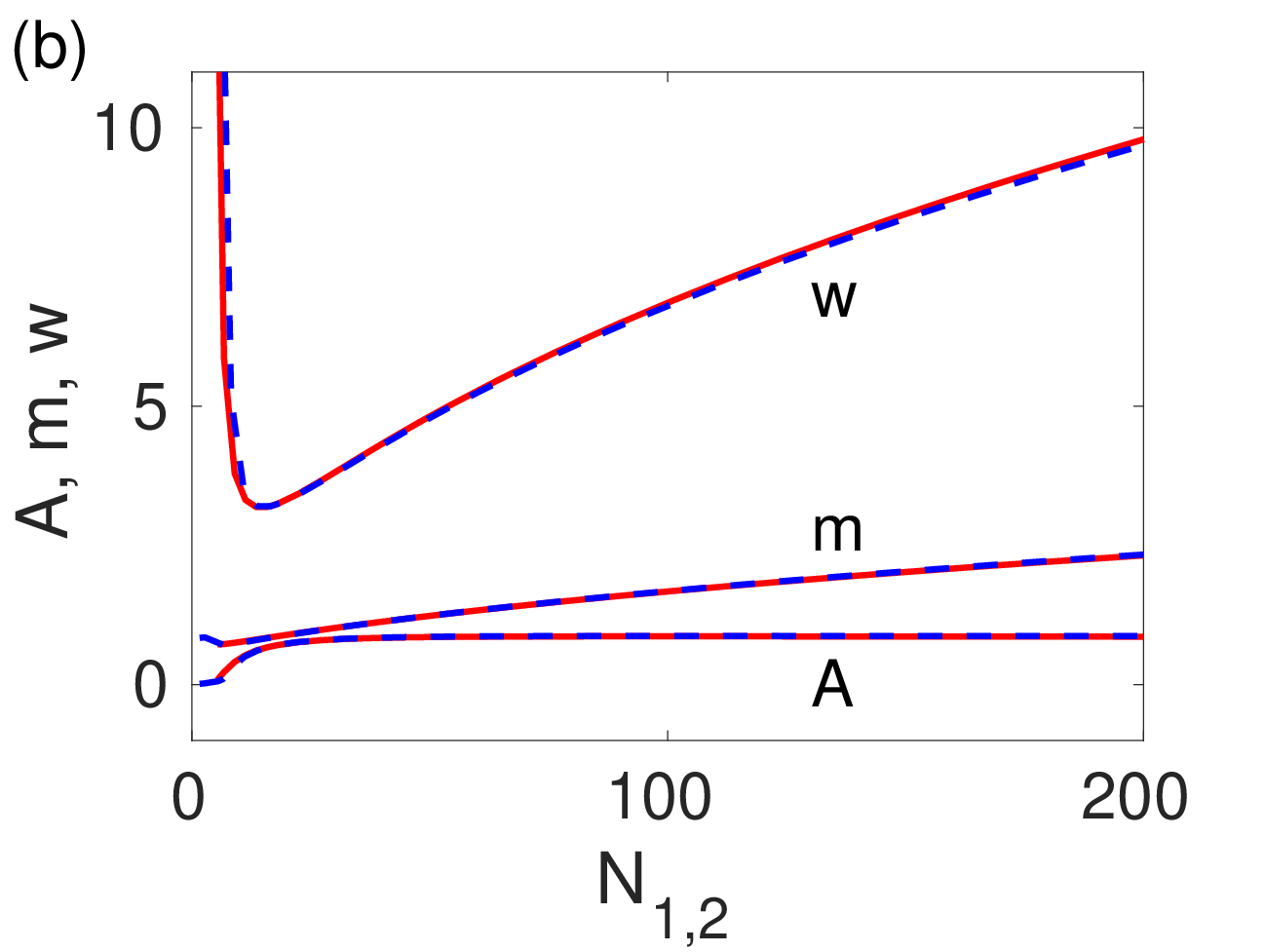}}
\caption{(a) Chemical potentials versus atom number for an initial population imbalance $Z_{0}=0.1$, obtained by imaginary-time propagation of Eq.~(\ref{eq:gpe}). The solid red curve shows the first component, $\mu_{1}(N_{1})$, the dashed blue curve shows the second component, $\mu_{2}(N_{2})$, the black curve indicates the total chemical potential, $\mu(N)$. (b) Droplet characteristic parameters amplitude $A$, power-law exponent $m$, and width $w\equiv 1/a$ as functions of atom number for both components at $Z_{0}=0.1$ (solid red: component 1; dashed blue: component 2). Other parameters are $q=1$, $g=1$ and $\kappa=0.01$.} 
\label{fig:muNParam}
\end{figure}

The stationary quantum droplet with $S=0$ parameters was obtained using the imaginary-time method. Figure~\ref{fig:muNParam} (a) shows the dependence of the chemical potentials on the total atom number $N$, obtained from imaginary-time simulations of Eq.~(\ref{eq:gpe}) for an initially imbalanced state with $Z_0=0.1$. The red solid curve represents the chemical potential $\mu_1(N_1)$ of the first component $u_1$, while the blue dashed curve shows $\mu_2(N_2)$ of the second component $u_2$. In the simulations, the component norms are set as
$$
N_1=\frac{N(1-Z_0)}{2},\qquad N_2=\frac{N(1+Z_0)}{2},
$$
so that $N=N_1+N_2$. The lower black solid curve displays the total chemical potential $\mu=\mu_1+\mu_2$ as a function of $N$.
According to the Vakhitov-Kolokolov stability criterion~\cite{VK}, the condition $d\mu/dN<0$ indicates dynamical stability of the QDs. This conclusion is supported by direct numerical simulations in which small initial deviations from equilibrium are introduced, confirming that the stationary states remain robust under weak perturbations.

	Figure~\ref{fig:muNParam}(b) summarises how the key parameters of the stationary quantum droplets: the peak amplitudes, effective widths, and the super-Gaussian shape index $m$ vary with the number of atoms in each component for an initially imbalanced configuration with $Z_0=0.1$. The red solid and blue dashed curves correspond to the first and second components, respectively. As the particle numbers increase, the peak amplitudes of both components approach saturation. In contrast, a growth of the widths becomes evident only at larger norms, reflecting the gradual expansion of the droplet while maintaining an almost constant central density.

\begin{figure}[htbp]
  \centerline{ \includegraphics[width=7.55cm]{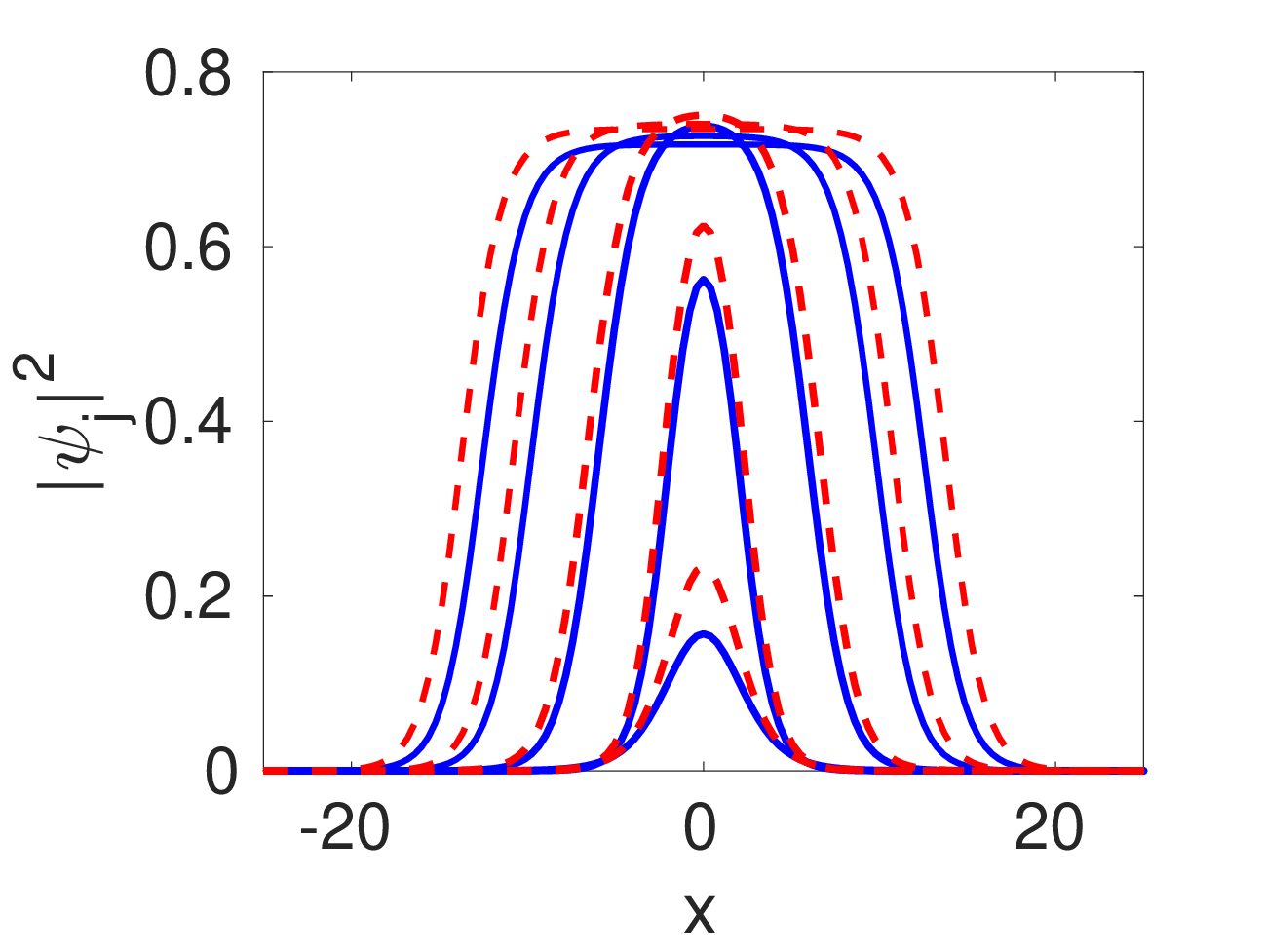}}
\caption{Density profiles $|\psi_j|^2$ ($j=1,2$) for an initial population imbalance $Z_0=0.1$. Solid blue curves denote component $j=1$ and dashed red curves denote component $j=2$. From the inner to the outer curves, the total atom numbers are $N=20,\,50,\,200,\,500$ and $800$. Parameters: $q=1$, $g=1$ and $\kappa=0.01$.}
\label{fig:denUV}
\end{figure}

The monotonic increase of the super-Gaussian index $m$ with atom number indicates a systematic reshaping of the density profile from a nearly Gaussian form at low norms to a flat-top structure at higher norms. This crossover is a standard signature of droplet formation in beyond-mean-field systems: the balance between the attractive mean-field contribution and the repulsive LHY term fixes the bulk density, while additional atoms primarily increase the droplet volume rather than the peak density. Consequently, the droplet behaves increasingly like an incompressible, liquid-like object with a well-defined interior and sharper edges, as also evidenced by the density profiles in Fig.~\ref{fig:denUV}. This trend is consistent with the characteristics reported in Ref.~\cite{PA-2016}.

Figure~\ref{fig:denUV} shows representative density profiles $|\psi_j(x)|^2$ of both components for several values of the total atom number $N$, with a small initial population imbalance $Z_0=0.1$. At low $N$, the stationary droplets are relatively dilute, and their profiles are close to Gaussian, reflecting the dominance of kinetic-energy smoothing and the absence of a well-developed bulk region. As $N$ increases, the droplets progressively acquire a flat-top shape with steeper edges. This evolution indicates the formation of a nearly uniform interior density (a ``bulk'' plateau) and a surface region, where the density rapidly drops to zero.

Such a crossover is a characteristic feature of self-bound quantum droplets stabilized by the competition between mean-field attraction and LHY-induced repulsion: once the equilibrium bulk density is established, additional atoms primarily increase the droplet size rather than the peak density. In other words, the droplets become increasingly liquid-like and effectively incompressible, with an expanding flat interior and a comparatively thin boundary layer. The near overlap of the two component profiles, despite the small imbalance, also suggests that the droplet remains close to a symmetric configuration, with the imbalance mainly manifesting as a slight difference in peak values and/or widths rather than a qualitative deformation of the overall shape.

\begin{figure}[htbp]
  \centerline{ \includegraphics[width=4.55cm]{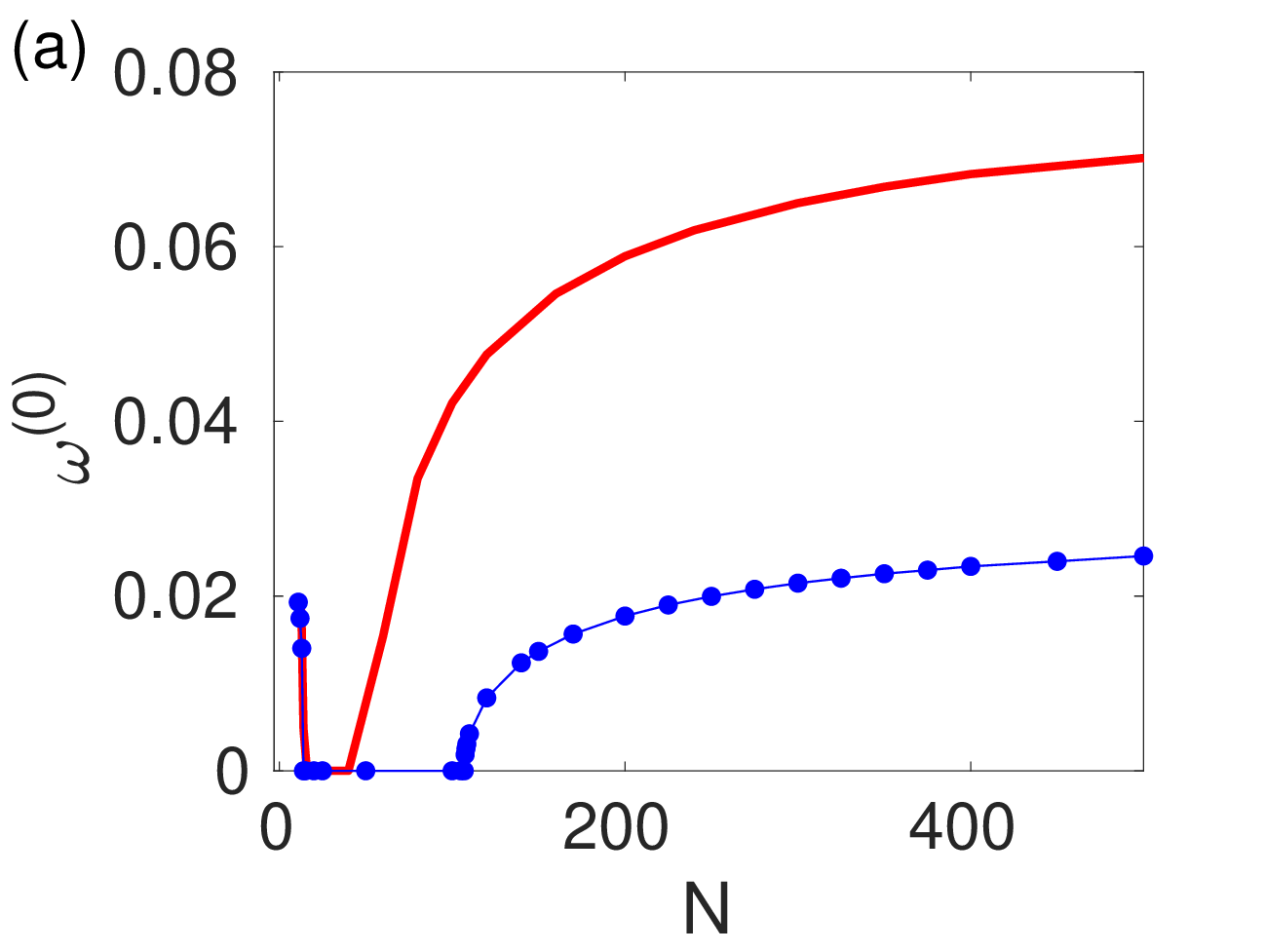}
  \includegraphics[width=4.4cm]{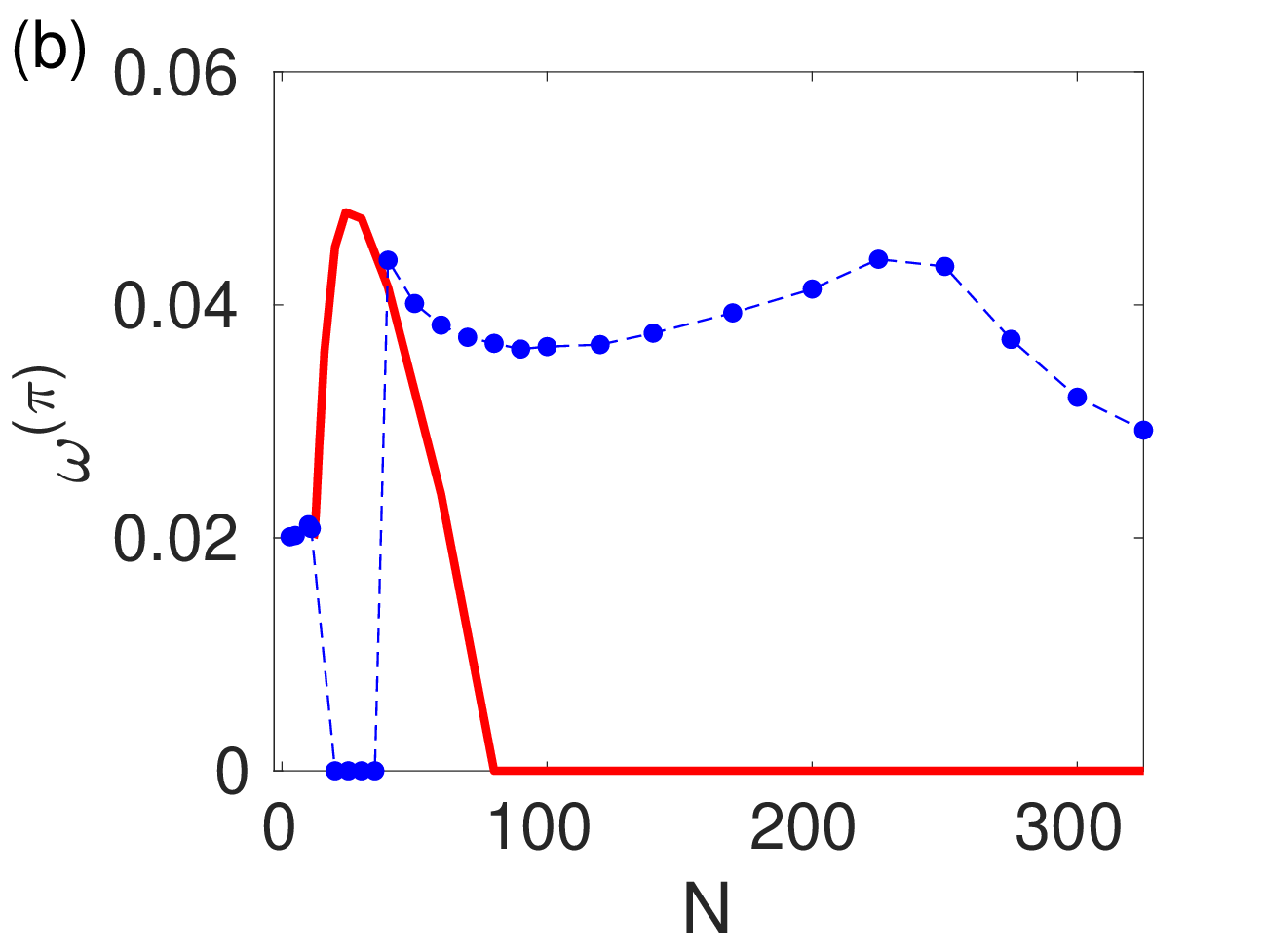} }
\caption{Dependence of the Josephson oscillation frequency on the total atom number $N$. Panel (a) shows the $\mathrm{zero}$-phase regime and panel (b) the $\pi$-phase regime. Points connected by lines denote results of direct numerical simulations. In both panels, the initial population imbalance is $Z_0=0.1$; parameters are $\kappa=0.01$ and $q=g=1$.}
\label{fig-freq}
\end{figure}

Figure~\ref{fig-freq} compares the Josephson oscillation frequency as a function of the total atom number $N$ for the zero- and $\pi$-phase modes. In panels (a) and (b), the red solid curves show the theoretical predictions obtained from Eqs.~(\ref{JFzero}) and (\ref{JFpi}), while the blue points connected by a line represent the frequencies extracted from direct numerical simulations of Eq.~(\ref{eq:gpe}). For sufficiently small $N$, the theoretical and numerical results are in close agreement, confirming the validity of the reduced Josephson description in the smaller norms (weakly nonlinear regime). 
As $N$ increases, the overall trend remains qualitatively similar, but noticeable quantitative deviations emerge. For fixed $(q, g, \kappa)$, Eq.~(\ref{JFzero}) predicts that the frequency becomes complex once the expression under the square root changes sign. In the frequency-norm plot, this interval is displayed as a zero-frequency segment, producing an apparent discontinuity. Physically, this indicates the loss of small-amplitude Josephson oscillations and the onset of the self-trapping (running-phase) regime. The numerical simulations exhibit a similar interval, although its boundaries are slightly shifted and typically broader than those predicted by the theoretical formula, which shows the limitations of the underlying variational assumptions. Importantly, this self-trapping window is not universal: by tuning $q$, $g$, and $\kappa$, the oscillatory regime can be restored (or extended) over the same range of $N$, and the numerical simulations confirm such behaviour.

Figure~\ref{fig-freq} (b) corresponds to the $\pi$-phase mode. Here, the variational prediction remains reliable only at small $N$. The simulations further indicate that, within the finite time interval where Josephson oscillations are expected, the two droplets may separate after several oscillation periods (typically $\sim$10--12, see Fig.~\ref{fig:dynamLargerQD} (b)), thereby ceasing to sustain a persistent periodic exchange. This behaviour can be attributed to the coupling contribution to the interaction energy, which can be found for $S=0,m=1$ as
\begin{eqnarray*}
E_{\mathrm{int}} \propto \kappa \iint dx\,dy \left(\psi_1^{\ast}\psi_2+\psi_2^{\ast}\psi_1\right) = \nonumber\\
\frac{\pi  \kappa A^2}{4 a^2}e^{-\frac{1}{4}a^2 (\Delta x^2 +\Delta y^2)} \, \cos(\theta),
\end{eqnarray*}
so that the effective interaction changes character depending on the relative phase difference, in the $\pi$-phase configuration, this may promote an effectively repulsive tendency, driving spatial separation (see Ref.~\cite{Otajonov2026}).

For larger $N$, discrepancies between the analytical and numerical frequencies become noticeable in both phase modes. The primary source is the simplifying ansatz used to derive the Josephson equations: both components were assumed to share identical, time-independent super-Gaussian parameters, $m_1=m_2=m$, and equal widths, $w_1=w_2=1/a$, while only the amplitudes were allowed to evolve in time. In the full model, however, increasing $N$ substantially modifies the droplet shape: the width and the super-Gaussian index $m$ vary strongly with $N$ (and may also vary dynamically during oscillations), whereas the peak amplitude tends to saturate as the droplet becomes more liquid-like and nearly incompressible. As a result, the reduced description progressively loses quantitative accuracy at high norms. A more refined variational approach, in which $w(t)$ and $m(t)$ are treated as dynamical variables and may differ between the two components, can improve the agreement; however, this comes at the expense of substantially greater analytical complexity.

\begin{figure}[htbp]
   \centerline{ \includegraphics[width=4.5cm]{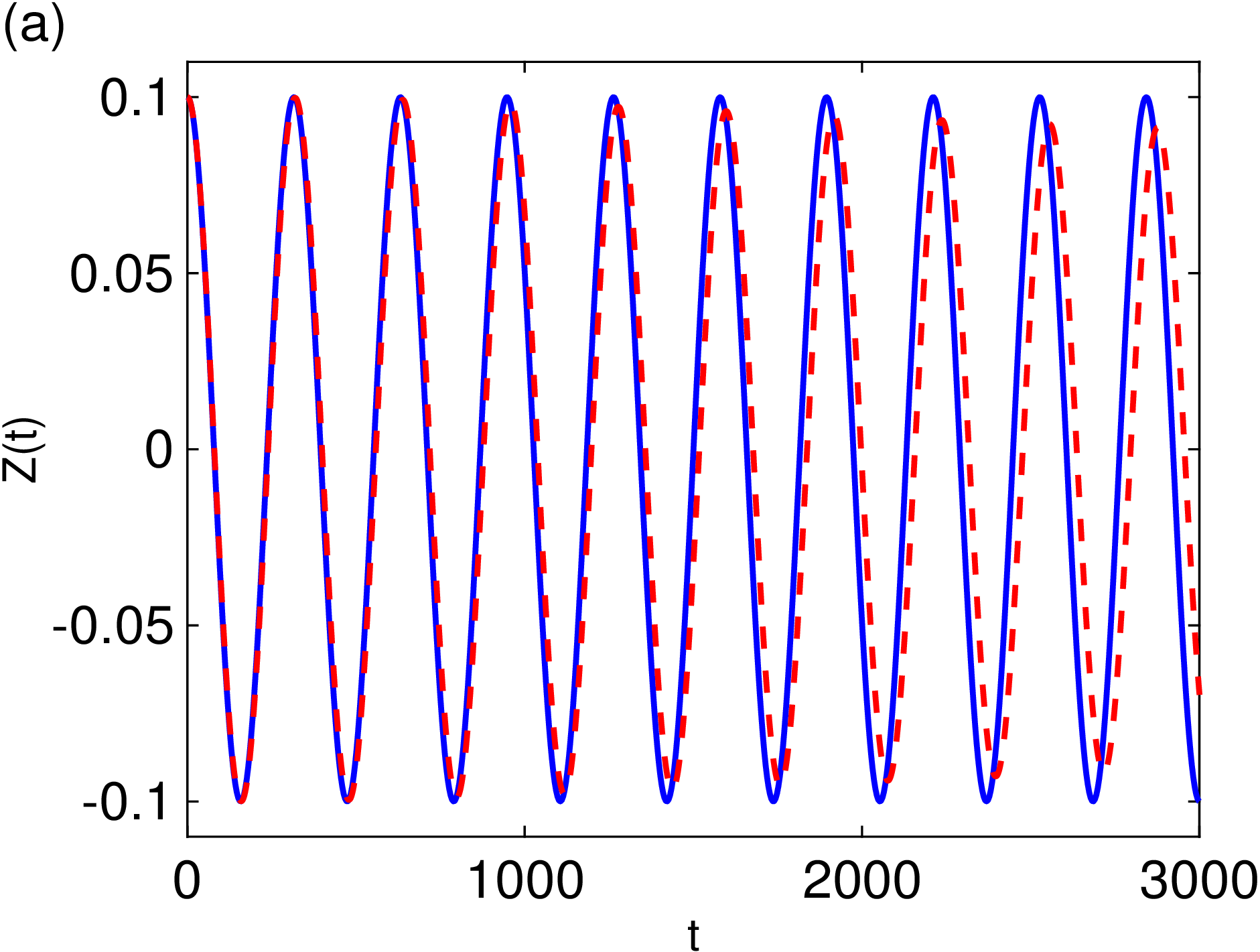} \hskip-0.1cm
      \includegraphics[width=4.5cm]{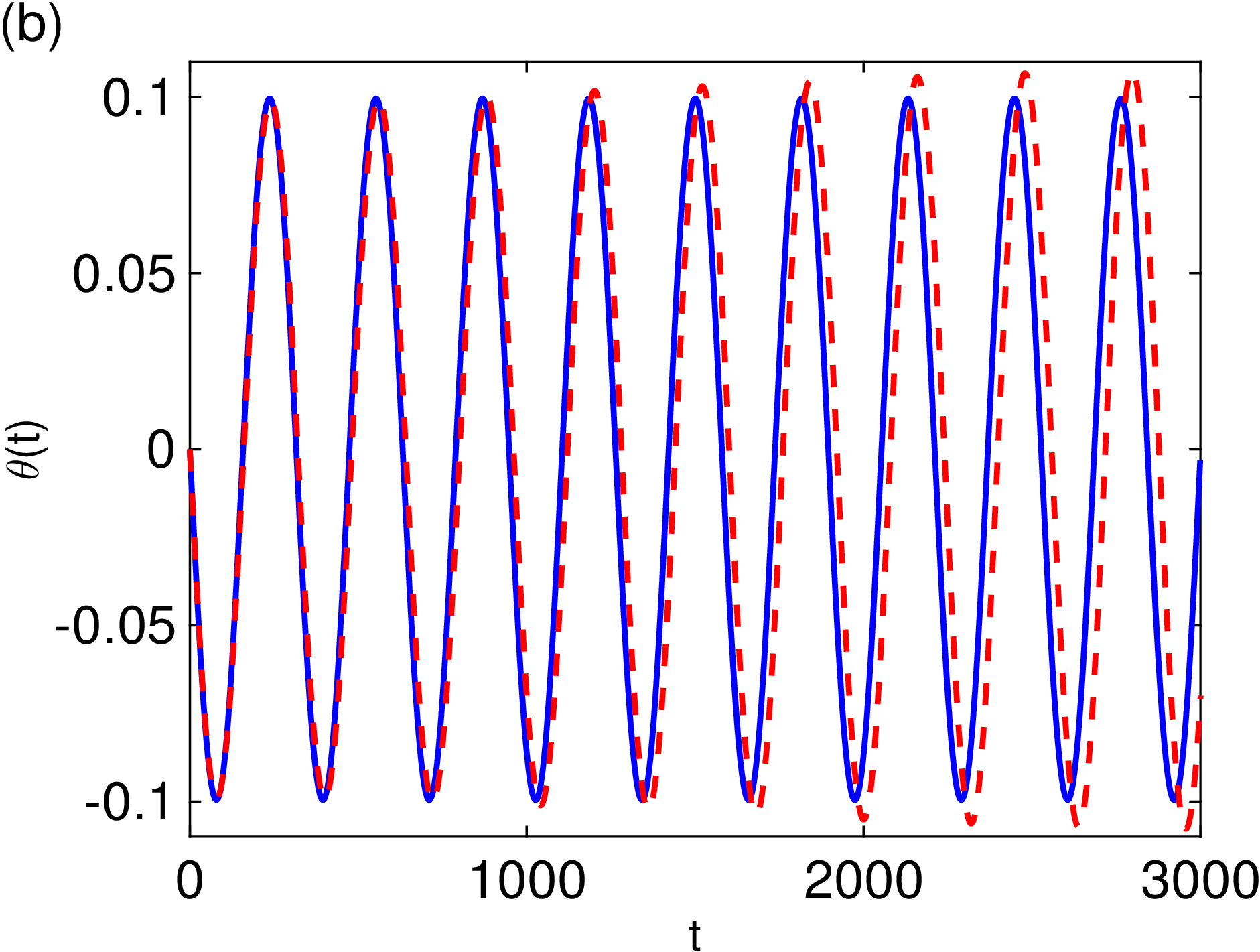}}
\caption{Typical time evolutions of the population imbalance $Z(t)$ and the relative phase $\theta(t)$ are shown for the small norm: (a) $Z(t)$ and (b) $\theta(t)$. In both panels, the blue solid curves depict the theoretical predictions obtained from Eq.~(\ref{JFzero}), while the red dashed curves show the corresponding results of direct numerical simulations of Eq.~(\ref{eq:gpe}). Parameters are $Z_0=0.1$, $N=4$, $\theta_0=0$, $q=1$, $g=1$, and $\kappa=0.01$.}
\label{fig:smallQDdynam}
\end{figure}
Figure~\ref{fig:smallQDdynam} shows representative time evolutions of the population imbalance $Z(t)$ and the relative phase $\theta(t)$. For relatively small norms (here, $N=4$), the results obtained from direct numerical simulations and the VA are in close agreement, indicating that the reduced Josephson description captures the essential dynamics in the small-norm regime. In particular, both approaches reproduce the oscillation period and amplitude of $Z(t)$, as well as the corresponding bounded phase dynamics $\theta(t)$, confirming that changes in the droplet shape remain sufficiently weak throughout the evolution.

Figure~\ref{fig:dynamLargerQD} (a, b) presents the time evolution of the relative population imbalance $Z(t)$ for two different initial phase configurations. Panel (a) corresponds to the zero-phase mode, where the system exhibits the standard Josephson-oscillation dynamics: in the direct numerical simulations, the two quantum droplets remain mutually bound over long evolution times and sustain a persistent, nearly periodic exchange of particles. 
\begin{figure}[htbp]
   \centerline{ \includegraphics[width=4.5cm]{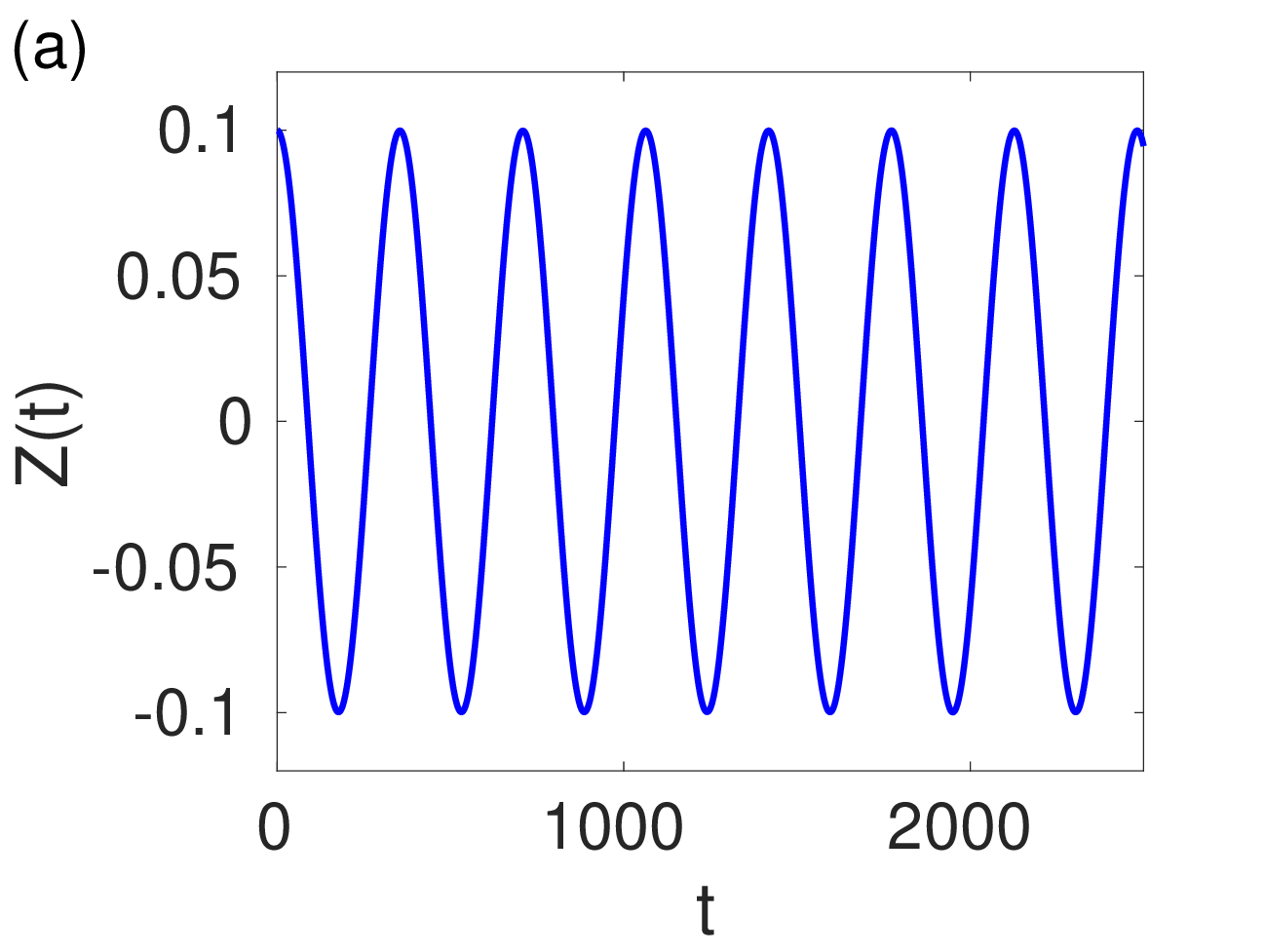} \hskip-0.1cm
      \includegraphics[width=4.5cm]{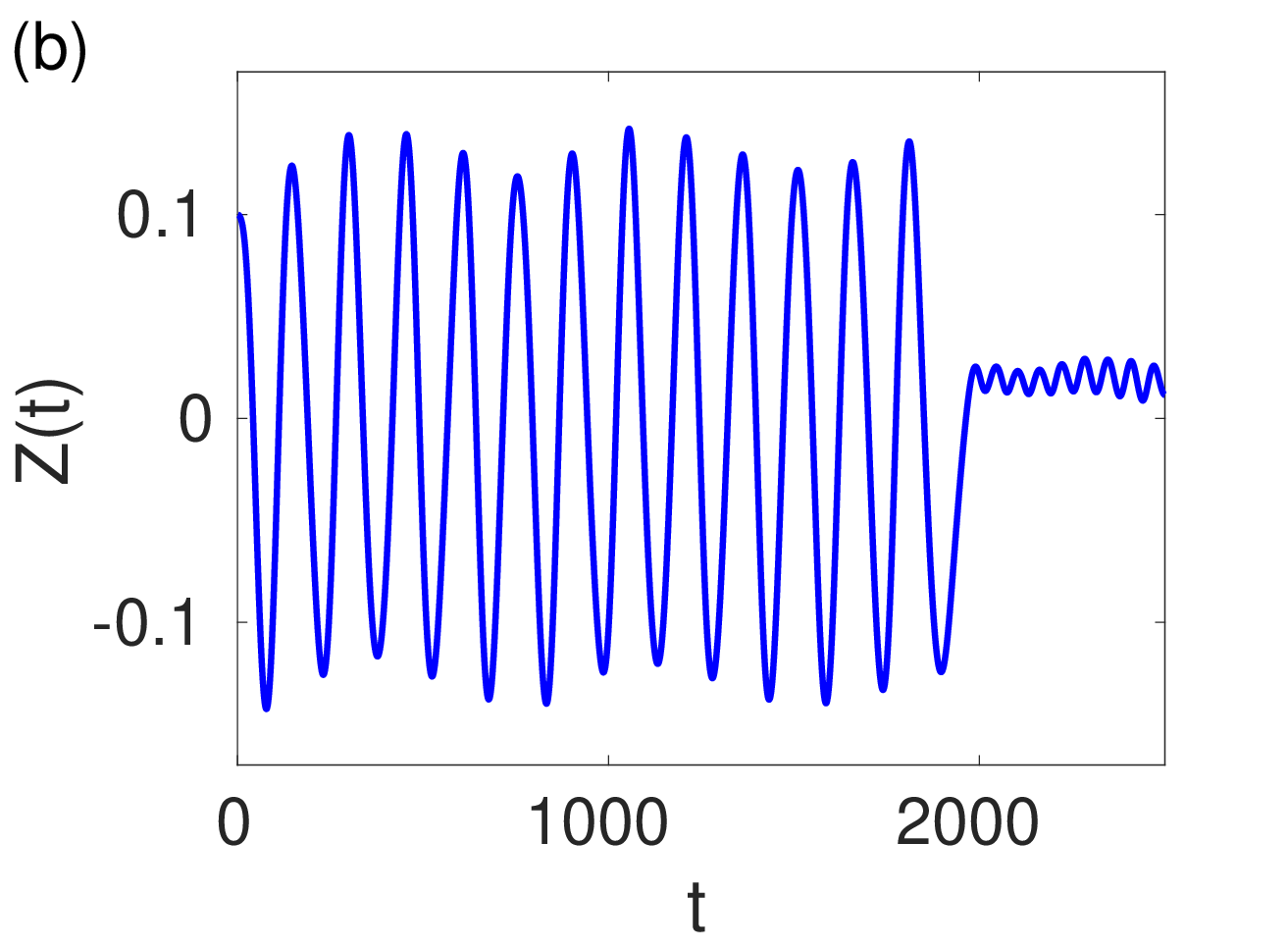}}
   \centerline{ \includegraphics[width=4.5cm]{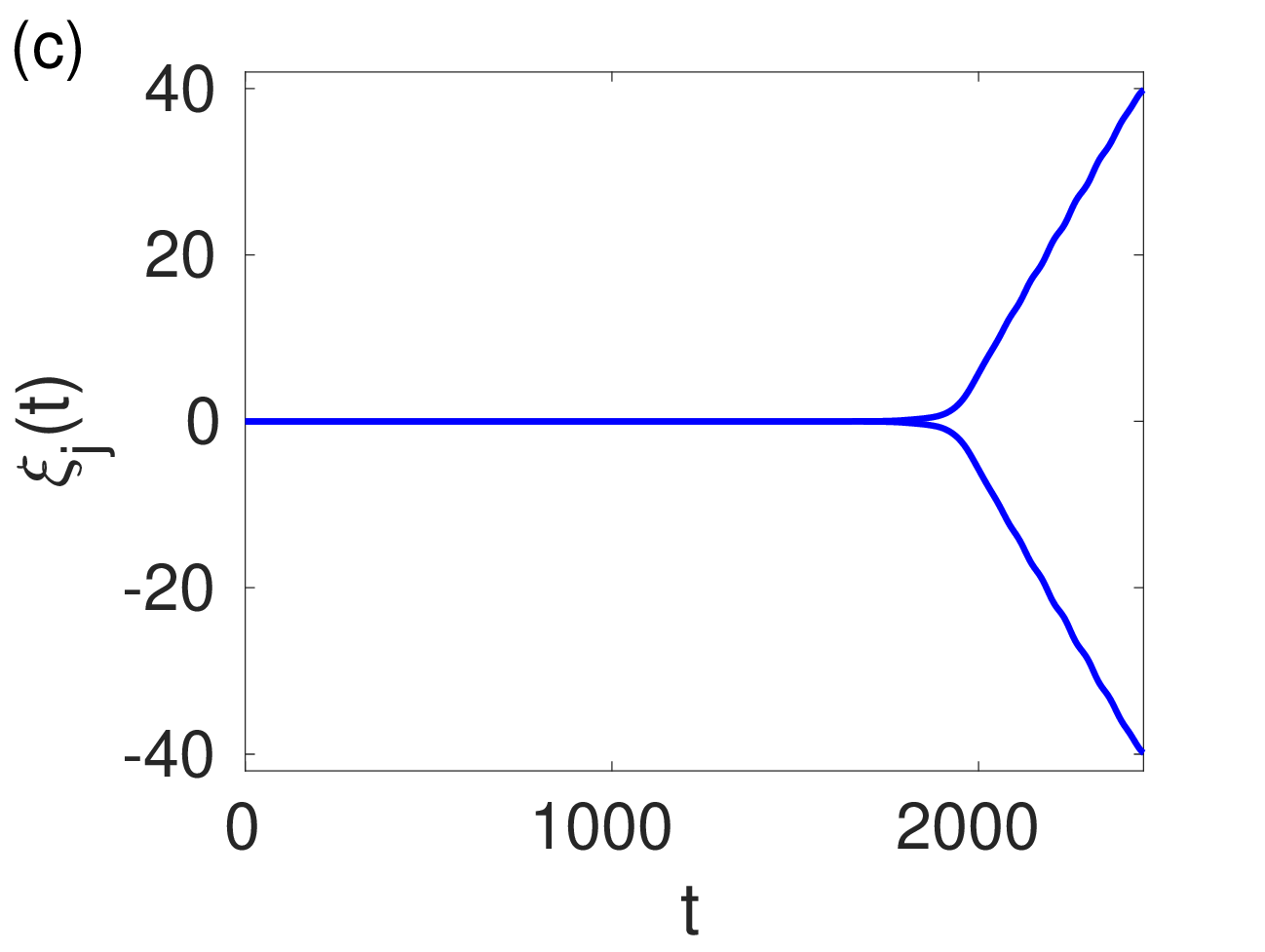} \hskip-0.1cm
      \includegraphics[width=4.5cm]{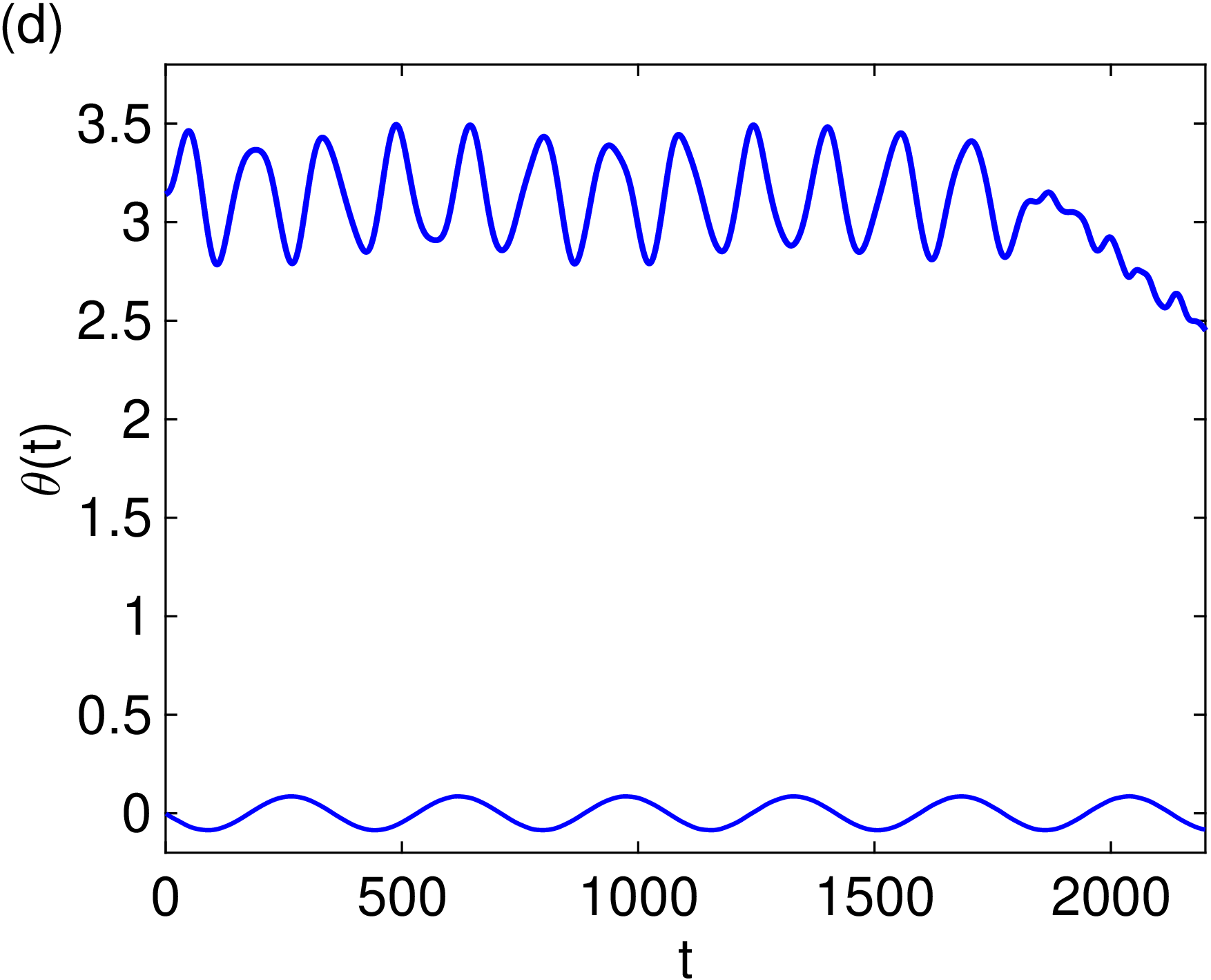}}
\caption{The time evolution of the population imbalance $Z(t)$ for (a) the zero-phase mode and (b) the $\pi$-phase modes are presented. For the $\pi$-phase configuration, the associated centre-of-mass dynamics, plotted in panel (c), indicate that after several Josephson oscillation periods the two quantum droplets drift apart and effectively decouple, leading to a breakdown of sustained population exchange. The corresponding relative-phase dynamics $\theta(t)$ is shown in panel (d): the upper curve corresponds to the zero-phase mode, while the lower curve corresponds to the $\pi$-phase mode. The simulations are performed for $Z_0=0.1$, $N=200$ with $q=1$, $g=1$, and $\kappa=0.01$. }
\label{fig:dynamLargerQD}
\end{figure}

	In contrast, panel (b) shows the $\pi$-phase regime. Here, after only a few Josephson oscillation periods, the droplets lose their bound configuration and separate, which terminates the coherent population exchange. This qualitative difference is consistent with the phase-sensitive nature of the inter-core coupling: because the coupling contribution to the interaction energy scales as $\sim \cos\theta$, the $\pi$-phase configuration can effectively promote repulsive behaviour, thereby driving spatial decoupling.
The associated centre-of-mass dynamics is plotted in Fig.~\ref{fig:dynamLargerQD} (c). One can clearly see that at approximately $t\approx 2000$ the oscillatory exchange ceases and the droplet centres begin to drift apart, indicating the onset of separation and effective suppression of tunnelling.Figure~\ref{fig:dynamLargerQD} (d) shows the corresponding evolution of the relative phase $\theta(t)$: the upper curve corresponds to the $\pi$-phase mode and the lower curve to the zero-phase mode. Notably, the zero-phase case maintains bounded phase oscillations consistent with sustained Josephson dynamics, whereas the $\pi$-phase case evolves toward a running or irregular phase behaviour as the droplets decouple.

\begin{figure}[htbp]
\centerline{%
\includegraphics[width=4.7cm]{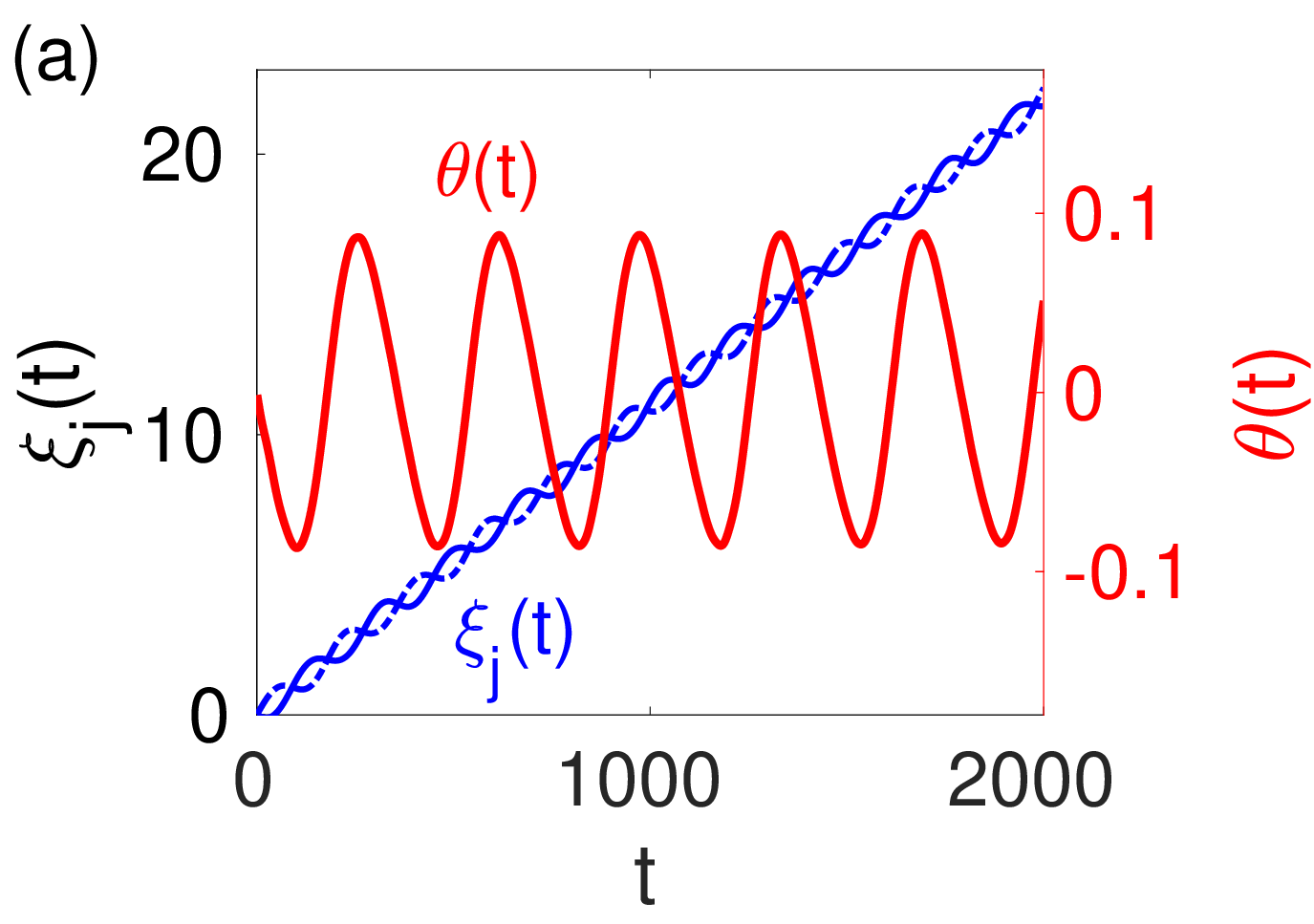}\hskip 0.2cm
\includegraphics[width=4.5cm]{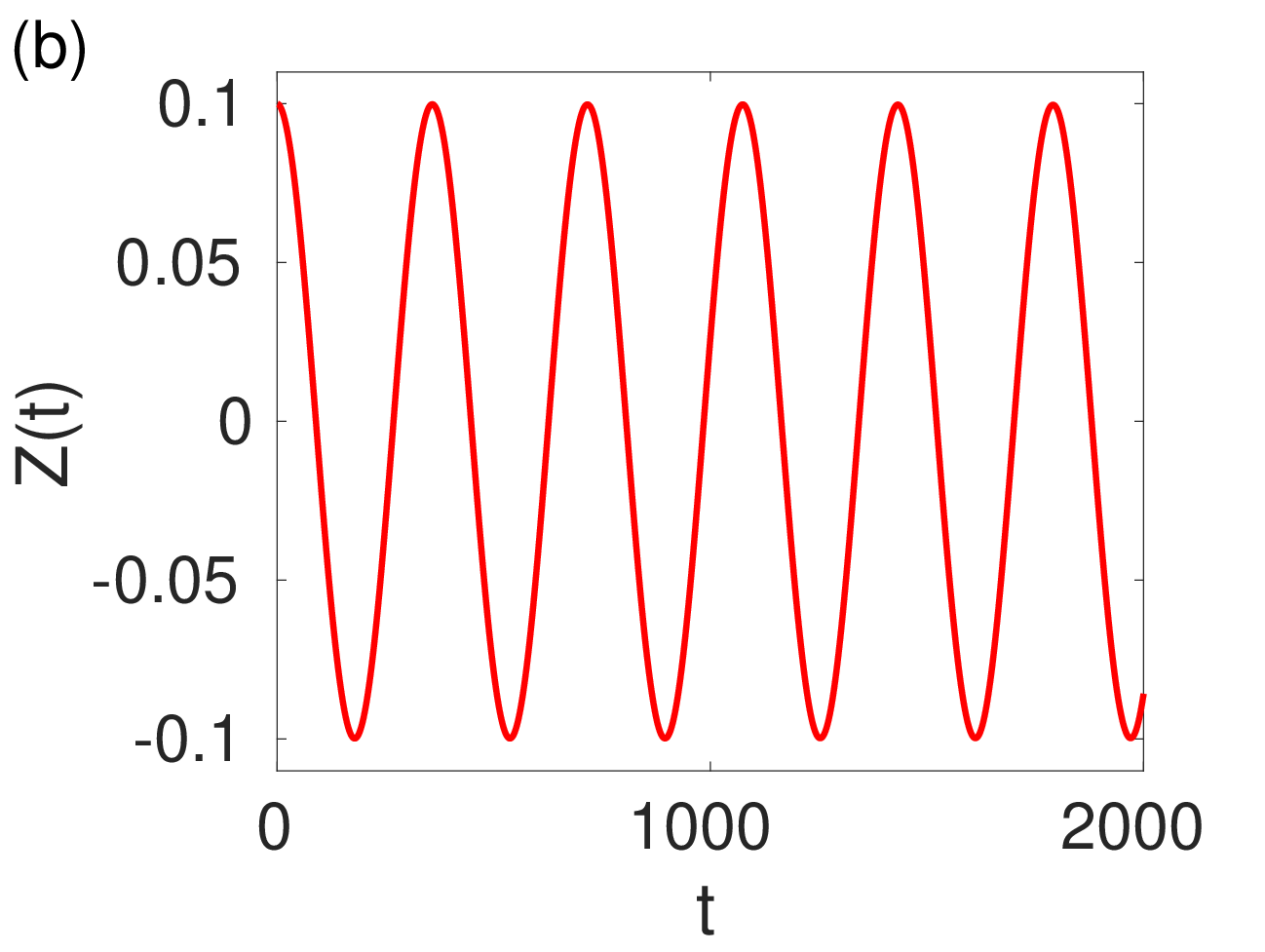}}
\caption{(a) Time evolution of the center-of-mass coordinates $\xi_j(t)$ (left axis) and the relative phase $\theta(t)$ (right axis). Solid and dash-dotted blue curves correspond to $\xi_1(t)$ and $\xi_2(t)$, respectively, while the solid red curve shows $\theta(t)$. (b) Dynamics of the population imbalance $Z(t)$ between the droplets.
In both panels, at $t=0$, the second component is imparted with a small momentum kick $k=0.02$ along the positive $x$-axis. Simulation parameters: $\theta_0=0$, $q=g=1$, $N=200$, and $\kappa=0.01$.}
\label{fig-AB}
\end{figure}

The Andreev-Bashkin effect (also termed nondissipative drag or entrainment) is a defining feature of multi-component superfluidity: a superflow in one component can induce a co-directed mass current in the other component even in the absence of dissipation or viscosity~\cite{Andreev1976, Nespolo2017}. In our system, an analogous entrainment mechanism may operate between the two quantum droplets. Specifically, if one droplet is set into motion, the intercomponent coupling can transfer momentum to the other droplet, so that both droplets propagate together as a bound, co-moving pair over an extended time interval. Figure~\ref{fig-AB} provides a representative illustration: at ($t=0$), the droplet of the second component (multiplied by $\exp(i k x)$) is imparted a small “kick” by imprinting a small phase gradient $k=0.02$ along the positive $x$ direction, while the first-component droplet is initially at rest. During the subsequent evolution, the moving (kicked) droplet entrains the initially stationary droplet, and the two centres of mass lock to a common velocity, demonstrating joint motion driven by nondissipative drag. The corresponding trajectories of the centres of mass, together with the evolution of the relative phase, are plotted in Fig.~\ref{fig-AB}(a).

The effectiveness of entrainment depends  on the degree of spatial overlap of the droplets and on the strength of intercomponent interactions. When the overlap is substantial and the coupling is sufficiently strong, the induced current can be appreciable and the co-moving state can be long-lived; in contrast, weak overlap or weak coupling suppresses the drag and allows the droplets to separate.

\section{Interactions of the Vortices}
\label{sec:VQDinteraction}

\begin{figure}[htbp]
   \centerline{ \includegraphics[width=4.2cm]{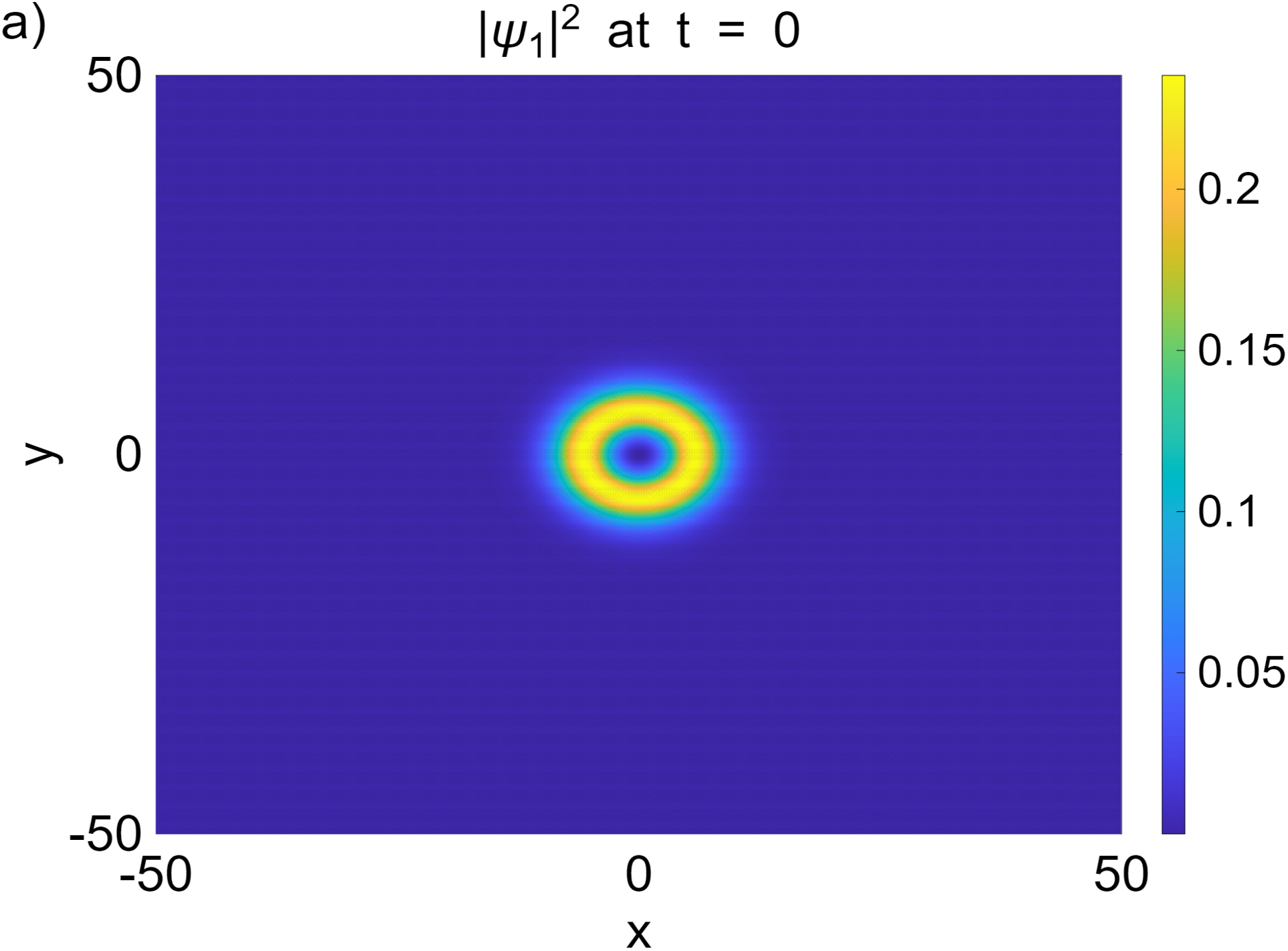} \hskip-0.1cm
      \includegraphics[width=4.4cm]{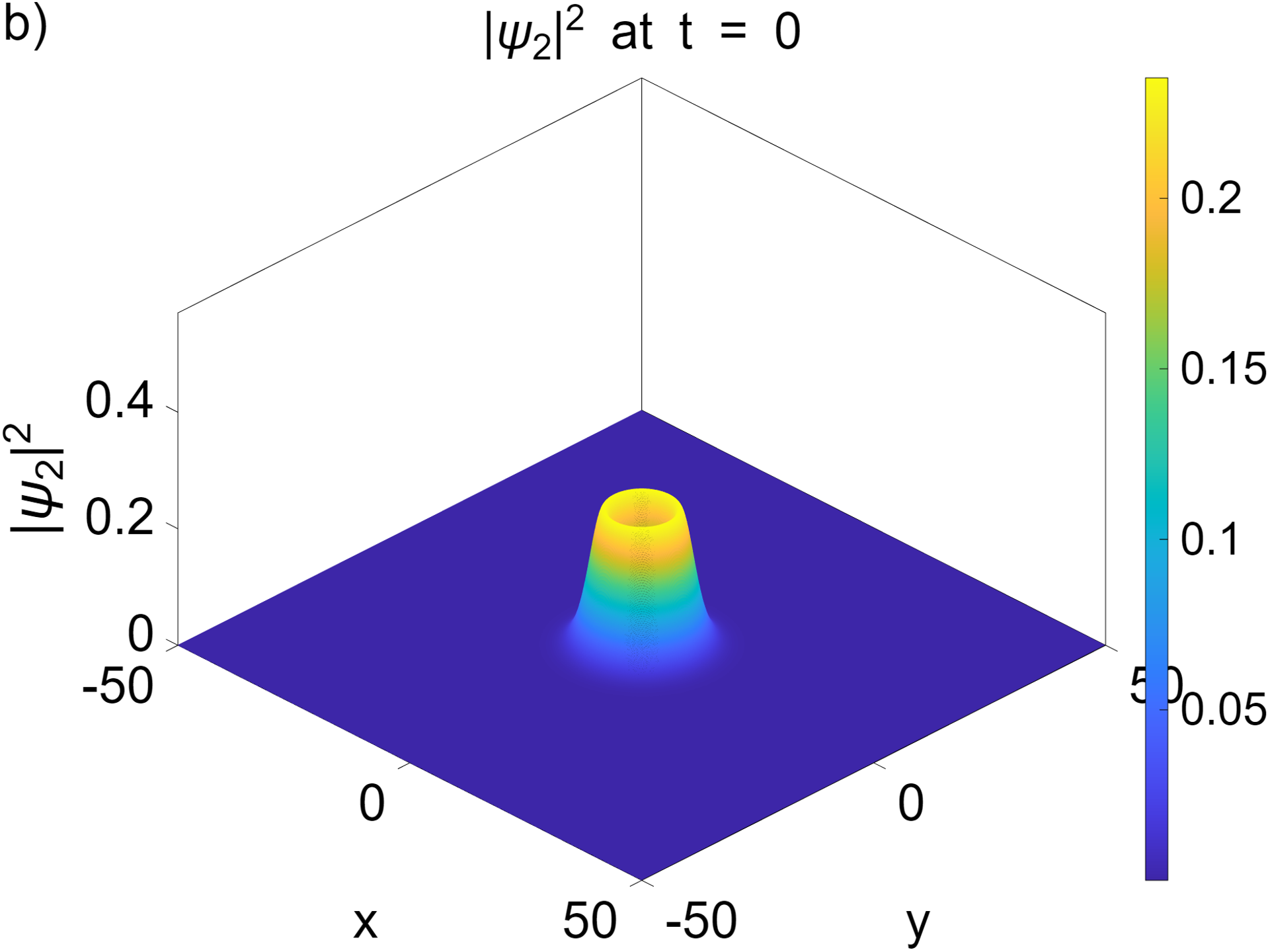}}
   \centerline{ \includegraphics[width=4.2cm]{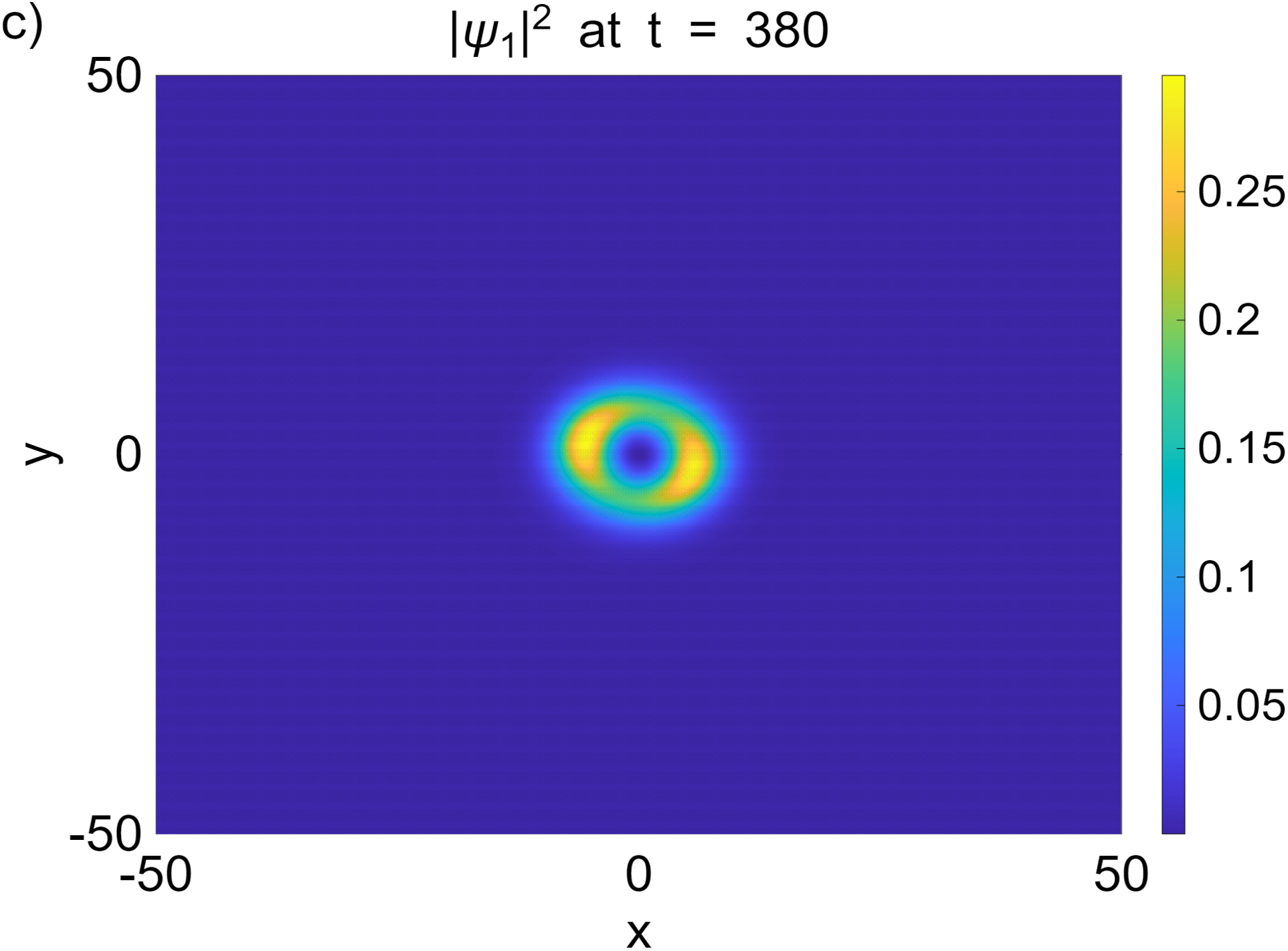} \hskip-0.1cm
      \includegraphics[width=4.4cm]{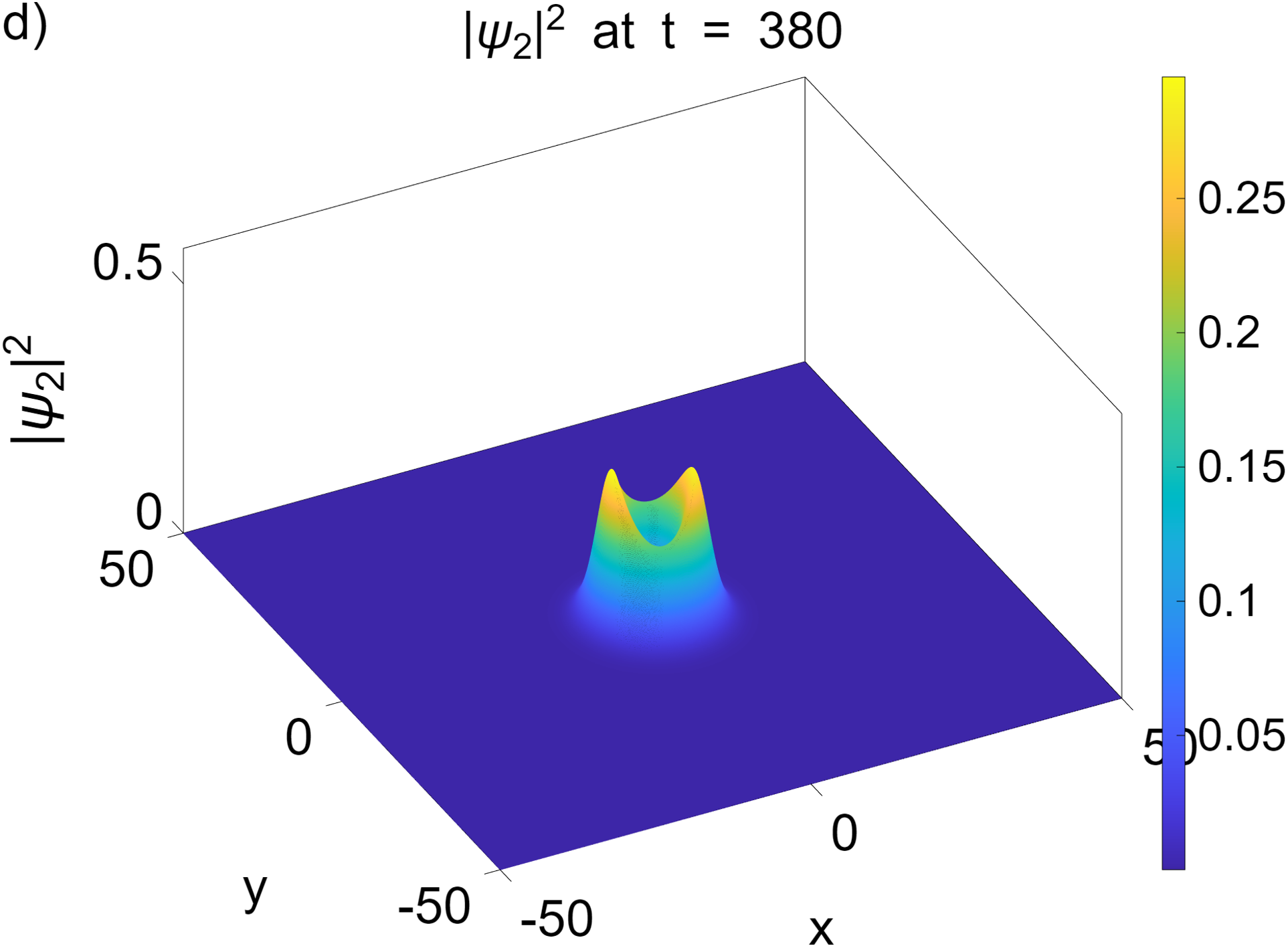}}
       \centerline{ \includegraphics[width=4.2cm]{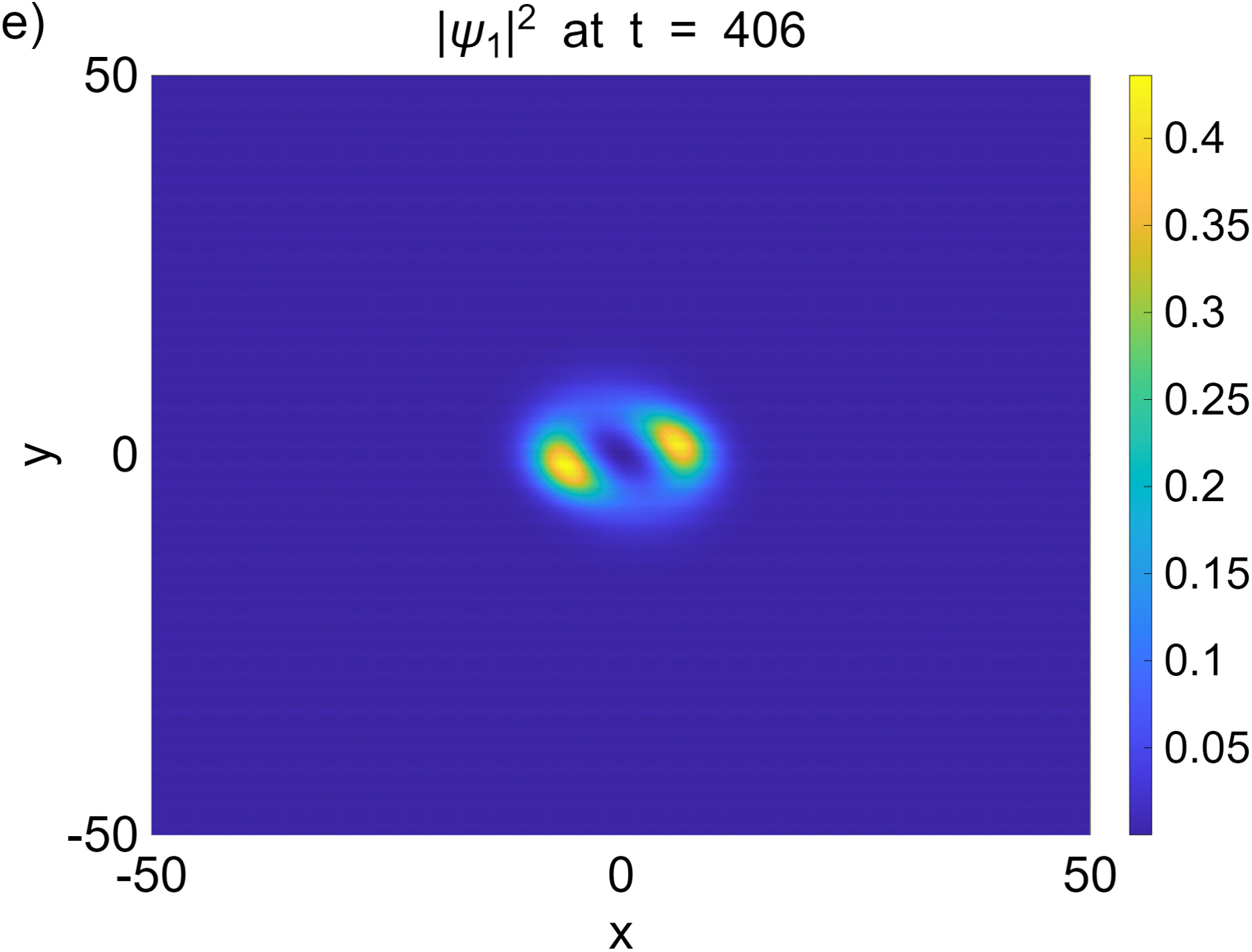} \hskip-0.1cm
      \includegraphics[width=4.4cm]{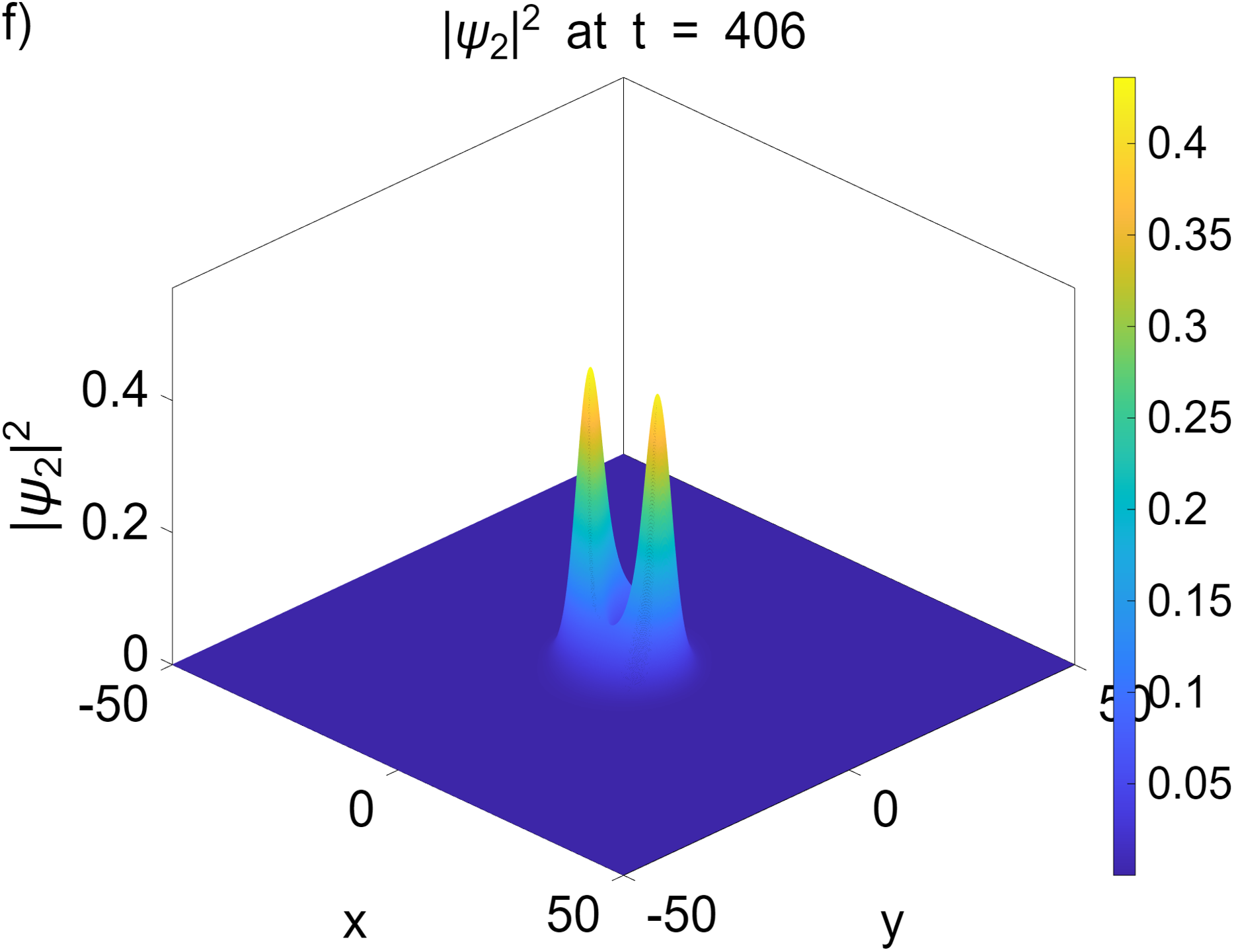}}
       \centerline{ \includegraphics[width=4.2cm]{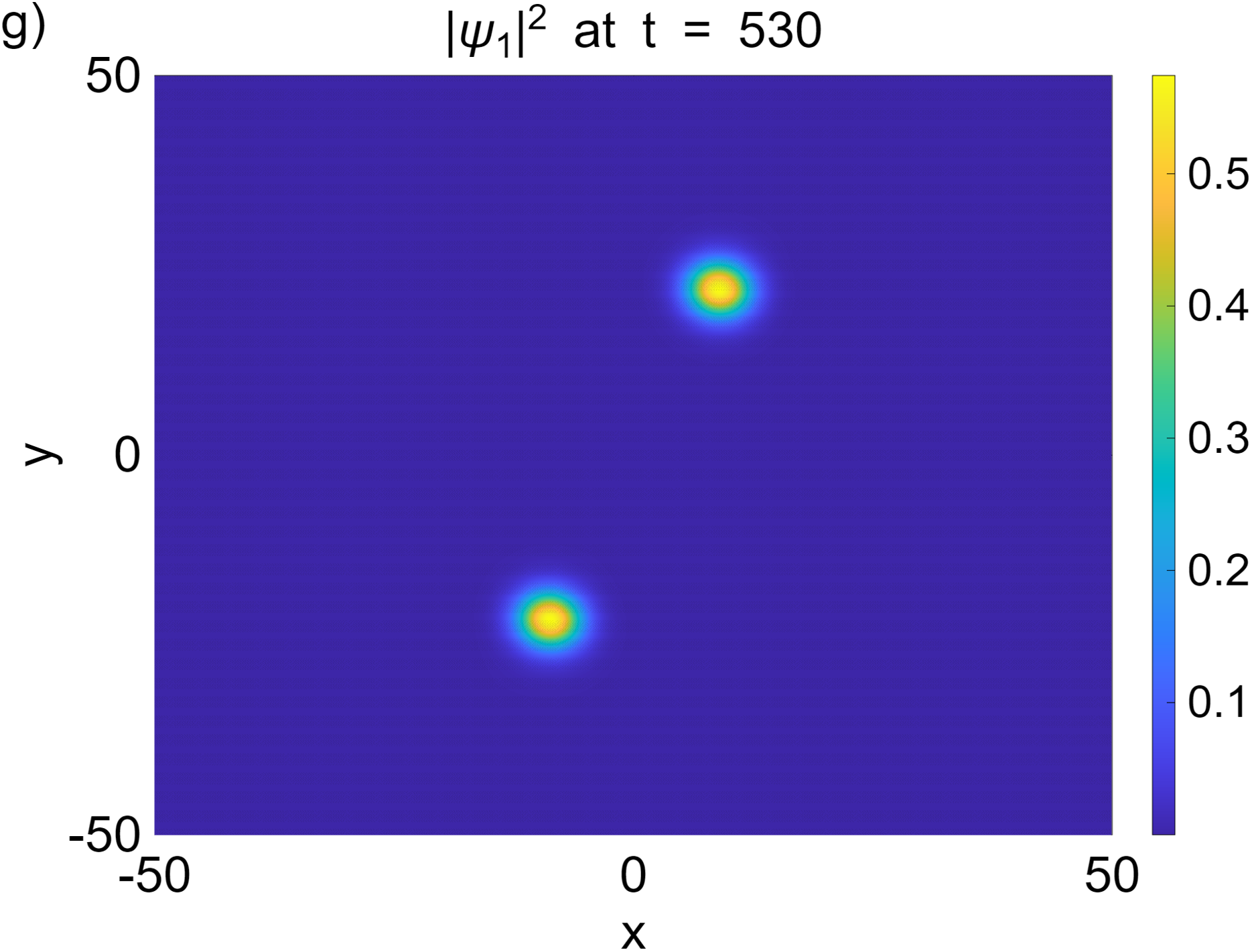} \hskip-0.1cm
      \includegraphics[width=4.4cm]{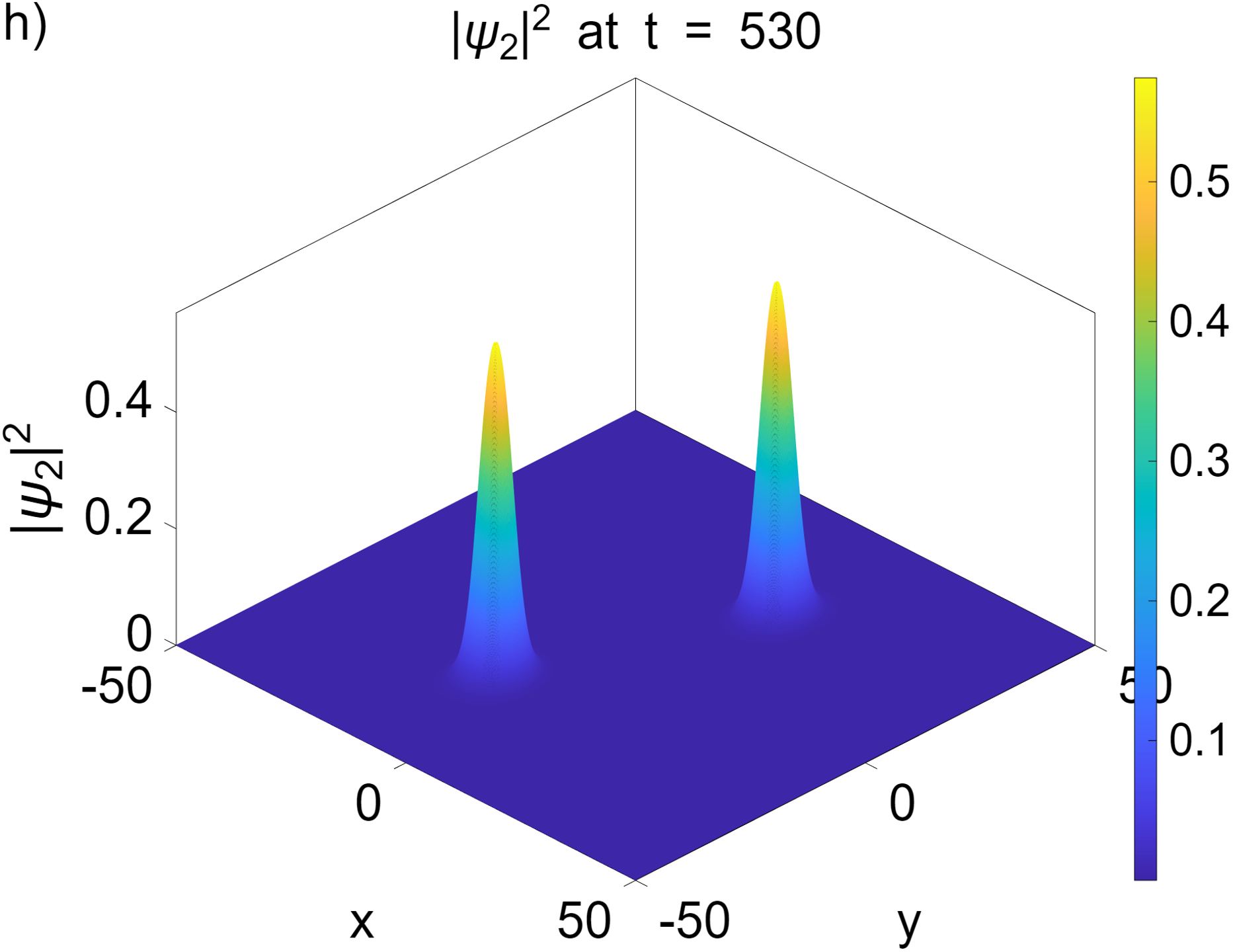}}
\caption{The density evolution of an unstable symmetric vortex state with winding number $S=1$ and initial norm $N=100$, obtained from direct simulations of Eq.~(\ref{eq:gpe}). In each column, the top and bottom rows show the density distributions of the first and second components, respectively, at the same evolution time. Panels (a,b) correspond to the initial state at $t=0$, while panels (c,d), (e,f), and (g,h) show the profiles at $t=380$, $t=406$, and $t=530$, respectively. Simulation parameters: $Z_0=0$, $\theta_0=0$, $q=g=1$, $N=100$, and $\kappa=0.01$. }
\label{fig:unstableQD}
\end{figure}

In this section, we examine Josephson-type transitions between vortex states. It is known from earlier works~\cite{Caplan2009, Li2018} that, in effectively self-attractive settings, vortex states may become unstable against the azimuthal (angular) modulational instability when the norm (equivalently, the atom number) is sufficiently small. Figure~\ref{fig:unstableQD} shows representative time evolutions of symmetric vortices with winding number $S=1$ taken from the unstable part of the parameter space, illustrating the onset of azimuthal instability and the ensuing fragmentation scenario. In each column, the first and second rows display the density distributions of the first and second components, respectively, at the same evolution time. At the initial stage, the vortices feature an axisymmetric ring-shaped density profile, as seen in panels (a) and (b). As the evolution proceeds, small azimuthal modulations emerge and grow into pronounced angular ripples, cf.\ panels (c)--(f). Ultimately, the ring breaks into a set of spatially separated localised fragments, as shown in panels (g) and (h).

In the present model, a vortex with topological charge (winding number) $S$ typically breaks up into $S+1$ (and, in some cases, $S+2$) fragments. Each fragment represents a fundamental zero-vorticity ($S=0$) droplet/soliton-like object. After the breakup, the fragments separate and propagate away from each other in opposite directions. Importantly, this scattering pattern is consistent with the zero total momentum of the initial symmetric vortex configuration, i.e., the fragments collectively conserve the vanishing centre-of-mass momentum. 

\begin{figure*}[t]
\includegraphics[width=4.5cm]{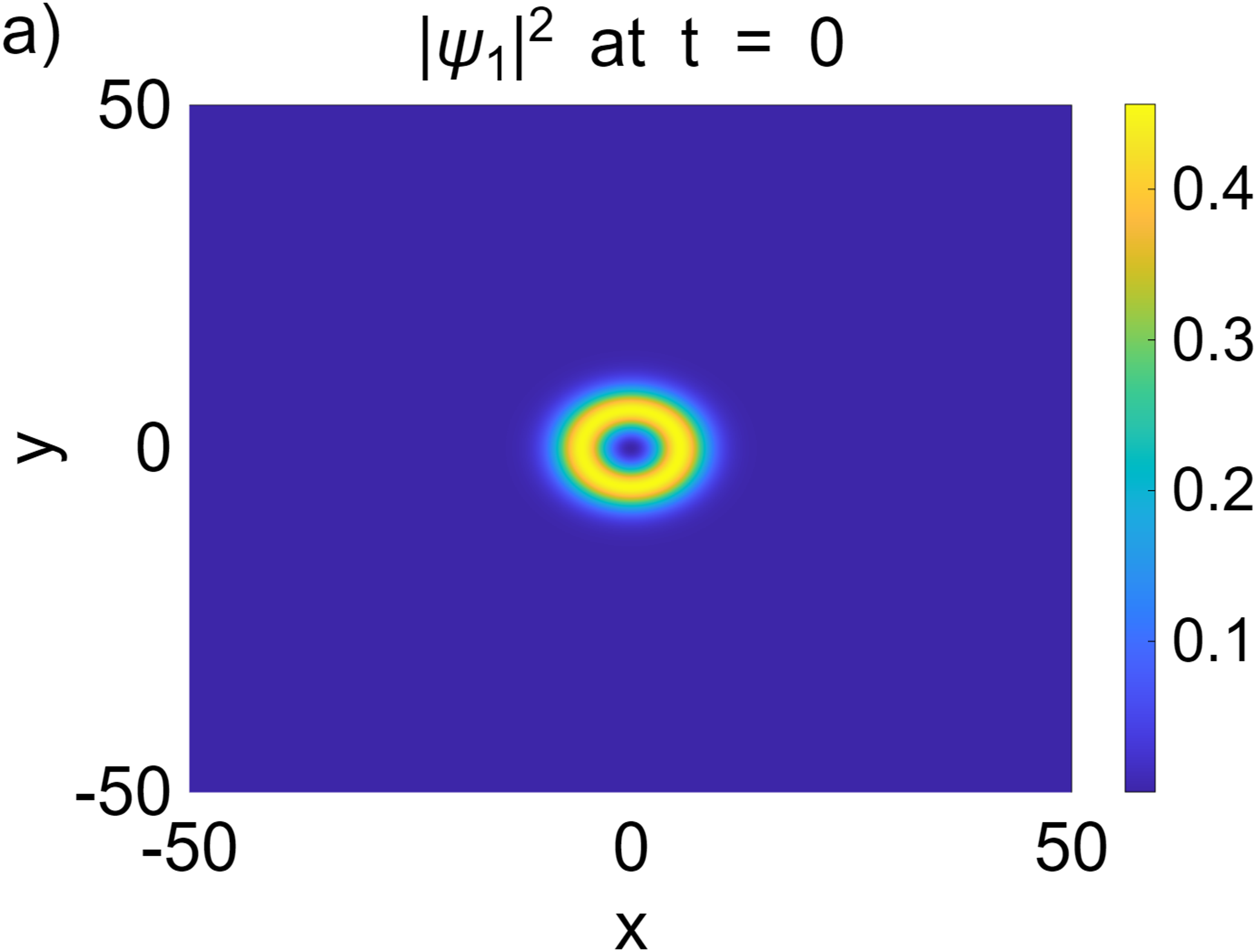}
\includegraphics[width=4.3cm]{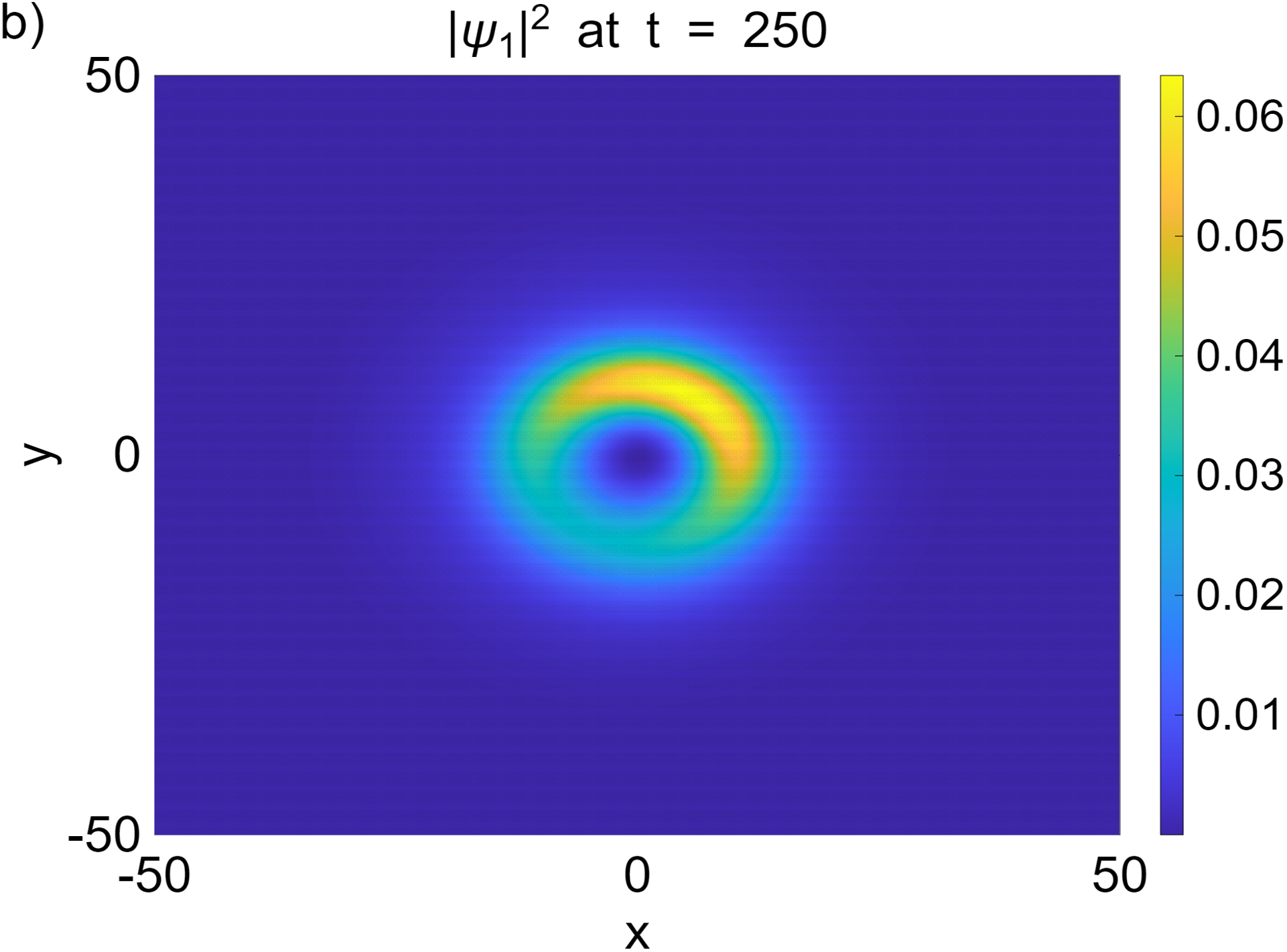}
\includegraphics[width=4.3cm]{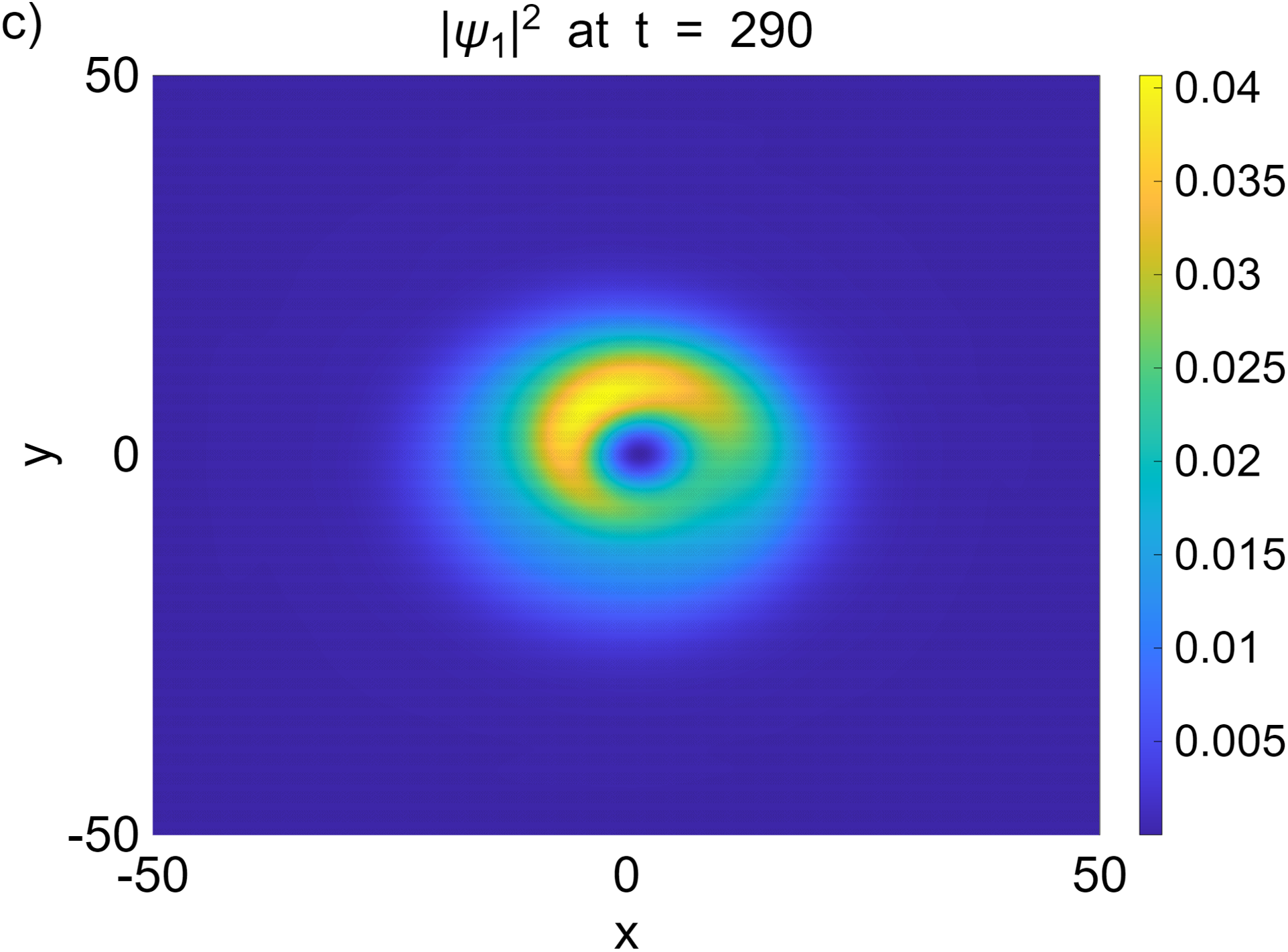}
\includegraphics[width=4.3cm]{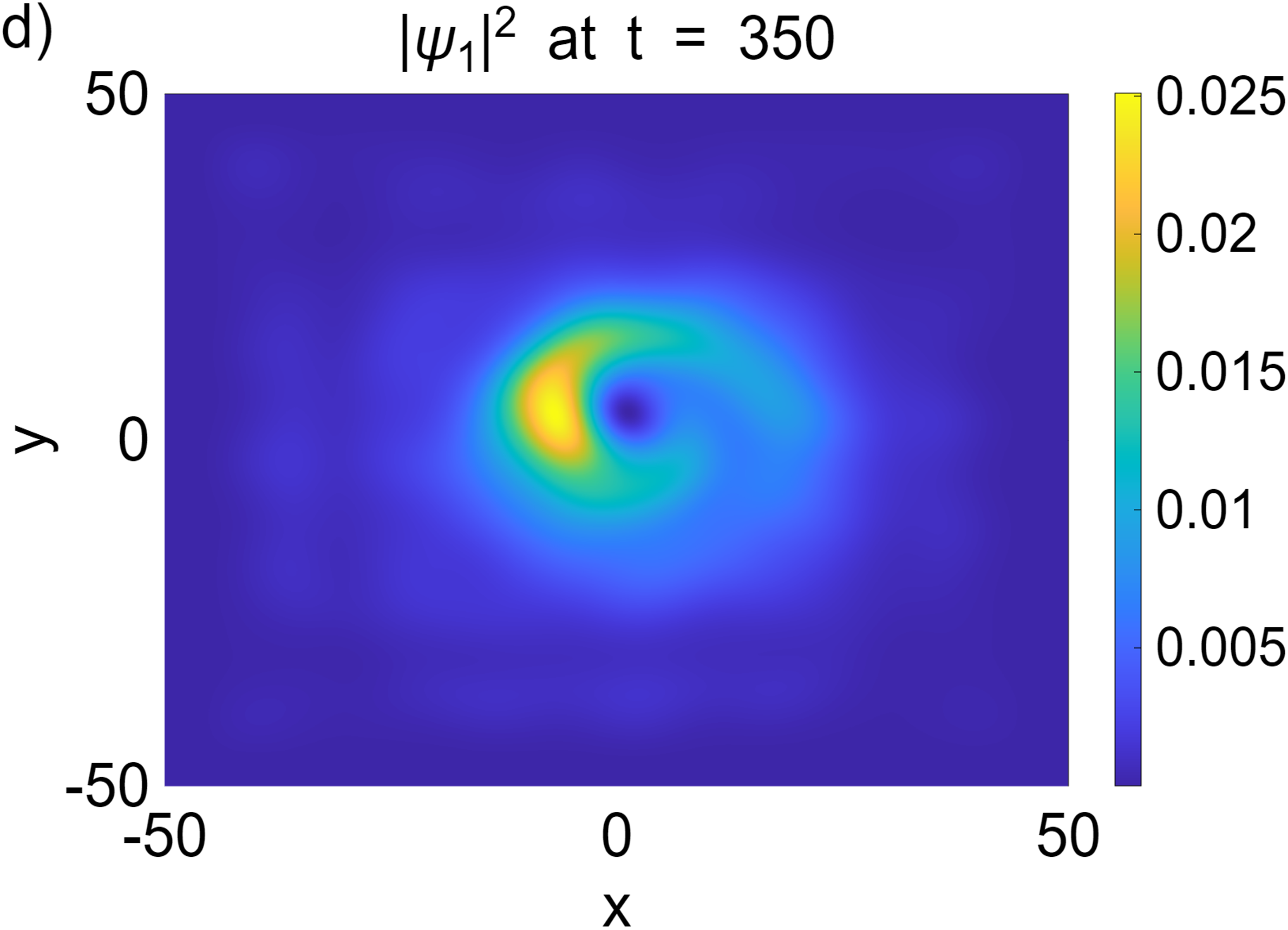}
\includegraphics[width=4.3cm]{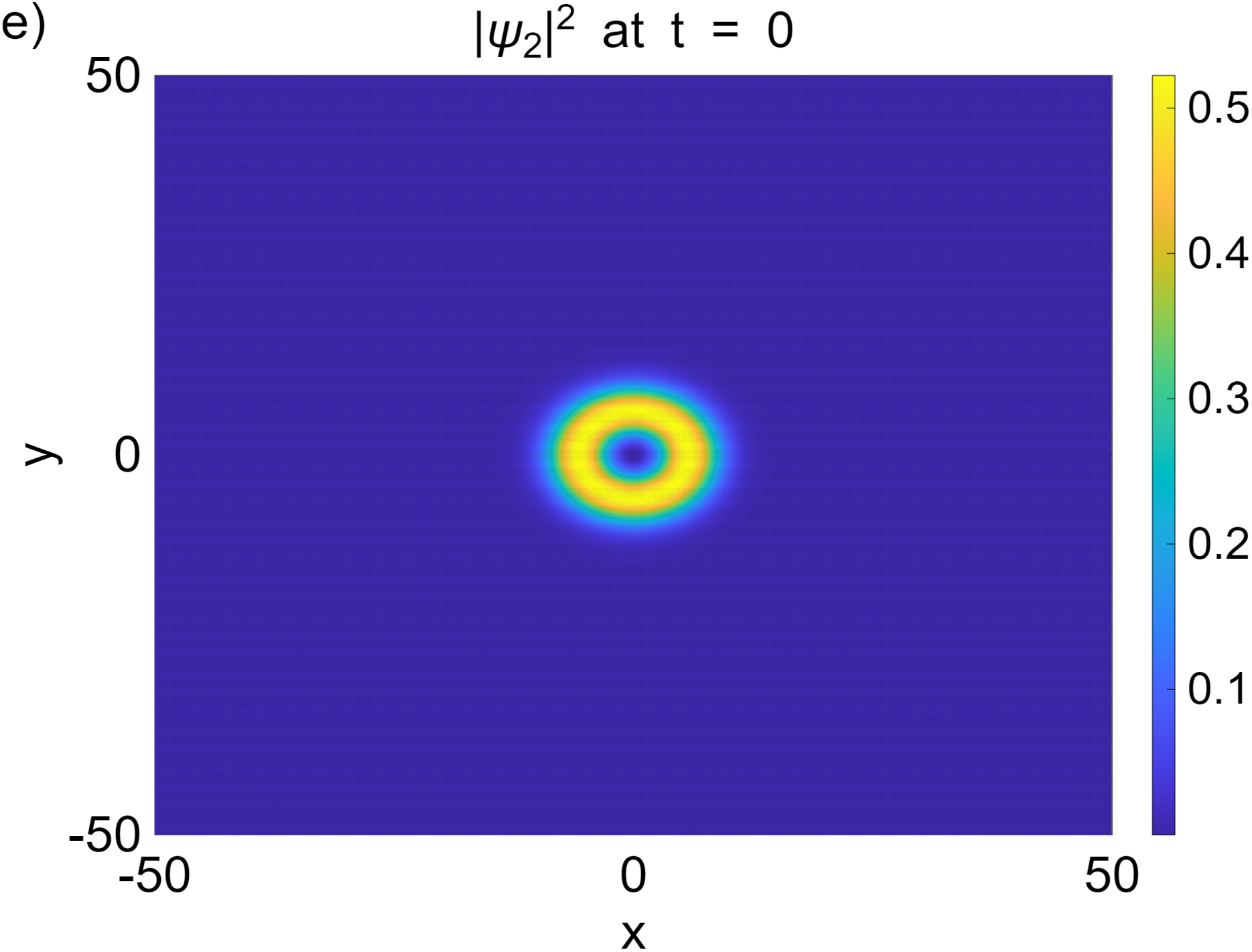}
\includegraphics[width=4.3cm]{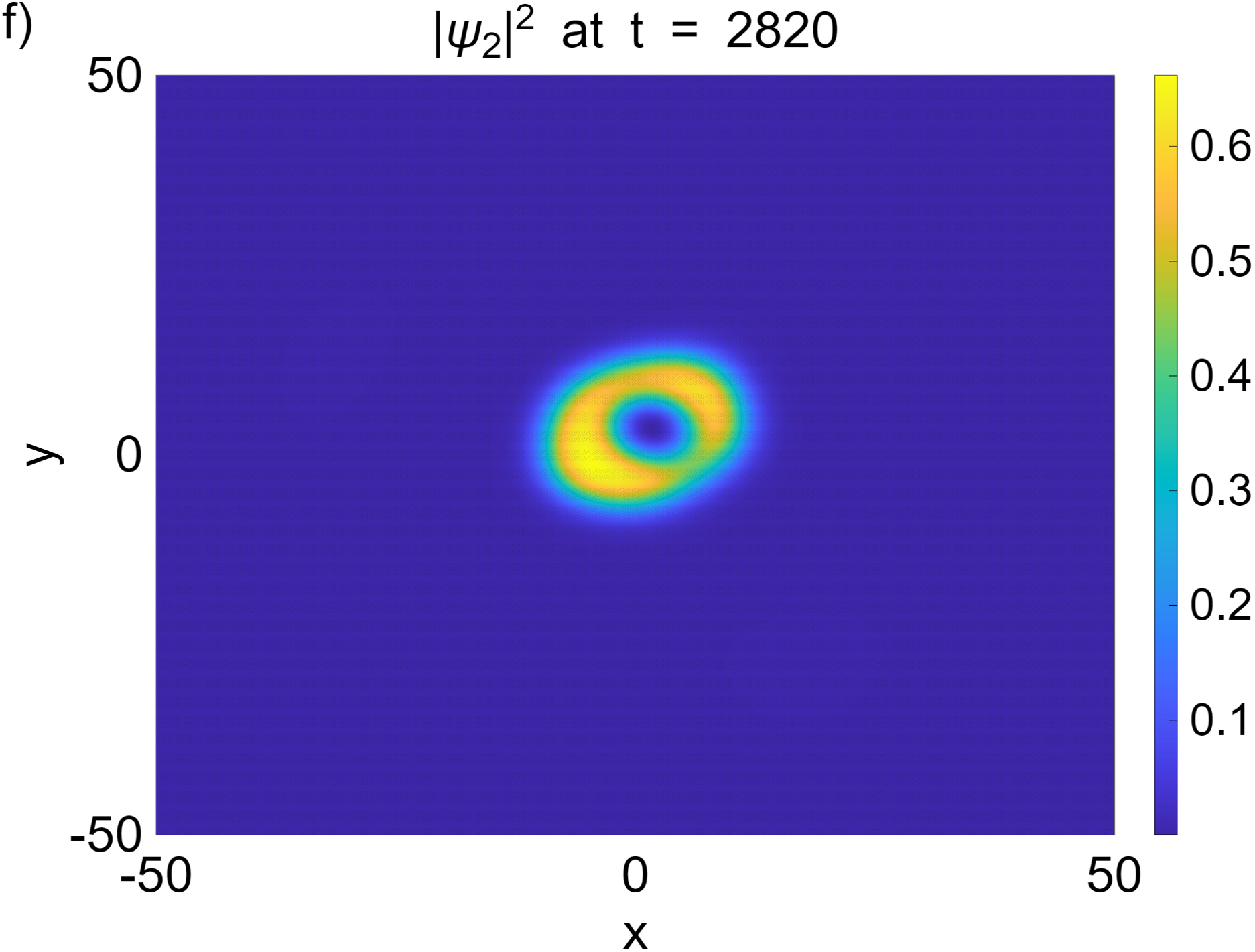}
\includegraphics[width=4.3cm]{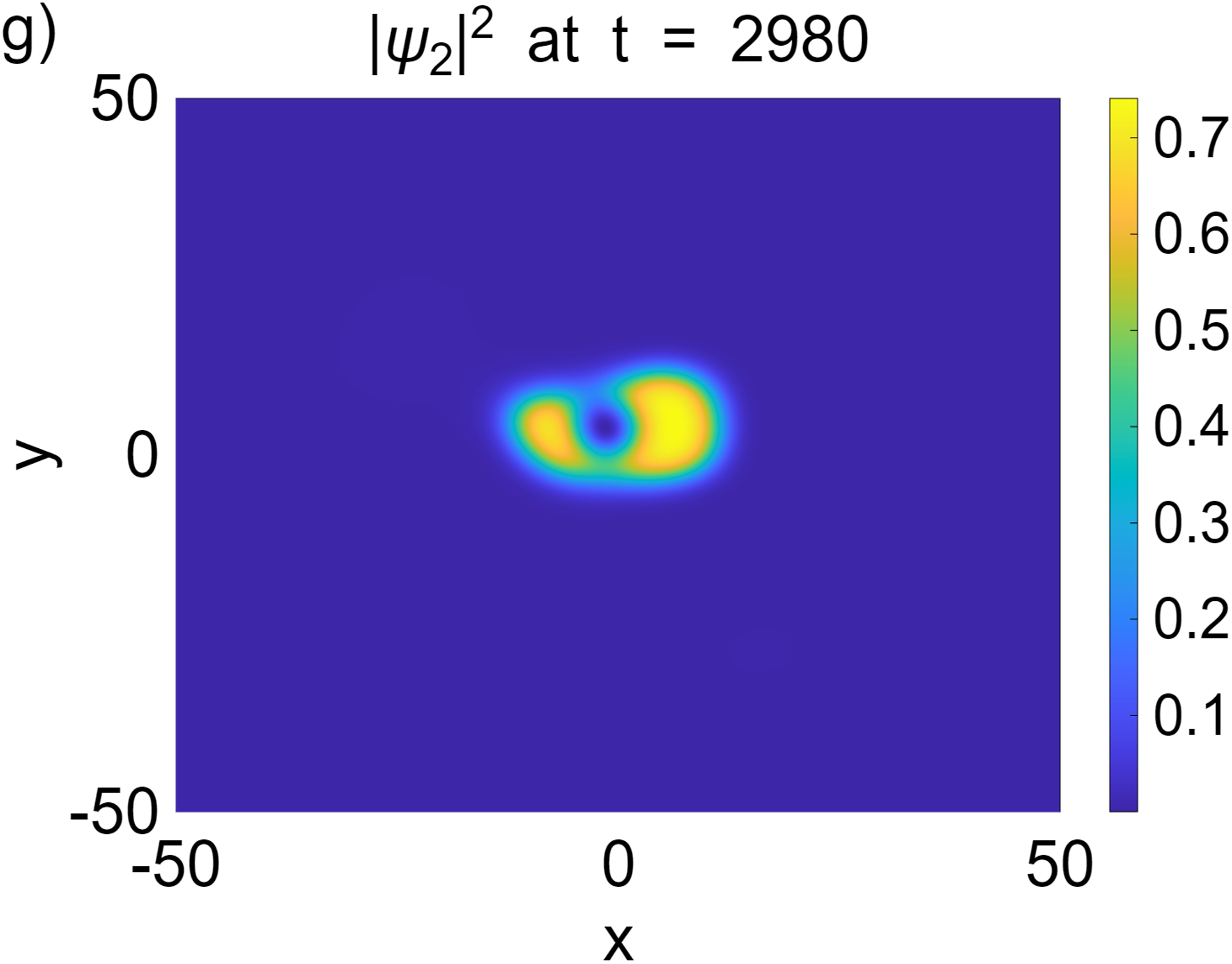}
\includegraphics[width=4.3cm]{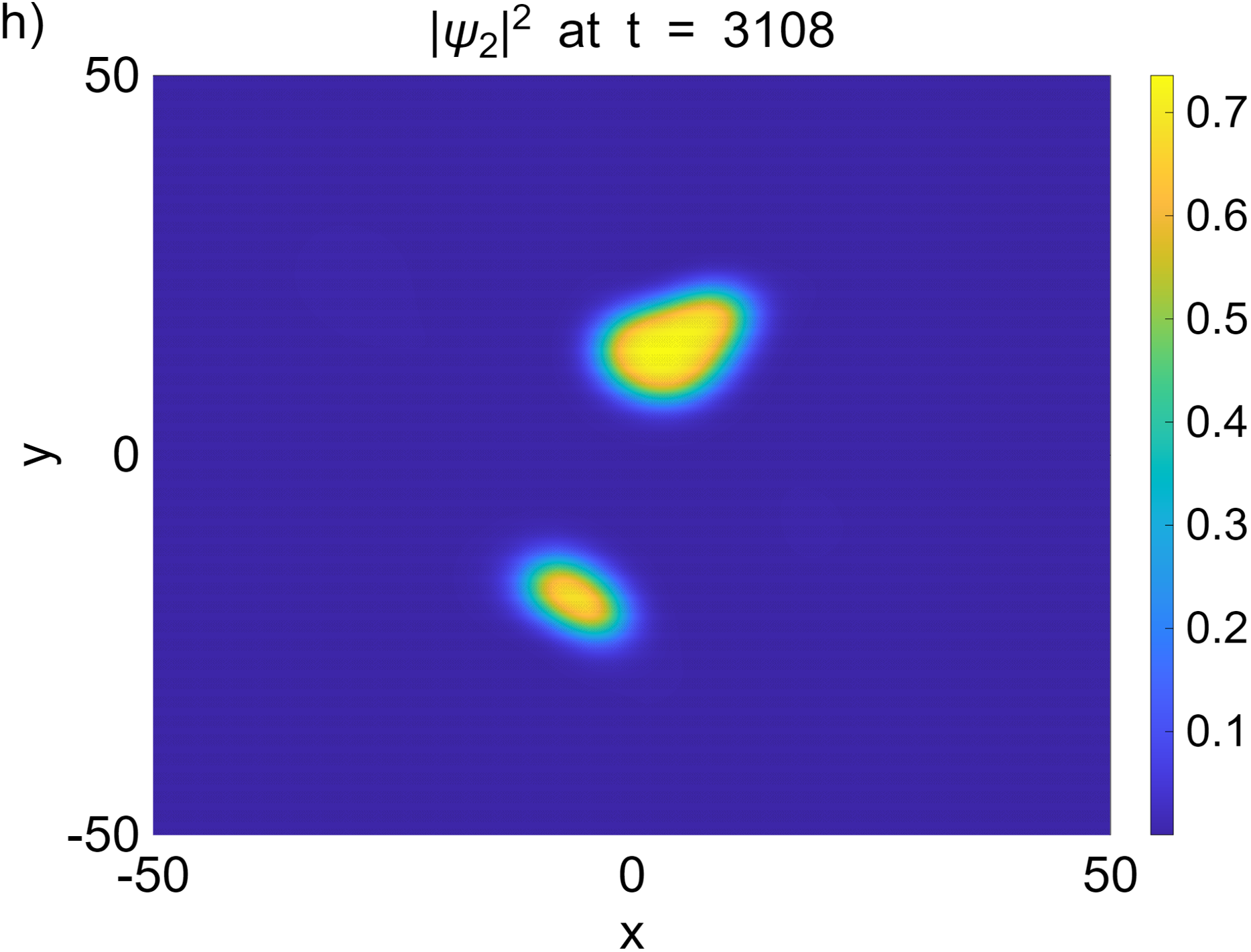}
\caption{Numerically obtained density evolution of an unstable asymmetric vortex with topological charge $S=1$ and initial norm $N=200$, illustrating a representative example of the crescent instability. The top and bottom rows display the density distributions $|\psi_1|^2$ and $|\psi_2|^2$, respectively, at selected evolution times. Panels (a)--(d) show the first component at $t=0$, $t=250$, $t=290$, and $t=350$. Panels (e)--(h) show the second component at $t=0$, $t=2820$, $t=2980$, and $t=3108$. Parameters are $Z_0=0$, $\theta_0=0$, $q=g=1$, and $\kappa=0.01$.}
\label{fig:CI}
\end{figure*}

Figure~\ref{fig:CI} illustrates the development of a crescent instability for an asymmetric two-component vortex with topological charge $S=1$ and total norm $N=200$ (corresponding to the self-trapping regime, $\kappa=0.01$). In the early stage of evolution, the component with the smaller number of particles (the weaker-density ring) loses its axial symmetry: its initially annular density distribution $|\psi_1|^2$ becomes strongly azimuthally modulated and transiently reshapes into a pronounced crescent-like pattern. Importantly, during this stage, the density remains connected (a single distorted lobe rather than separated fragments). The surface-tension-like restoring forces of the droplet and the coherent inter component coupling can partially rebuild the ring, leading to several cycles of ``ring $\leftrightarrow$ crescent'' deformation. After a few such cycles, the repeated excitation of azimuthal modes and the ongoing exchange of norm between the components drive the weak component into irregular dynamics with the radiation of small-amplitude waves. 

The second component, which carries a larger number of the norm and therefore has a more robust flat-top/ring structure, preserves its annular shape for a substantially longer time. Nevertheless, at later times it undergoes the more destructive splitting instability: the ring breaks into spatially separated fragments and the vortex core disappears. For a vortex of charge $S$, the dominant azimuthal modulational instability typically produces $S+1$ fragments; thus, for $S=1$, the outcome is splitting into two fundamental (zero-vorticity) localised states.

There are qualitative differences between the crescent and splitting instabilities. The crescent instability is primarily a symmetry-breaking deformation of the ring into a single off-centred lobe (an azimuthal mode, often the lowest one), without immediate breakup into separate pieces; it may be oscillatory and can exhibit partial reversibility.
On the other hand, the splitting instability is an irreversible fragmentation process in which the annulus breaks into multiple distinct density peaks, destroying the phase singularity associated with the vortex.

In the strongly asymmetric (self-trapped) regime, the weak component is effectively ``soft'': its lower density reduces nonlinear self-stabilisation, making it highly sensitive to azimuthal modulational perturbations. At the same time, the linear coupling $\kappa$ and the relative-phase dynamics continuously transfer norm and imprint phase gradients between components, which efficiently excite azimuthal surface modes of the ring. The competition between these symmetry-breaking azimuthal modes and the restoring tendency of the droplet to maintain a compact boundary leads to the observed cyclic ring-to-crescent deformations. As perturbations accumulate, the oscillatory deformation becomes irregular, while the more strongly populated component resists deformation longer but eventually succumbs to the intrinsic vortex-splitting instability into $S+1$ fragments.
 
\begin{figure}[htbp]
\centerline{
\includegraphics[width=6.7cm]{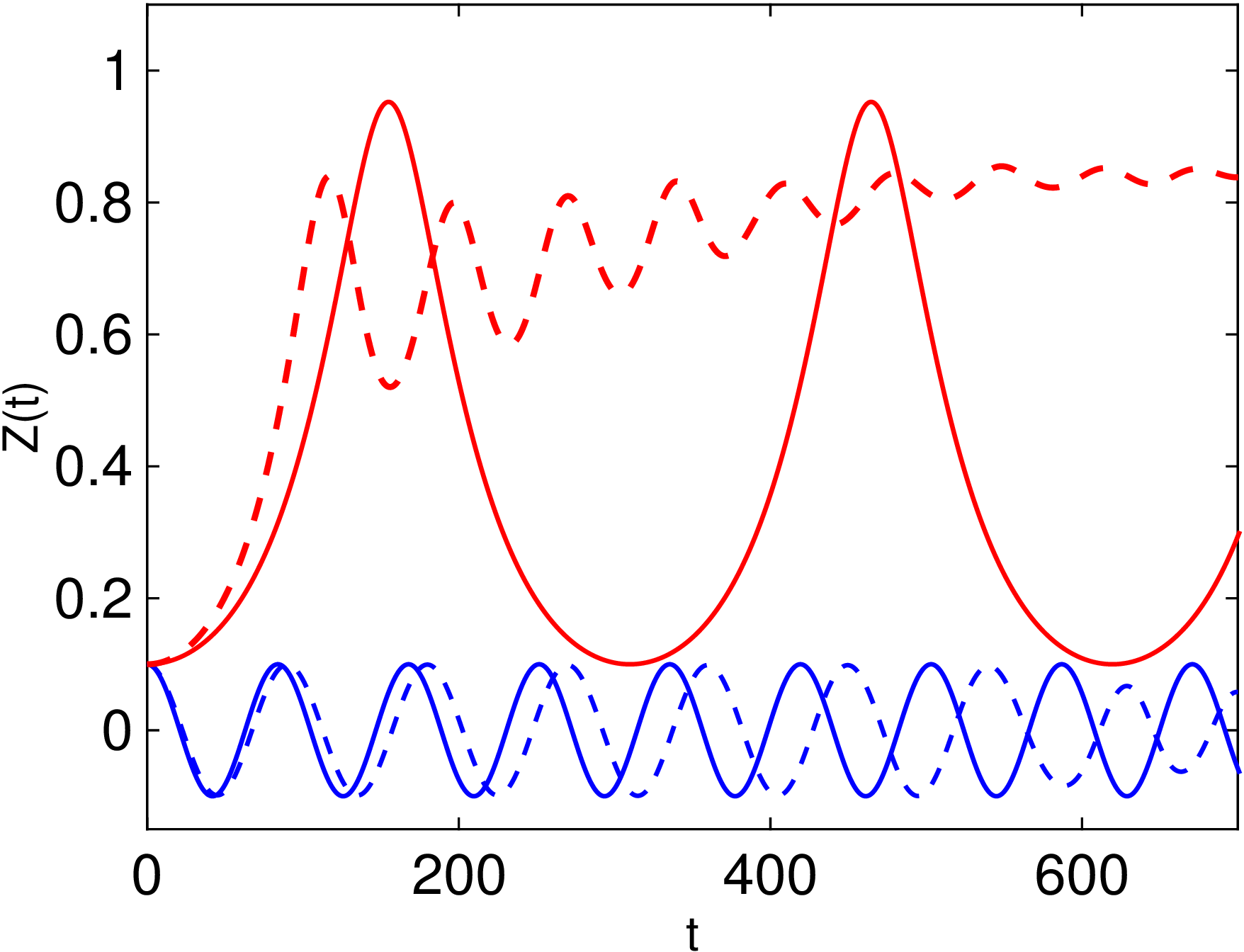}}
\caption{Time evolution of the population (relative) imbalance for unstable vortices with norm $N=200$. The upper set of curves corresponds to the weaker linear coupling, $\kappa=0.01$, placing the system in the macroscopic self-trapping regime. The VA prediction obtained from Eq.~\eqref{eq:QD_JosepEq} is shown by the red solid line, while the red dashed line shows the direct numerical simulations. In the simulations, the vortex develops a crescent-type instability (see Fig.~\ref{fig:CI}). 
The lower set of curves corresponds to a stronger coupling, $\kappa=0.05$, where the dynamics exhibits Josephson oscillations between unstable asymmetric vortices. The VA result is plotted by the blue solid line and the numerical simulations by the blue dashed line. In the simulations, vortex splitting sets in at $t\approx 600$, a typical manifestation of the underlying instability. Parameters are $Z_0=0.1$, $\theta_0=0$, and $q=g=1$. While the VA cannot capture the onset of splitting or crescent instabilities, it reproduces the two qualitative regimes, Josephson oscillations and macroscopic self-trapping and provides a reasonable description of the dynamics.}
\label{fig:dynZunableQD}
\end{figure}
If the instability time scale of an asymmetric vortex state is much longer than the characteristic Josephson period, coherent Josephson exchange can be observed even for dynamically unstable vortices with relatively small norm. This timescale separation can be tuned by varying the linear tunnelling coupling $\kappa$ and/or the initial imbalance $Z_0$, see Fig.~\ref{fig:dynZunableQD}.
The upper set of curves corresponds to weaker coupling, $\kappa=0.01$, where the system operates in the macroscopic self-trapping regime: the population imbalance remains strongly biased, and most atoms stay localised in one vortex core. The variational-approximation prediction from Eq.~\eqref{eq:QD_JosepEq} is shown by the red solid line, while the red dashed line shows direct numerical simulations. In the simulations, the less populated component develops a crescent-type (azimuthal symmetry-breaking) instability, whereas the more populated component eventually exhibits vortex splitting (see Fig.~\ref{fig:CI}). Consistent with self-trapping, the intercomponent transfer is incomplete: the imbalance does not reverse sign, and no sustained back-and-forth oscillations are established.

The lower set of curves corresponds to stronger coupling, $\kappa=0.05$, for which the dynamics displays Josephson oscillations between the two unstable asymmetric vortex states. The VA result is shown by the blue solid line and the numerical simulation by the blue dashed line. In the simulations, vortex splitting sets in at $t\approx 600$, signalling the onset of the underlying dynamical instability. While the VA cannot capture crescent deformations or splitting, it reproduces the two qualitative regimes, Josephson oscillations and macroscopic self-trapping, and provides a reasonable description of the pre-instability dynamics.

We also observe a systematic trend that the lifetime of a vortex state defined as the time interval during which it remains close to an axisymmetric rotating configuration before the onset of clearly visible azimuthal distortion and fragmentation increases with the particle number (norm). 
This behaviour is consistent with the effective stabilisation of the azimuthal excitation spectrum at larger norms, which delays (or suppresses) the development of the azimuthal modulational instability~\cite{Caplan2009}.

\begin{figure}[htbp]
\centerline{
\includegraphics[width=4.7cm]{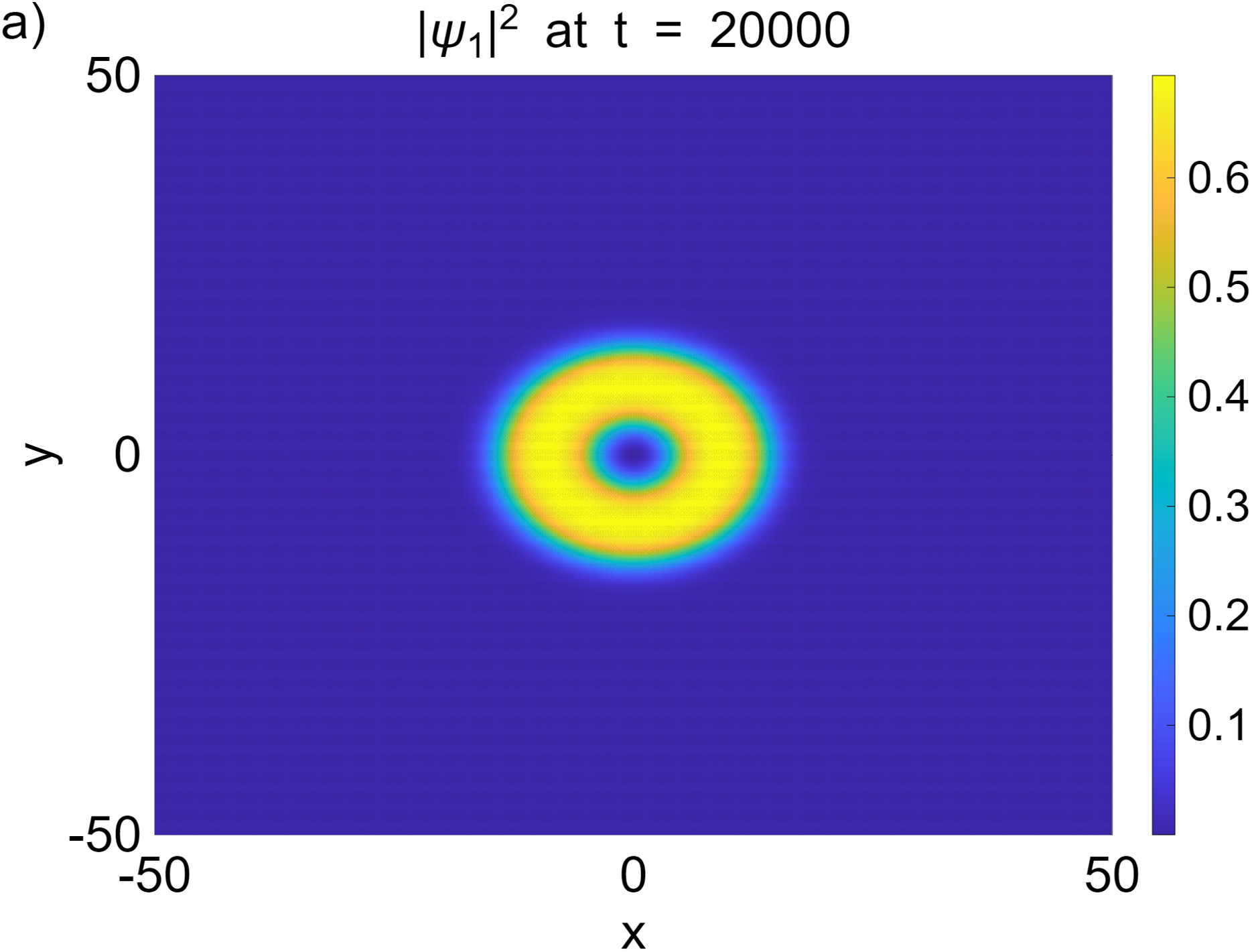} 
\includegraphics[width=4.5cm]{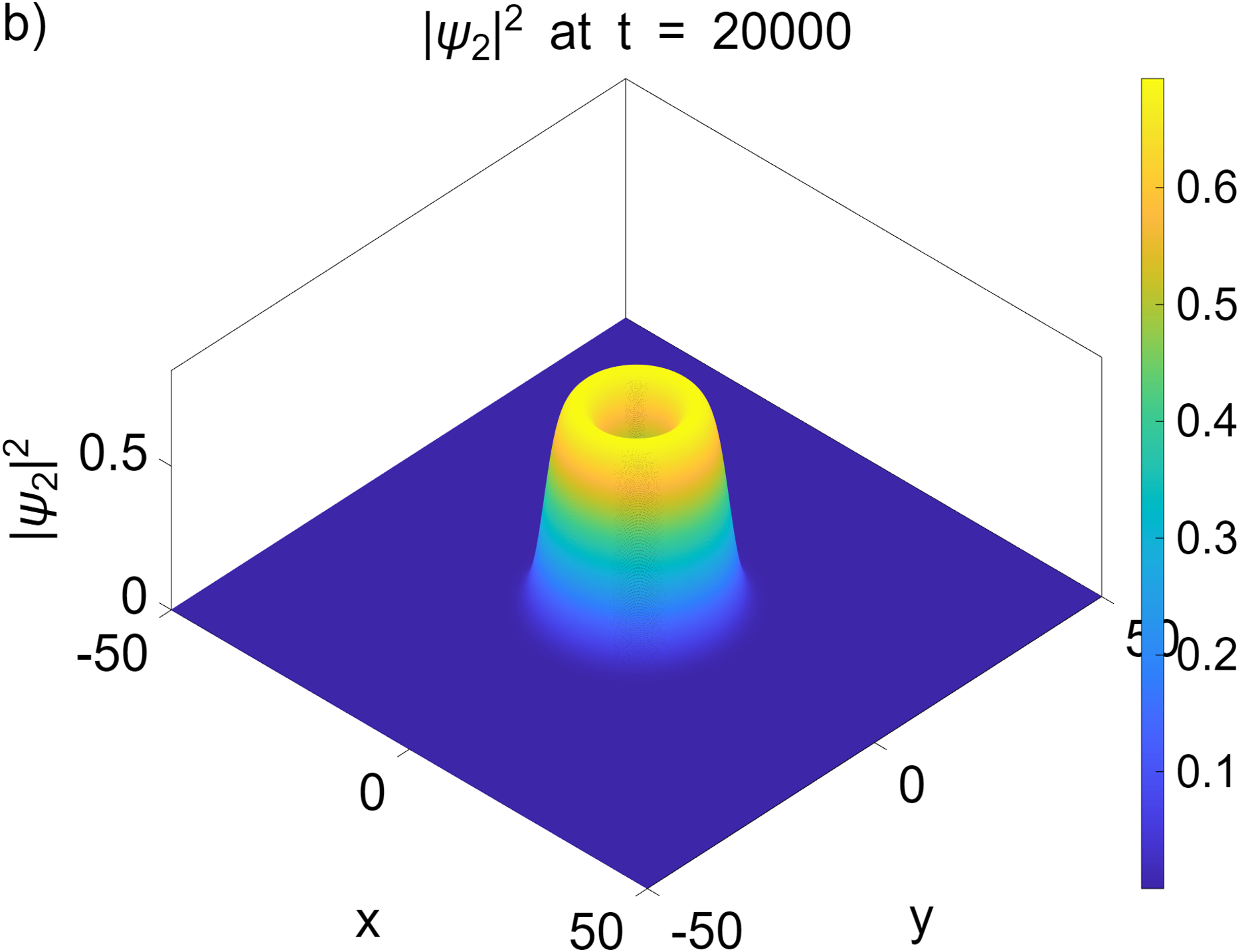}}
\caption{The density evolution of a stable symmetric vortex state with winding number $S=1$ and norm $N=800$. Panels (a) and (b) display the density distributions of the first and second components, respectively. The remaining parameters are $Z_0=\theta_{0}=0$, $q=g=1$, and $\kappa=0.01$.}
\label{fig:StableQD}
\end{figure}

In the analysis of Josephson-type transitions between vortex states, we therefore focus on dynamically robust vortex solutions chosen well inside the stability domain. In practice, this means selecting vortices with sufficiently large norms, for which direct time-dependent simulations demonstrate long-term persistence under small perturbations, without the onset of azimuthal deformation or fragmentation on the considered time scales. A representative example of the evolution of the density distribution for such a stable vortex state, shown at $t=2\times 10^{4}$, is displayed in Fig.~\ref{fig:StableQD}.

\begin{figure}[htbp]
\centerline{%
\includegraphics[width=4.7cm]{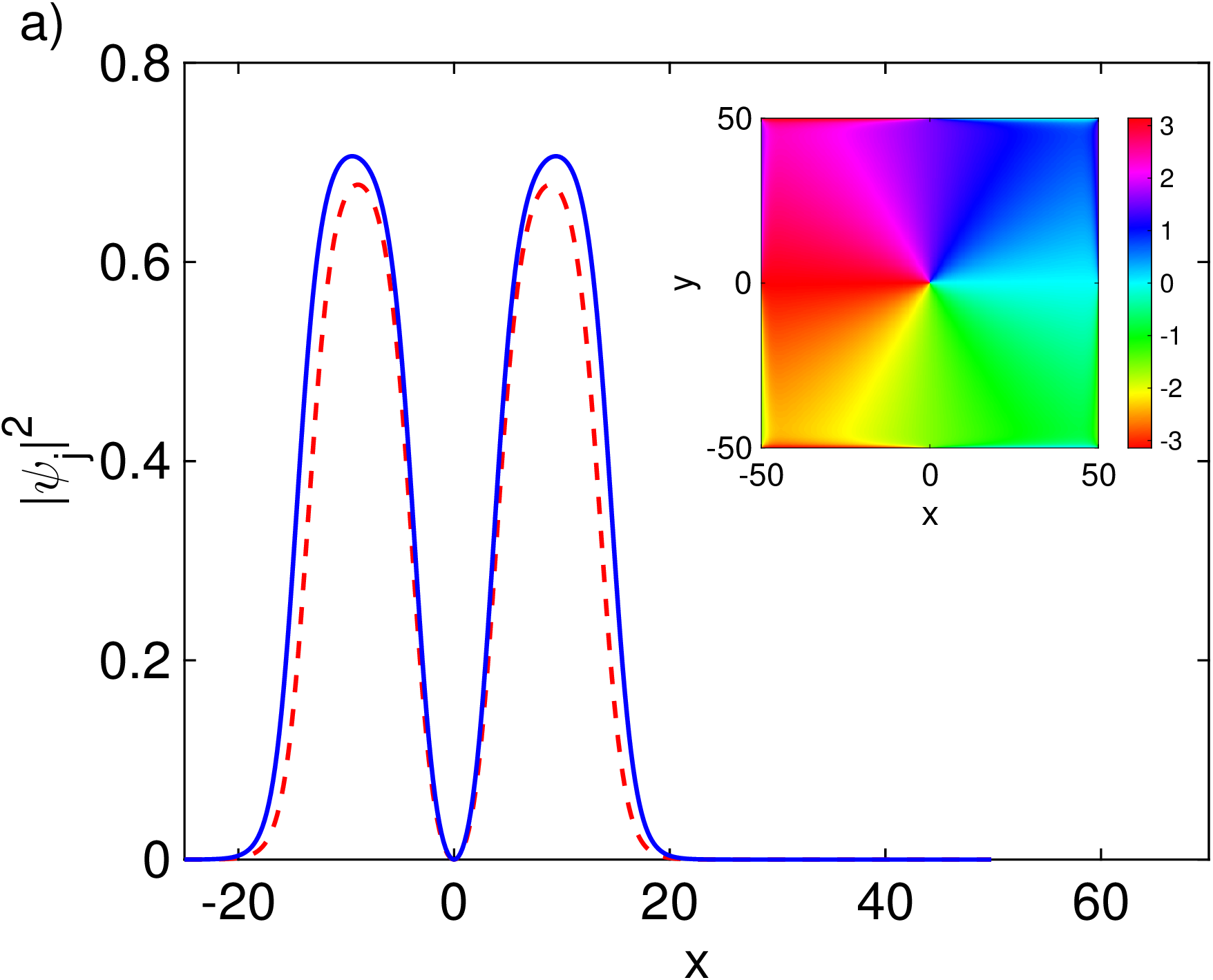}\hskip 0.2cm
\includegraphics[width=4.5cm]{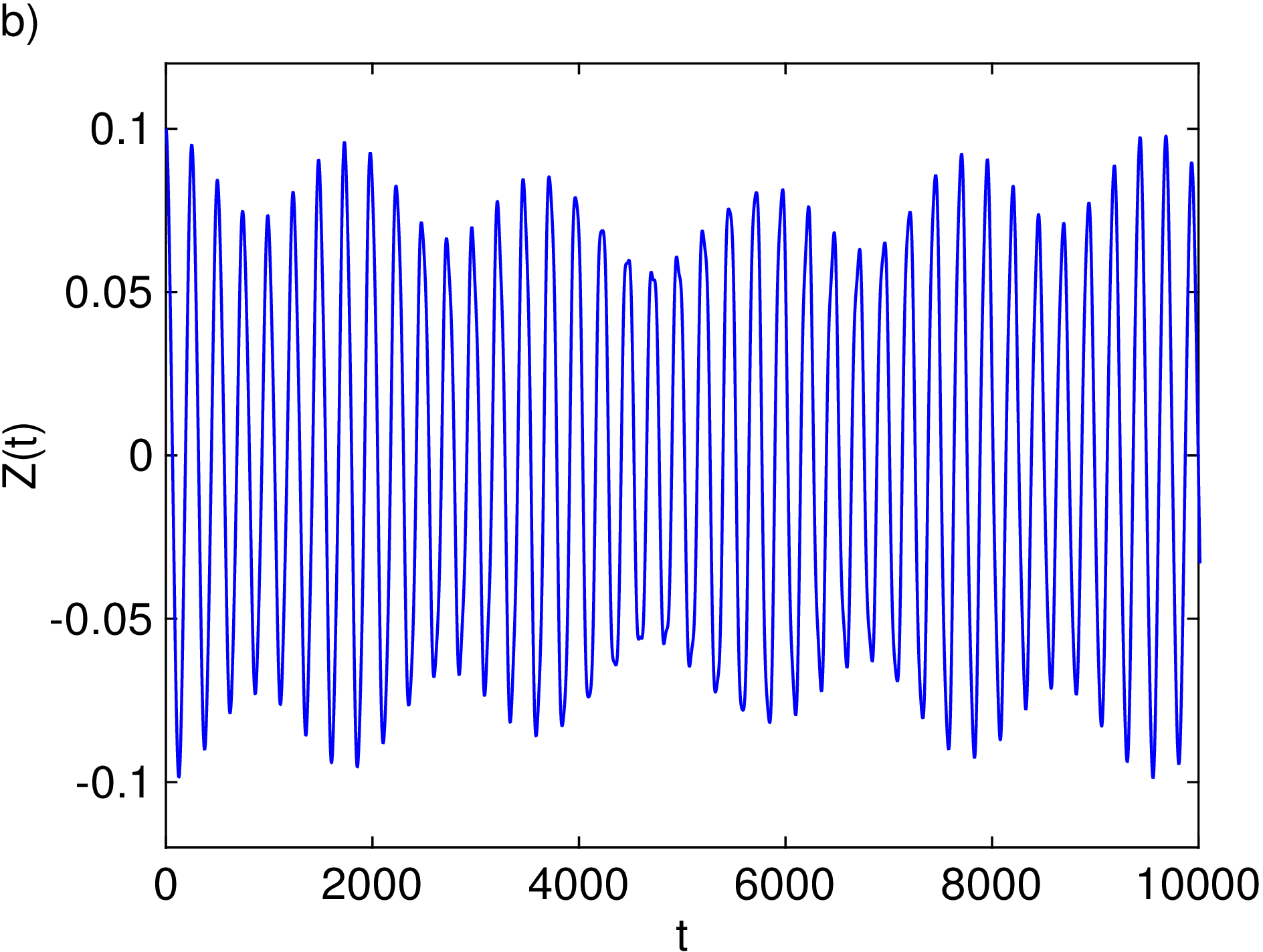}}
\includegraphics[width=4.7cm]{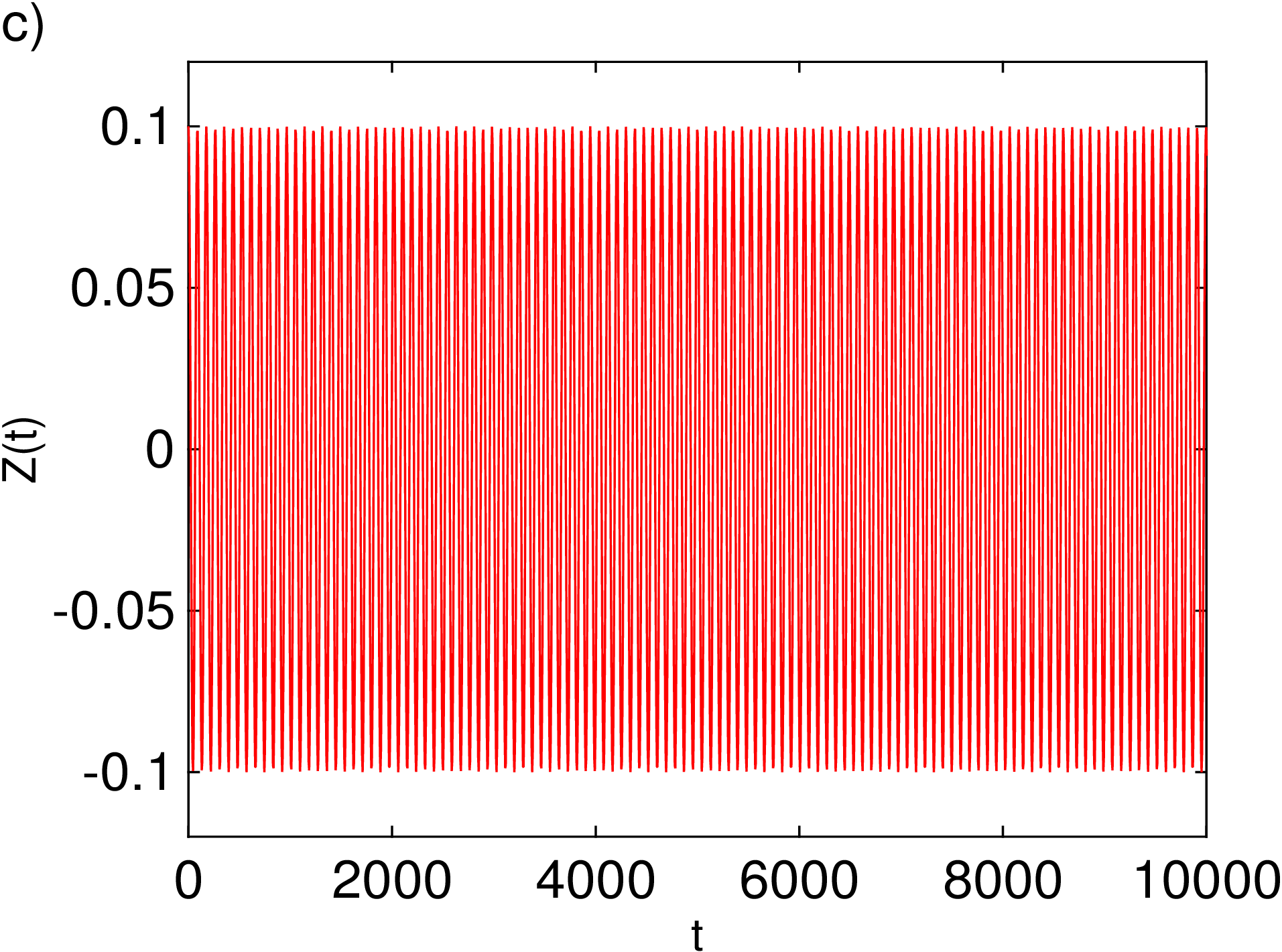}
\caption{(a) Density cross-sections corresponding to Fig.~\ref{fig:StableQD}, computed for a small initial population imbalance $Z_{0}=0.1$. The blue solid curve shows the first component, while the dashed curve shows the second component. The inset displays the associated phase distribution, confirming the presence of an $S=1$ vortex. (b) Numerically simulated time evolution of the population imbalance $Z(t)$ between the two vortex-carrying components. (c) Corresponding evolution of $Z(t)$ predicted by the VA. Parameters are $\theta_{0}=0$, $q=g=1$, $N=800$, and $\kappa=0.01$.}
\label{fig:S1DensAndZ}
\end{figure}

For Josephson-type population transfer between vortex states to be observable, the initial condition must include a slight asymmetry between the components. Accordingly, starting from the vortex solution that remains stable in long-time simulations (Fig.~\ref{fig:StableQD}), we impose a small initial population imbalance $Z_{0}=0.1$, and then analyse the ensuing Josephson oscillations between the two components.
The cross-sections of the density profiles $|\psi_j|^{2}$ at $t=0$ are displayed in Fig.~\ref{fig:S1DensAndZ}(a). The component carrying the larger norm $N_2=N(1+Z_0)/2$ corresponds to the higher-density curve (the upper blue solid line). In contrast, the component with the smaller norm $N_1=N(1-Z_0)/2$ is represented by the lower red dashed line. The inset shows the associated phase pattern, which unambiguously confirms the vorticity $S=1$, the phase winds by $2\pi$ around the vortex pivot located at the zeros of the local amplitude. 

\begin{figure}[htbp]
\centerline{%
\includegraphics[width=4.3cm]{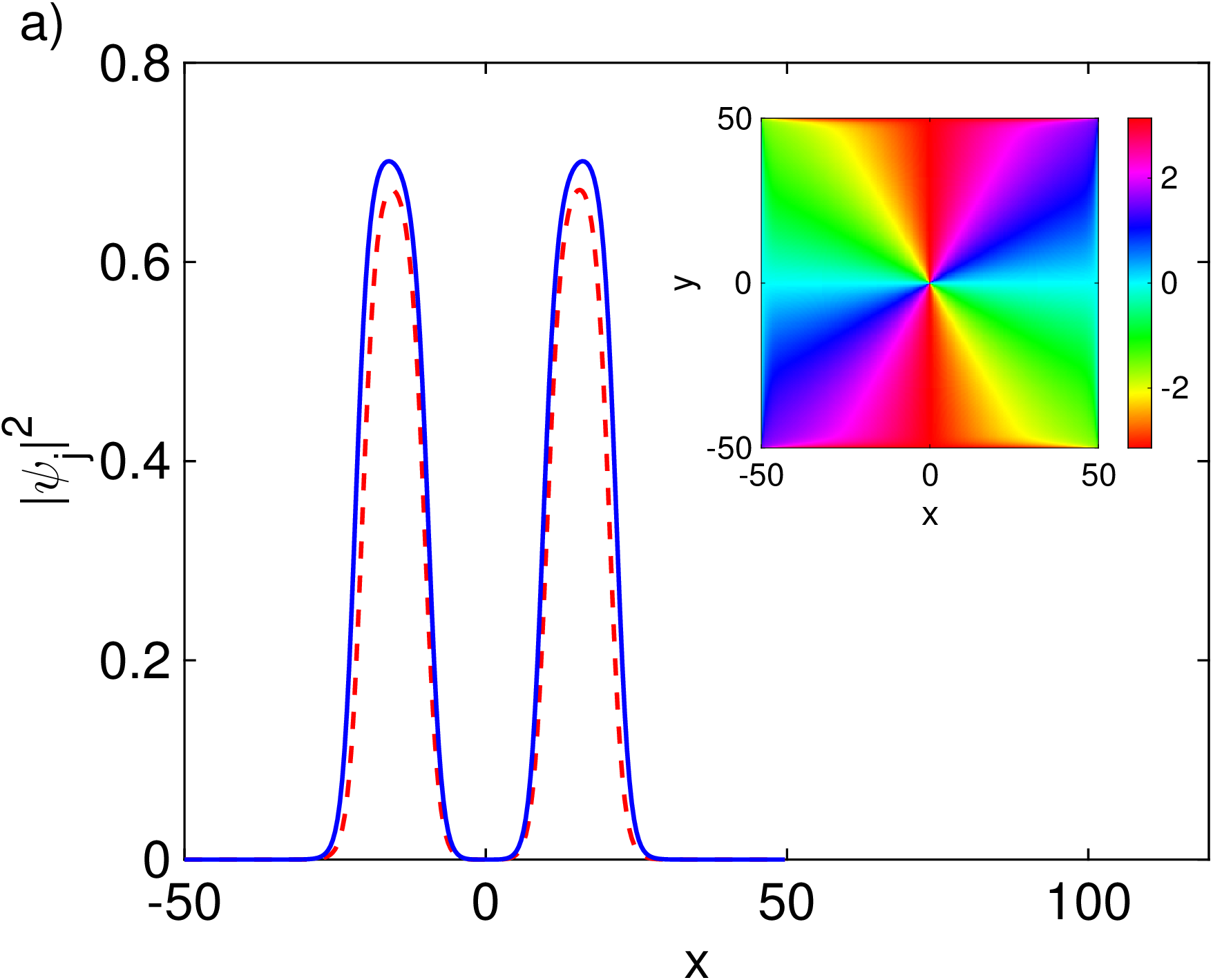}\hskip 0.2cm
\includegraphics[width=4.5cm]{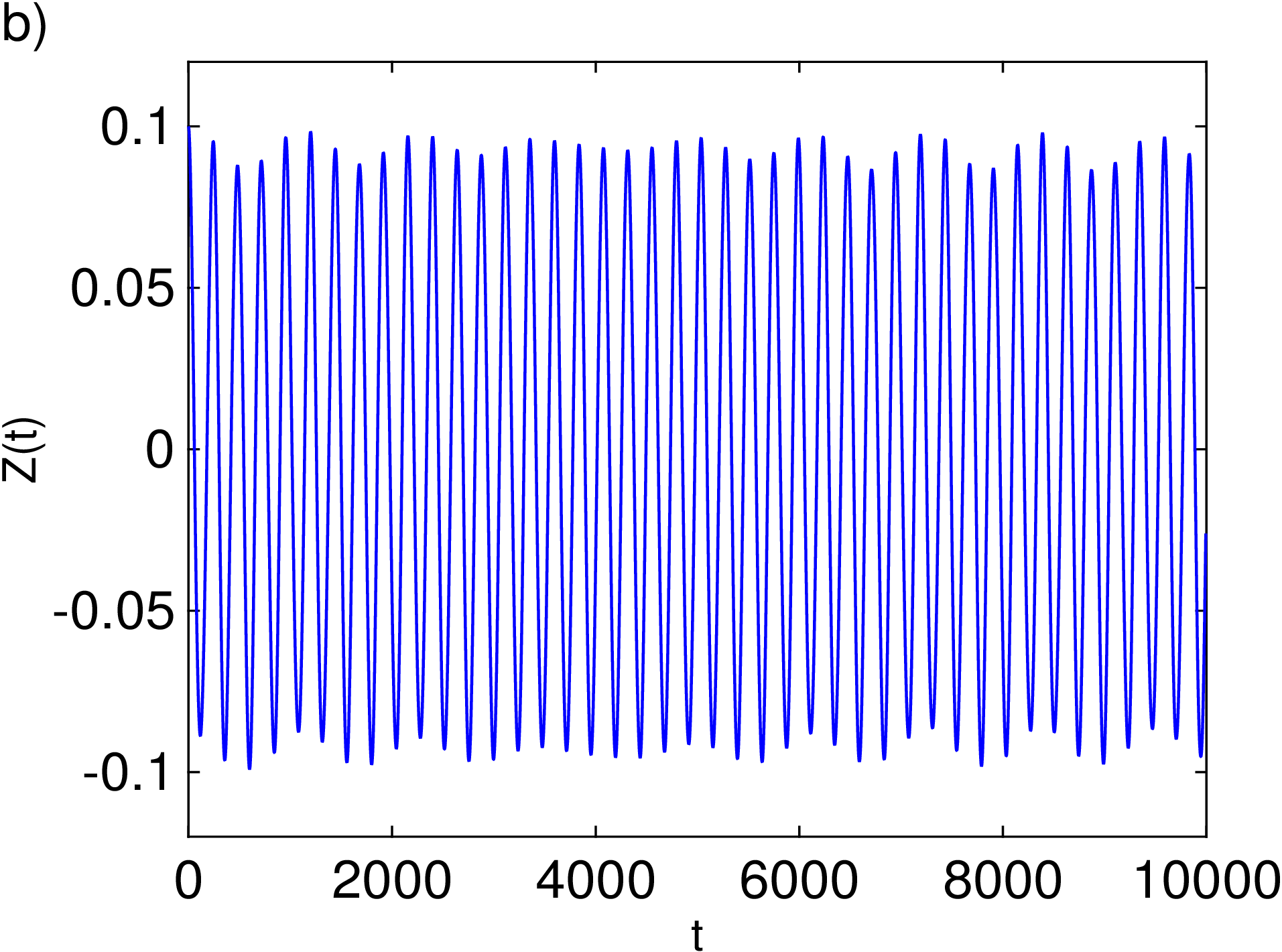}}
\centerline{\includegraphics[width=4.3cm]{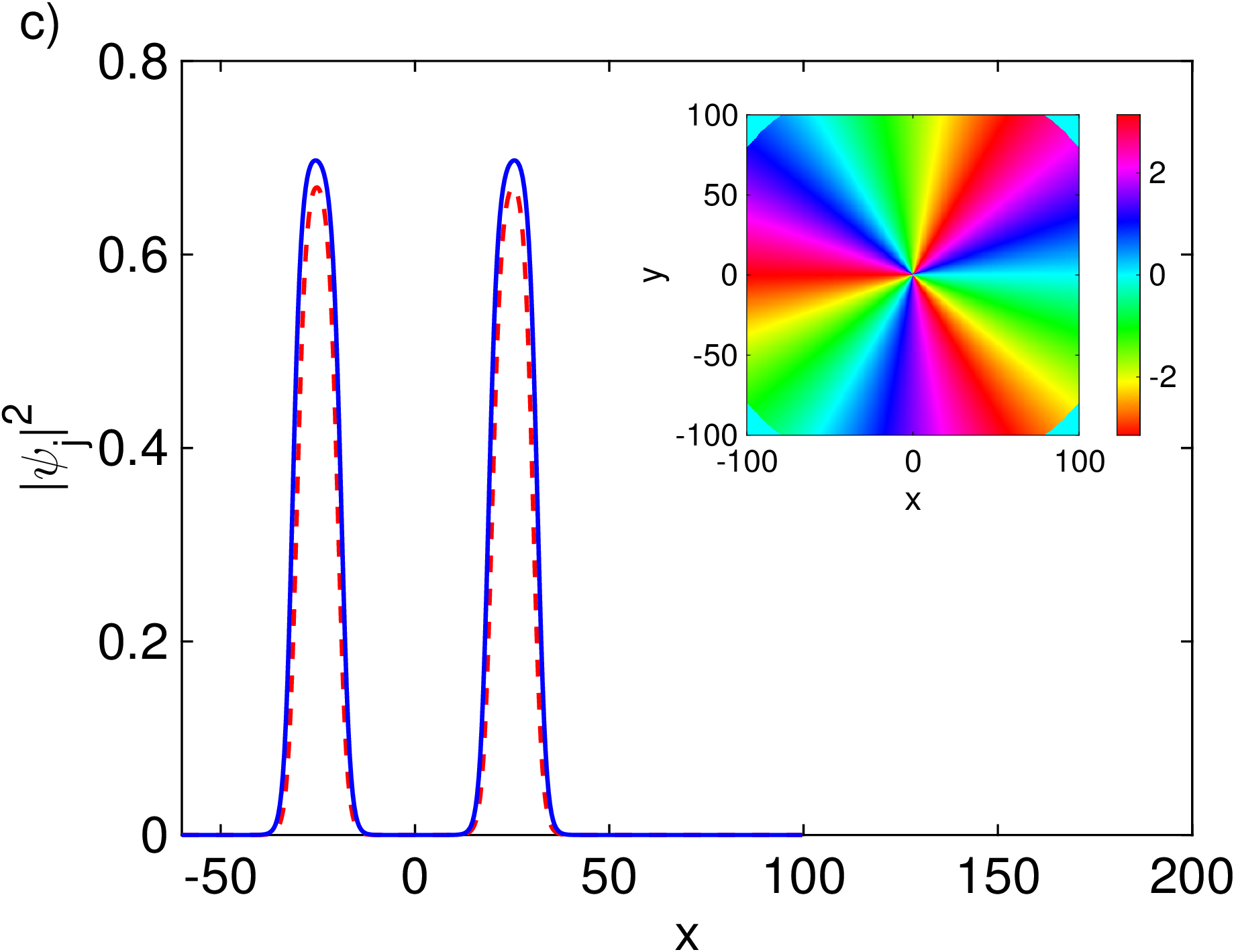}\hskip 0.2cm
\includegraphics[width=4.5cm]{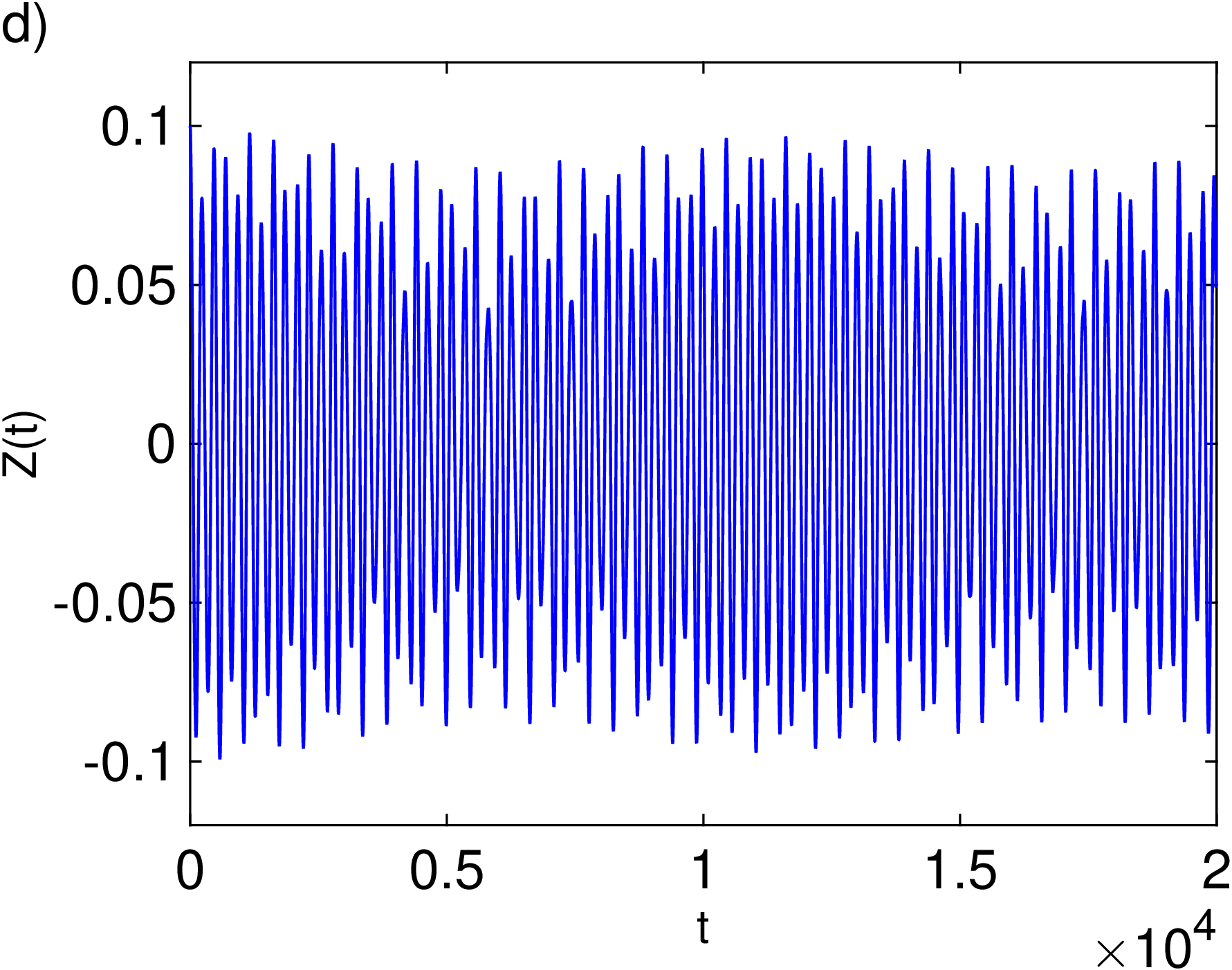}}
\caption{Josephson population transfer between higher-charge vortex states. Panels (a, b) correspond to $S=2$ with $N=1500$, while panels (c, d) correspond to $S=3$ with $N=2500$. Panels (a, c) show density cross-sections, where solid blue and dashed red curves represent $|\psi_1|^2$ and $|\psi_2|^2$, respectively; the insets display the corresponding phase maps, confirming the vorticity via phase winding around pivots located at zeros of the local amplitude. Panels (b, d) show the time evolution of the population imbalance $Z(t)$. Other parameters are $Z_0=0.1$, $\theta_0=0$, $\kappa=0.01$ and $q=g=1$.}
\label{fig-VorS2S3}
\end{figure}

Figure~\ref{fig:S1DensAndZ}(b) shows the numerically computed time evolution of the population imbalance $Z(t)$, which quantifies the Josephson-type exchange between the two $S=1$ vortex states hosted by the two-core condensate. The simulations reveal long-lived, nearly periodic particle transfer: the population flows alternately from the first core to the second and back, leading to sustained oscillations of $Z(t)$ about zero. The oscillation amplitude remains comparable to the imposed initial asymmetry, with $Z(t)$ repeatedly changing sign and reaching values on the order of $\pm Z_0$. This behaviour is characteristic of the Josephson-oscillation regime (as opposed to macroscopic self-trapping, where $Z(t)$ would remain biased and typically would not cross zero). 
Figure~\ref{fig:S1DensAndZ}(c) presents the corresponding imbalance dynamics predicted by the variational approximation. The VA reproduces the existence of coherent Josephson oscillations and captures their qualitative features; however, the oscillation period differs noticeably from that obtained in the direct simulations, with the VA period being off by a factor of several. This discrepancy reflects the reduced nature of the VA, which treats the dynamics in terms of a few symmetrical variational variables and cannot fully account for the deformation of the vortex density profiles.
Josephson population transfer is also observed between vortex states with topological charges $S=2$ and $3$ (see Fig.~\ref{fig-VorS2S3}), and the same approach can be extended to higher-charge vortices. 

	Just as in the droplet setting, we explore the manifestation of the Andreev-Bashkin (nondissipative-drag) effect for vortex states. In the vortex context, an analogous entrainment mechanism can operate between two coupled vortices, both in the symmetric configuration and in the presence of a controlled asymmetry. In particular, for asymmetric vortex pairs, the Andreev-Bashkin drag and Josephson-type population transfer may occur simultaneously. 
\begin{figure}[htbp]
\centerline{%
\includegraphics[width=4.5cm]{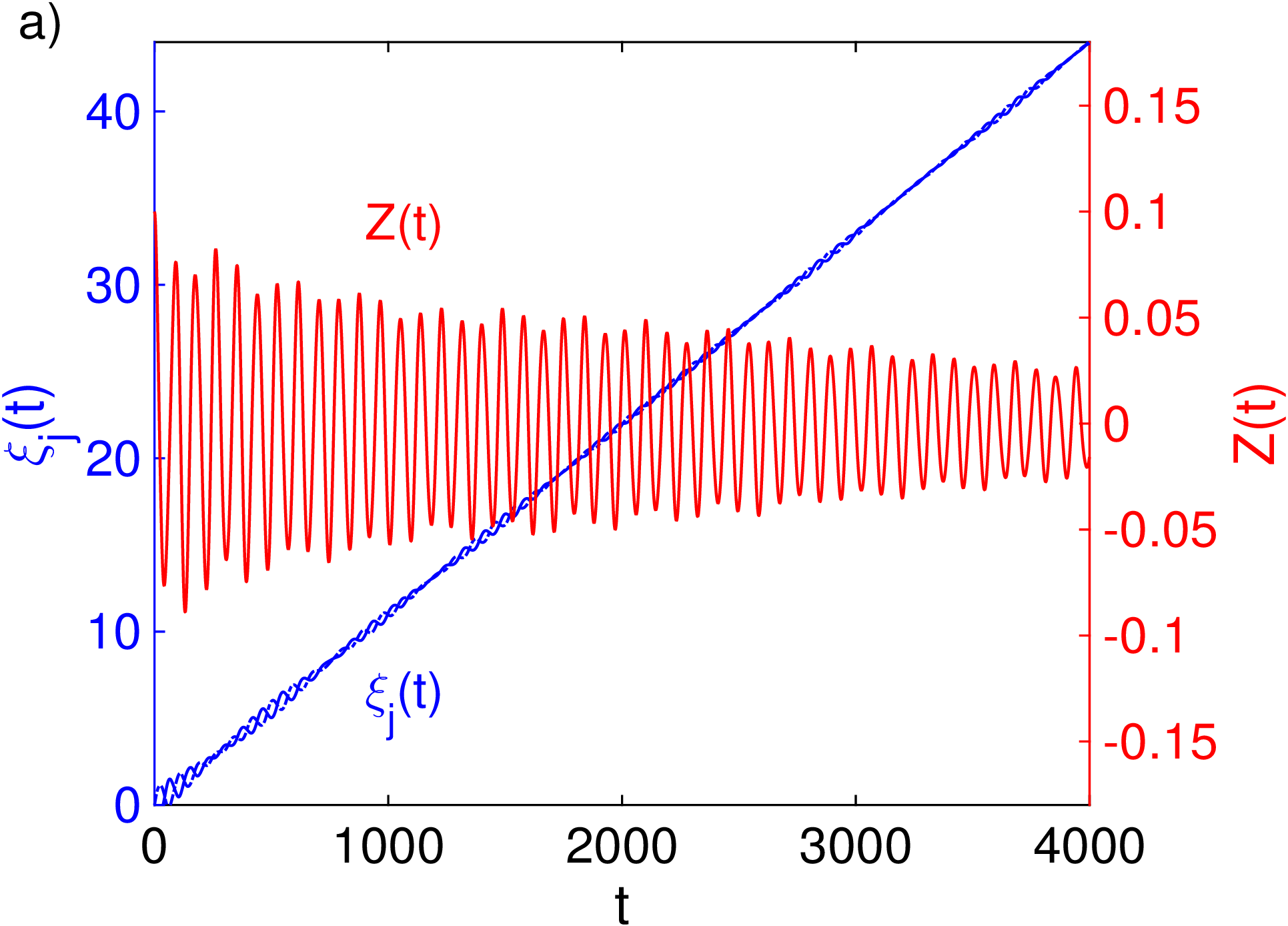}
\includegraphics[width=4.5cm]{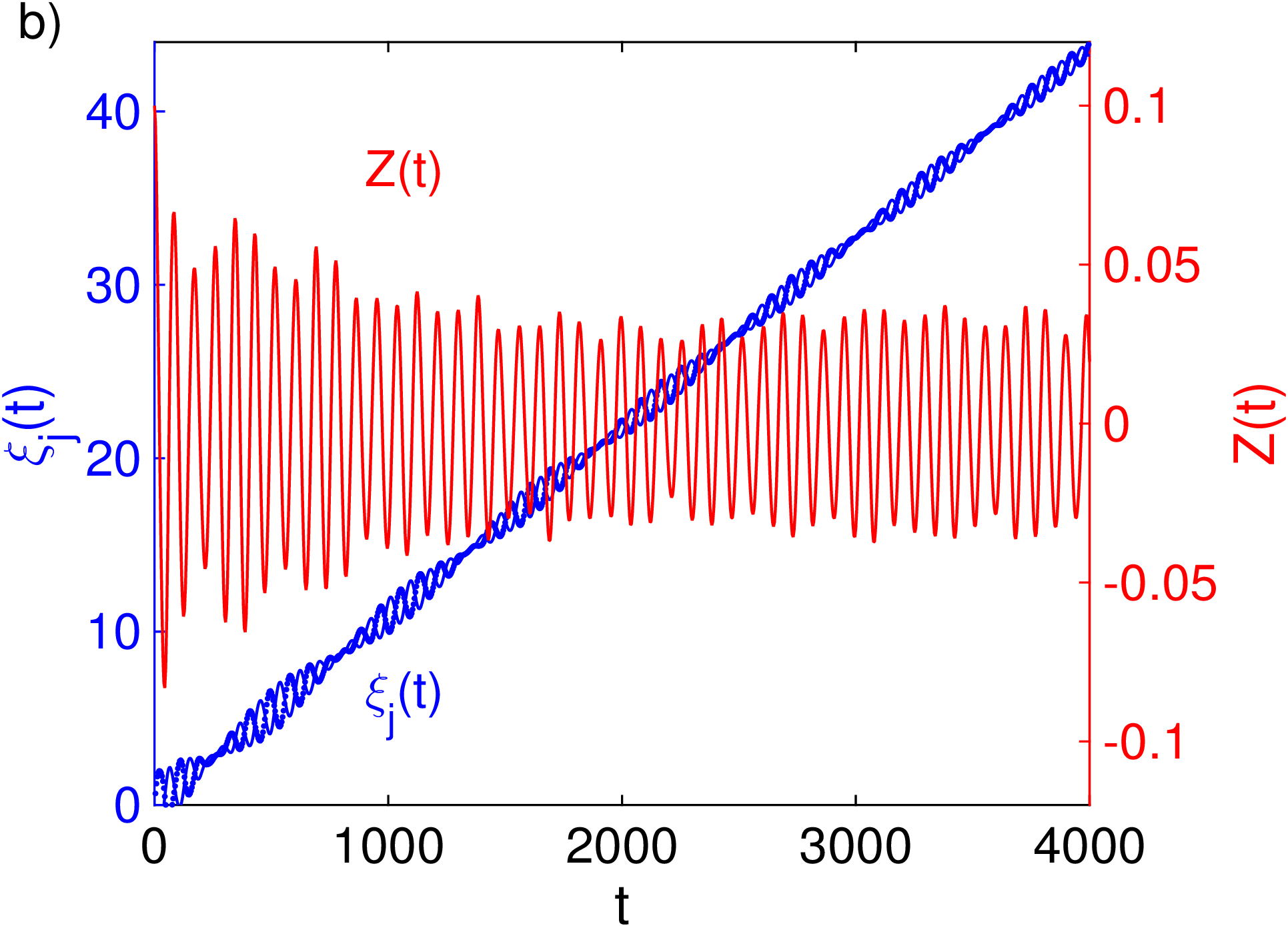}}
\caption{Time evolution of the centre-of-mass coordinates $\xi_j(t)$ (left axis) and the population imbalance $Z(t)$ (right axis) under a weak initial momentum kick applied to the second component. In both panels, the blue solid and blue dashed (dotted) curves show $\xi_1(t)$ and $\xi_2(t)$, respectively, while the red solid curve corresponds to $Z(t)$. At $t=0$, component $j=2$ is given a small kick $k=0.02$ along the positive $x$ direction. Panel (a) corresponds to vortices with $S=1$ and $N=800$. Panel (b) corresponds to vortices with $S=2$ and norm $N=1500$. Other parameters are $Z_0=0.1$, $\theta_0=0$, $q=g=1$ and $\kappa=0.03$.}
\label{fig:S1AB}
\end{figure}
A representative example is shown in Fig.~\ref{fig:S1AB} for vortices with $S=1$ and $2$. At the initial moment, the vortex in the second component is given a weak momentum kick by imprinting a small phase gradient, $\psi_2 \to \psi_2 \exp(i k x)$, with $k=0.02$ along the positive $x$ direction, while the vortex in the first component is initially at rest. During the subsequent evolution, the moving (kicked) vortex entrains the initially stationary one: the two centres of mass rapidly adjust and then propagate with (approximately) the same velocity, demonstrating joint motion driven by nondissipative drag. This behavior is shown in Fig.~\ref{fig:S1AB} by the trajectories of the centers of mass, where the solid and dashed blue curves represent $\xi_1(t)$ and $\xi_2(t)$ (left vertical axis), together with the simultaneously evolving population imbalance $Z(t)$ shown by the solid red curve (right vertical axis). 

The strength of the entrainment is primarily controlled by the degree of spatial overlap between the vortices. In our two-core setting, the overlap is effectively regulated by the linear coupling $\kappa$: for sufficiently large $\kappa$, the induced drag current is strong enough to lock the vortices into a long-lived co-moving state. Conversely, when the overlap is weak (small $\kappa$), the drag diminishes, and the vortices tend to decouple, allowing their centres of mass to separate over time.

\section{Experimental estimations of the parameters}
\label{sec:estimations}

In this section, we estimate the predicted effects under experimentally relevant conditions. As a concrete example, we consider a two-component condensate of $^{39}\mathrm{K}$ atoms prepared in two hyperfine states, e.g., $\lvert F=1,m_F=0\rangle$ (component $j=1$) and $\lvert F=1,m_F=-1\rangle$ (component $j=2$), with atomic mass $m_0=6.49\times 10^{-26}\,\mathrm{kg}$, that are loaded in a symmetrical two-cored trap. The intra-species interactions are taken equal, $a_{11}=a_{22}\equiv a=50\,a_0$, while the inter-species scattering length is chosen as $a_{12}=-1.1\, a_{11}$ (with $a_0$ the Bohr radius), corresponding to weak intra-species repulsion ($a_{11}, a_{22}>0$) and inter-species attraction ($a_{12}<0$). This is precisely the regime $\delta a=-|a_{12}|+\sqrt{a_{11}a_{22}}<0$ in which the residual mean-field attraction can be suppressed and become of the same order as the LHY term to support quantum droplets~\citep{Cabrera2018, Cheiney2018}. 

We assume tight transverse confinement with frequency $\omega_z=2\pi\times 1000\,\mathrm{Hz}$, which sets the characteristic transverse length scale (harmonic-oscillator length)
$l_z=\sqrt{\frac{\hbar}{m_0\omega_0}}\approx 0.51~\mu\mathrm{m}.$
In our normalization (with $q=g=1$), the corresponding characteristic scales are $r_s=l_z \approx 0.51~\mu\mathrm{m}$, $t_s=\omega_0^{-1}\approx 0.16~\mathrm{ms}$, and $\psi_s \approx 4.582 \times 10^{7}\,\mathrm{m}^{-1/2}$. Accordingly, the computational domain size $L=L_x=L_y=200$ corresponds to a transverse extent of $L_{\rm phys}\approx 200 \,r_s \simeq 102~\mu\mathrm{m}$, while the dimensionless evolution time $t=500$ corresponds to $T\simeq 500\,t_s\approx 79.6~\mathrm{ms}$, which is well within typical experimental observation times. We can estimate that the dimensionless Josephson frequency of 0.02 corresponds to a frequency of $20$ Hz, which is consistent with experimental values~\cite{Raghavan1999, Albiez}.
With these scalings, the dimensionless norm $N=100$ corresponds to a physical particle number $N_{\mathrm{real}}\approx 5.43\times 10^{4}$, and the characteristic droplet size is $\sim 9.95~\mu\mathrm{m}$ (as extracted from the simulated density profile).

It is also worth emphasising that the corresponding physical scales are readily tunable in realistic setups. In particular, modest adjustments of the $s$-wave scattering lengths (via magnetic Feshbach resonances) and of the transverse confinement frequency $\omega_0$ (by changing the trap strength) allow one to vary the effective mean-field and LHY coefficients, as well as the characteristic length and time scales.

\section{Conclusions}
\label{sec:Conc}

In this paper, we have investigated the dynamics of the binary Bose gas confined in a symmetric dual-core, pancake-shaped trap, which takes into account the effect of the beyond-mean-field quantum fluctuations. For the spatially uniform state, the corresponding effective Hamiltonian has been derived in terms of the population imbalance and relative phase, and analytical expressions for the Josephson oscillation frequencies in the zero- and $\pi$-phase modes have been obtained.

Explicit separatrix conditions have been derived for both the zero- and $\pi$-phase modes, expressed in terms of the critical linear coupling (or equivalently the critical initial imbalance), which distinguish macroscopic self-trapping/self-localisation (unbounded running-phase trajectories) from Josephson oscillations (closed orbits with periodic population exchange).
It is further determined how the interplay of the LHY contribution, two-body interactions, and the linear coupling controls the emergence of symmetry-breaking bifurcations. In the zero-phase branch, it was found that two pitchfork bifurcations occur sequentially: a supercritical bifurcation at smaller norms followed by a subcritical one at larger norms, producing bistability and hysteresis in the bifurcation diagram. In the $\pi$-phase branch, a single subcritical pitchfork bifurcation is identified.

For the inhomogeneous BEC, Josephson dynamics of quantum droplets in the same two-core geometry have been investigated. A variational approximation is developed for the coupled droplets. The stationary droplet parameters are obtained from imaginary-time simulations, and their dynamical robustness is verified against small perturbations. Within this framework, analytical estimates are derived for the Josephson frequencies in both the zero- and $\pi$-phase droplet regimes and are validated by direct time-dependent simulations. For smaller atom numbers, good agreement between the variational predictions and numerical results was found. For larger atom numbers, substantial deviations (sometimes by orders of magnitude) from the variationally predicted frequencies may occur. This reflects the increasing role of shape deformations and nontrivial interdroplet interactions. Therefore, in this regime, we mainly rely on direct numerical simulations, using the variational predictions only for qualitative guidance. In the $\pi$-phase droplet tunnelling regime, it is observed that after roughly $10$--$12$ oscillation periods, the droplets separate and drift apart in opposite directions. This behaviour can be explained by analysing the effective droplet-droplet interaction energy. The numerical simulations also reveal signatures consistent with the Andreev-Bashkin nondissipative drag, manifested as entrainment of one component's motion by the other.

Finally, Josephson phenomena for vortex states have been addressed: symmetric vortices are first analysed, and it is shown that, at low particle numbers, they are sensitive to an azimuthal modulational instability that breaks the initial axial symmetry and drives fragmentation. In particular, a vortex with winding number $S$ typically splits into $S+1$ (and occasionally $S+2$) localised fundamental fragments. For unstable vortices, the lifetime before breakup is found to increase with particle number, indicating a progressive suppression of the azimuthal instability as the number of atoms grows. At sufficiently large norms, long-time simulations confirm that vortex states remain robust against small perturbations, defining a well-pronounced stability domain. Within this stable regime, the existence of Josephson-type population transfer between vortices with charges $S=1,2$, and $3$ is demonstrated numerically. Andreev-Bashkin-type entrainment between moving vortices is also identified. It is also found that unstable asymmetric vortices can undergo a crescent-like (azimuthal symmetry-breaking) instability. Although such deformations and vortex splitting cannot be captured, the variational approximation developed for quantum droplets and vortex states remains a valuable qualitative tool for exploring the macroscopic self-trapping and Josephson-oscillation regimes. In future work, it will be interesting to explore Josephson dynamics of higher-charge vortices, extend the analysis to the LHY-fluid regime~\cite{LHYfluid}, and consider analogous coupled-core geometries in other physical settings, for example, in dipolar systems~\cite{Masalaeva2026}.

\section{Acknowledgments}
This work has been supported by the State Budget of the Republic of Uzbekistan (Grant No. 2026 year award).

\appendix
\section{Derivation of the Model Equation}
\label{appen}

Let us consider a three-dimensional two-component Bose-Einstein condensate confined in a double-pancake trap and including the beyond-mean-field Lee-Huang-Yang correction. The coupled Gross-Pitaevskii equations for the two components are written as
\begin{eqnarray}
\nonumber
&i \hbar \partial_T \Psi_{1}
=-\cfrac{\hbar^2}{2 m_0}\left(\partial_Z^2 + \nabla_{\rho}^2\right)\Psi_{1}
+ \left[ V_{\mathrm{ext}} + g_{11} |\Psi_1|^2 +  \right. 
\\ \nonumber
& \left. g_{12} |\Psi_2|^2 + \gamma_{\mathrm{QF}} \left(g_{11} |\Psi_1|^2 + g_{22} |\Psi_2|^2\right)^{3/2} \right]\Psi_1 ,
\\ \nonumber
&i \hbar \partial_T \Psi_{2} = -\cfrac{\hbar^2}{2 m_0}\left(\partial_Z^2 + \nabla_{\rho}^2\right)\Psi_{2}
+ \left[ V_{\mathrm{ext}} + g_{22} |\Psi_2|^2 + \right.
\\ 
& \left. g_{21} |\Psi_1|^2  + \gamma_{\mathrm{QF}}
\left(g_{11} |\Psi_1|^2 + g_{22} |\Psi_2|^2\right)^{3/2}
\right]\Psi_2 .
\label{eq:3D_GPE_12}
\end{eqnarray}
Here $\Psi_{1,2}(\rho,Z,T)$ are the macroscopic wave functions of the two components, normalized to the corresponding particle numbers, $m_0$ is the atomic mass, $\gamma_{\mathrm{QF}}=4 m_0^{3/2} g^2 / (3 \pi^2 \hbar^3 2^{5/2})$, and $\rho=(X,Y)$ denotes the in-plane coordinate. The operator $\nabla_\rho^2$ acts in the transverse $(X,Y)$-plane, while the $Z$-direction is assumed to be strongly confined. The coefficients $g_{jj}>0$ describe the intra-species repulsion, whereas $g_{12}=g_{21}$ account for the inter-species interaction. The last term represents the Lee-Huang-Yang correction, which is repulsive and originates from quantum fluctuations beyond the mean-field approximation. 

The external potential $V_{\mathrm{ext}}(Z,\rho)$ is assumed to have a double-well structure along the $Z$-direction, giving rise to two parallel disk-shaped condensates weakly coupled by tunneling through the barrier. Our goal is to derive an effective quasi-two-dimensional description for the dynamics in the two pancakes.

We focus on the symmetric configuration,
$$
\Psi_1=\Psi_2=\Psi,\qquad
g_{11}=g_{22}=g,\qquad
g_{12}=g_{21}<0,
$$
and introduce the standard combination
$$
\delta g = g_{12}+\sqrt{g_{11}g_{22}}=g_{12}+g,
\qquad
|\delta g|\ll g.
$$

For $\delta g<0$, the mean-field contribution is effectively attractive.
Under these assumptions, Eqs.~(\ref{eq:3D_GPE_12}) reduce to the single effective equation
\begin{eqnarray}\label{eq:single3Dgpe}
&i \hbar \partial_T \Psi
= -\frac{\hbar^2}{2 m_0}\left(\partial_Z^2 + \nabla_{\rho}^2\right)\Psi
+  \\
\nonumber
&\left[
V_{\mathrm{ext}}
+\delta g |\Psi|^2
+2^{3/2} g^{3/2} \gamma_{\mathrm{QF}} |\Psi|^3
\right]\Psi .
\end{eqnarray}
%
The competition between mean-field and quantum fluctuations contributions is responsible for self-trapping and droplet formation.
We assume strong confinement along the $Z$-axis, so that the nonlinear energy scale remains much smaller than the excitation energy of the transverse harmonic confinement. A sufficient condition is
$$
g n \ll \hbar \omega_Z ,
$$
where $n$ is the characteristic three-dimensional density and $\omega_Z$ is the trapping frequency in the $Z$-direction. This inequality ensures that the condensate remains close to the lowest localized mode in each well and that higher excited states in the longitudinal direction can be neglected. 

To derive an effective two-dimensional model, we employ the standard two-mode approximation~\cite{Shchesnovich2004}
\begin{equation}
\Psi(\rho,Z,T)=\Psi_L(\rho,T)\Phi_L(Z)+\Psi_R(\rho,T)\Phi_R(Z),
\label{eq:2modeAnsatz}
\end{equation}
where $\Phi_L(Z)$ and $\Phi_R(Z)$ are the localized ground-state wave functions of the left and right wells, respectively. These functions are assumed to be real, normalized, and weakly overlapping:
$$
\int_{-\infty}^{+\infty}\Phi_L^2(Z)\,dZ=
\int_{-\infty}^{+\infty}\Phi_R^2(Z)\,dZ=1,
$$
$$
\int_{-\infty}^{+\infty}\Phi_L(Z)\Phi_R(Z)\,dZ \approx 0.
$$
The functions $\Psi_L(\rho,T)$ and $\Psi_R(\rho,T)$ then describe the slowly varying in-plane dynamics in the left and right pancakes. 
Substituting ansatz~(\ref{eq:2modeAnsatz}) into Eq.~(\ref{eq:single3Dgpe}) and projecting onto $\Phi_L$ and $\Phi_R$, we arrive at the coupled quasi-two-dimensional equations
\begin{eqnarray}
\nonumber
i\hbar \partial_T \psi_L
&=
-\frac{\hbar^2}{2m_0}\nabla_\rho^2 \Psi_L
+\bar{\delta g}\,|\Psi_L|^2\Psi_L +\\
\nonumber
&\Gamma_{\mathrm{QF}}\,|\Psi_L|^3\Psi_L +E_L \Psi_L
-\bar{\kappa}\,\Psi_R,
\\ \nonumber
i\hbar \partial_T \Psi_R
&=-\frac{\hbar^2}{2m_0}\nabla_\rho^2 \Psi_R
+\bar{\delta g}\,|\Psi_R|^2\Psi_R + \\
&\Gamma_{\mathrm{QF}}\,|\Psi_R|^3\Psi_R +E_R \Psi_R
-\bar{\kappa}\,\Psi_L.
\label{eq:2D_coupled_R}
\end{eqnarray}
The terms $E_L$ and $E_R$ are the onsite energies associated with the localized states in each well, while $\bar{\kappa}$ is the linear tunneling coefficient. The sign of the coupling term is chosen so that $\bar{\kappa}>0$ corresponds to conventional Josephson-type tunneling between the two-core BEC.

The effective nonlinear coefficients are given by
\begin{equation}
\bar{\delta g}=\delta g \int_{-\infty}^{+\infty}\Phi_L^4(Z)\,dZ,
\label{eq:gbar}
\end{equation}
\begin{equation}
\Gamma_{\mathrm{QF}}= 2^{3/2} g^{3/2}\gamma_{\mathrm{QF}}
\int_{-\infty}^{+\infty}\Phi_L^5(Z)\,dZ,
\label{eq:GammaQF}
\end{equation}
while the tunneling constant is
\begin{equation}
\bar{\kappa} = -\int_{-\infty}^{+\infty}
\left[ \frac{\hbar^2}{2m_0}\,\Phi_{L,Z}\Phi_{R,Z}
+ \Phi_L(Z)\,V_{\mathrm{ext}}(Z)\,\Phi_R(Z) \right] dZ.
\label{eq:Kappa}
\end{equation}
The onsite energies are
\begin{equation}
E_j = \int_{-\infty}^{+\infty}
\left[
\frac{\hbar^2}{2m_0}\left(\partial_Z \Phi_j\right)^2
+ V_{\mathrm{ext}}(Z)\Phi_j^2(Z) \right] dZ,
\,\, j=L,R.
\label{eq:Ej}
\end{equation}
For a symmetric double well one has $E_L=E_R=E_0$. Since this common energy merely generates a global phase rotation, it can be eliminated by the transformation
$$
\Psi_j(\rho,T)=\tilde{\Psi}_j(\rho,T)\exp\left(-\frac{iE_0 T}{\hbar}\right),
\,\, j=L,R.
$$
Dropping the tildes afterwards, Eqs.~(\ref{eq:2D_coupled_R}) reduce to
\begin{eqnarray}
\nonumber 
i\hbar \partial_T \Psi_L
&=-\cfrac{\hbar^2}{2m_0}\nabla_\rho^2 \Psi_L
+\bar{\delta g}\,|\Psi_L|^2\Psi_L \\ \nonumber
& +\Gamma_{\mathrm{QF}}\,|\Psi_L|^3\Psi_L -\bar{\kappa}\,\Psi_R,
\\ \nonumber 
i\hbar \partial_T \Psi_R
&= -\cfrac{\hbar^2}{2m_0}\nabla_\rho^2 \Psi_R
+\bar{\delta g}\,|\Psi_R|^2\Psi_R \\ 
&+\Gamma_{\mathrm{QF}}\,|\Psi_R|^3\Psi_R-\bar{\kappa}\,\Psi_L .
\label{eq:reduced2D}
\end{eqnarray}
This system describes two linearly-coupled quasi-two-dimensional condensates with competing residual mean-field attraction and fluctuation-induced repulsion. It can therefore support a rich variety of symmetric, antisymmetric, and asymmetric self-trapped states, including droplet-like localized modes. In this equation, replacing $L$ and $R$ with $1$ and $2$, respectively, yields Eq.~(\ref{eq:DimGPE}).

As a concrete approximation for the localized longitudinal modes, one may choose shifted Gaussian functions,
\begin{eqnarray}
\nonumber
\Phi_L(Z)
&=&
\frac{1}{\pi^{1/4}l_z^{1/2}}
\exp\left[-\frac{(Z+Z_0)^2}{2l_z^2}\right],
\\
\Phi_R(Z)
&=&
\frac{1}{\pi^{1/4}l_z^{1/2}}
\exp\left[-\frac{(Z-Z_0)^2}{2l_z^2}\right],
\label{eq:shiftedGauss}
\end{eqnarray}
where $l_z=\sqrt{\hbar/m_0\omega_Z}$ is the harmonic-oscillator length. The parameter $2Z_0$ gives the distance between the centers of the two pancakes. In the well-separated limit $Z_0\gg l_z$, the overlap between the localized modes becomes exponentially small:
$$
\int_{-\infty}^{+\infty}\Phi_L(Z)\Phi_R(Z)\,dZ
= e^{-Z_0^2/l_z^2} \ll 1.
$$
This justifies the two-mode approximation and the neglect of nonlinear cross-overlap terms.
For the Gaussian ansatz~(\ref{eq:shiftedGauss}), one can find
\begin{equation}
\bar{\delta g}
=
\frac{\delta g}{\sqrt{2\pi}\,l_z},
\qquad
\Gamma_{\mathrm{QF}}
=
\frac{4}{\sqrt{5}}\,
\pi^{-3/4}
g^{3/2}\gamma_{\mathrm{QF}}\,l_z^{-3/2}.
\label{eq:effective_coefficients}
\end{equation}
Thus, after integrating out the confined direction, both the effective mean-field and Lee-Huang-Yang nonlinearities are renormalized by the transverse localization length $l_z$. In particular, stronger confinement enhances the effective two-dimensional nonlinear interaction strengths.

\end{document}